\documentclass[pra,aps,showpacs,twocolumn,superscriptaddress]{revtex4}
\usepackage{amsmath}
\usepackage{graphicx}

\newcommand{\ket}[1]{|#1\rangle}

\newcommand{\eq}{\begin{equation}}
\newcommand{\fine}{\end{equation}}

\newcommand{\op}[1]{\hat{#1}}

\begin{document}

\title{Wigner-function theory and decoherence of the quantum-injected optical parametric
amplifier}

\author{Nicol\`{o} Spagnolo}
\affiliation{Dipartimento di Fisica, ``Sapienza'' Universit\'{a} di Roma, piazzale 
Aldo Moro 5, I-00185 Roma, Italy}
\affiliation{Consorzio Nazionale Interuniversitario per le Scienze Fisiche della Materia,
piazzale Aldo Moro 5, I-00185 Roma, Italy}

\author{Chiara Vitelli}
\affiliation{Dipartimento di Fisica, ``Sapienza'' Universit\'{a} di Roma, piazzale 
Aldo Moro 5, I-00185 Roma, Italy}
\affiliation{Consorzio Nazionale Interuniversitario per le Scienze Fisiche della Materia,
piazzale Aldo Moro 5, I-00185 Roma, Italy}

\author{Tiziano De Angelis}
\affiliation{Dipartimento di Fisica, ``Sapienza'' Universit\'{a} di Roma, piazzale 
Aldo Moro 5, I-00185 Roma, Italy}

\author{Fabio Sciarrino}
\affiliation{Dipartimento di Fisica, ``Sapienza'' Universit\'{a} di Roma, piazzale 
Aldo Moro 5, I-00185 Roma, Italy}
\affiliation{Consorzio Nazionale Interuniversitario per le Scienze Fisiche della Materia,
piazzale Aldo Moro 5, I-00185 Roma, Italy}

\author{Francesco De Martini}
\affiliation{Dipartimento di Fisica, ``Sapienza'' Universit\'{a} di Roma, piazzale 
Aldo Moro 5, I-00185 Roma, Italy}
\affiliation{Accademia Nazionale dei Lincei, via della Lungara 10, I-00165 Roma,
Italy}

\begin{abstract}

Recent experimental results demonstrated the generation of a quantum
superpositon (MQS), involving a number of photons in excess of $%
5\times10^{4} $, which showed a high resilience to losses. In order to
perform a complete analysis on the effects of de-coherence on this
multiphoton fields, obtained through the Quantum Injected Optical Parametric
Amplifier (QIOPA), we invesigate theoretically the evolution of the Wigner
functions associated to these states in lossy conditions. Recognizing the
presence of negative regions in the W-representation as an evidence of
non-classicality, we focus our analysis on this feature. A close
comparison with the MQS based on coherent $|\alpha\rangle$ states allows to
identify differences and analogies.

\end{abstract}

\maketitle

\section{Introduction}

In the last decades the physical implementation of macroscopic quantum
superpositions (MQS) involving a large number of particles has attracted a
great deal of attention. Indeed it was generally understood that the
experimental realization of a MQS is very difficult and in several instances
practically impossible owing to the extremely short persistence of quantum
coherence, i.e., of \ the extremely rapid decoherence due to the
entanglement established between the macroscopic system and the environment 
\cite{Niel00,Zure91,Zure03,Zure07}. Formally, the irreversible decay
towards a probabilistic classical mixture is implied theoretically by the
tracing operation of the overall MQS state over the environmental variables 
\cite{Dur02,Dur04}. In the framework of quantum information different
schemes based on optical systems have been undertaken to generate and to
detect the MQS condition. A Cavity - QED scheme based on the interaction
between Rydberg atoms and a high-Q cavity has lead to the indirect
observation of Schr\"{o}dinger Cat states and of their temporal evolutions.
In this case the microwave MQS\ field stored in the cavity can be addressed
indirectly by injecting in the cavity, in a controlled way, resonant or
non-resonant atoms as ad hoc \textquotedblright measurement
mouses\textquotedblright\ \cite{Brun92,Raim01}. A different approach able to
generate freely propagating beams adopts photon-subtracted squeezed states;
experimental implementations of quantum states with an average number of
photons of around four have been reported both in the pulsed and
continuous wave regimes \cite{Neer06,Ourj06,Ourj07,Ourj09}. These states
exhibit non-gaussian characteristics and open new perspectives for quantum
computing based on continuous-variable (CV) systems, entanglement
distillation protocols \cite{Eise02,Dong08}, loophole free tests of Bell's
inequality.

In the last few years a novel \textquotedblright quantum
injected\textquotedblright\ optical parametric amplification (QI-OPA)\
process has been realized in order to establish the entanglement between a
single photon and a multiphoton state given by an average of many thousands
of photons, a Schr\"{o}dinger Cat involving a "macroscopic field". 
Precisely, in a high-gain QI-OPA "phase-covariant" cloning machine
the multiphoton fields were generated by an optical amplifier system bearing
a high nonlinear (NL)\ gain $g$ and seeded by a single-photon belonging to
an EPR entangled pair \cite{DeMa98,DeMa98a,DeMa05a,DeMa05,Scia05}.

While a first theoretical insight on the dynamical features of the QIOPA
macrostates and a thorough experimental characterization of the quantum
correlations were recently reported \cite{Naga07,DeMa08}, a complete quantum
phase-space analysis able to recognize the persistence of the QI-OPA\
properties in a decohering environment is still lacking \cite{Schl01,Wign32}. 
Among the different representation of quantum states in the
continuous-variables space \cite{Cahi69}, the Wigner quasi-probability
representation has been widely exploited as an evidence of non-classical
properties, such as squeezing \cite{Wall95} and EPR non-locality \cite%
{Bana98}. In particular, the presence of negative quasi-probability regions
has been considered as a consequence of the quantum superposition of
distinct physical states \cite{Bart44}.

In the present paper we investigate the Wigner functions associated to
multi-photon states generated by optical parametric amplification of
microscopic single photon states. We focus our interest on the effects of
de-coherence on the macro-states and on the emergence of the "classical"
regime in the amplification of initially pure quantum states. The Wigner
functions of these QI-OPA\ generated states in presence of losses are
analyzed in comparison with the paradigmatic example of the superposition of
coherent, Glauber's states, $|\alpha \rangle $.

The paper is structured as follows. In Section II, we introduce the
conceptual scheme and describe the evolution of the system both in the
Heisenberg and Schr\"{o}dinger picture. Section III\ is devoted to the
calculation of the Wigner function of the QI-OPA amplified field. We first
consider a single-mode amplifier, which is analogous to the case of
photon-subtracted squeezed vacuum. Then we derive a compact expression of
the Wigner function in the case of a two-mode amplifier in the
"collinear" case, i.e. for common $k-$vectors of the amplified output
fields. In Section IV, we introduce, for the collinear case, a
decoherence model apt to simulate the de-cohering losses affecting the
evolution of the macrostates density matrix. This evolution is then compared
to the case of the coherent $|\alpha \rangle $ MQS. Section V is devoted to
a brief review of the features of coherent states superpositions (CSS).
Hence in Section VI we derive an explicit analytic expression of the Wigner
functions for the QI-OPA amplified states in presence of decoherence. The
negative value of the W-representation in the origin of the phase-space for
such states above a certain value of the \textquotedblright 
system-enviroment \textquotedblright\ interaction parameter, is an evidence 
of the persistance of quantum properties in presence of decoherence. 
This aspect is then compared with the
case of the $|\alpha \rangle $ states superposition. Then in Section VII we study a
complementary approach to enlighten the resilience to losses of the QI-OPA MQS
with respect to the $|\alpha \rangle $ states superposition. Precisely, we
define a coherence parameter, based on the concept of state-distance in
Hilbert spaces, and we study its decrease as a function of the losses
parameter for both classes of MQS. Finally, in Section VIII we analyze a different
configuration based on a non-collinear optical parametric amplifier in the
quantum injected regime, i.e. a \emph{universal quantum cloning machine}. We calculate
the Wigner function associated to the states generated by this device in absence
and in presence of decoherence, focusing on the persistence of quantum
properties after the propagation over a lossy channel.

\section{Optical parametric amplification of a single photon state}

\begin{widetext}

\begin{figure}[ht]
\centering
\includegraphics[width=0.7\textwidth]{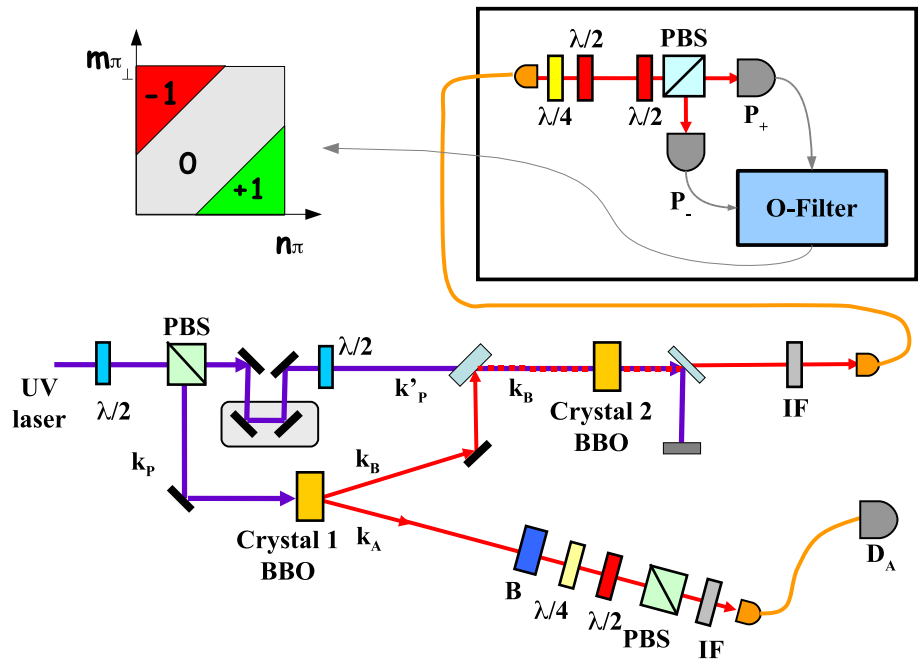}
\caption{Scheme of the experimental setup. The main UV laser beam 
provides the OPA excitation field beam at $\lambda = 397.5 nm$. A type II
BBO (Beta Barium Borate) crystal (crystal 1: C1) generates pair of photons with $\lambda = 795 nm$. 
In virtue of the EPR non-local correlations established between the modes 
$\mathbf{k}_{A}$ and $\mathbf{k}_{B}$, the preparation of a single-photon on 
mode $\mathbf{k}_{B}$ with  polarization state $\vec{\pi}_{\varphi}$ is conditionally 
determined by detecting a single-photon  after proper polarization analysis on 
the mode $\mathbf{k}_{A}$ (polarizing beamsplitter (PBS), $\lambda/2$ and 
$\lambda/4$ waveplates, Soleil-Babinet compensator (B), interferential filter (IF)). 
The photon belonging to $\mathbf{k}_{B}$, together with the pump laser beam 
$\mathbf{k}_{p}'$,  is fed into an high gain optical parametric amplifier consisting 
of a NL crystal 2 (C2), cut for collinear type-II phase matching. The fields are 
coupled to single-mode fibers. For more details refer to \cite{DeMa08}.
\textbf{Right Inset}: Measurement apparatus. After fiber polarization compensation,
the field is analyzed by two photomultipliers $\left\{ P_{+}, P_{-} \right\}$ and then
discriminated through an O-Filter device.
\textbf{Left inset}: Action of the O-Filter in the photon number space: the ($\pm 1$) 
outcomes are assigned whether $n_{\pi} - m_{\pi_{\bot}}>k$ or $m_{\pi_{\bot}} - n_{\pi}>k$, 
where $k$ is a tunable threshold condition. The central region leads to an inconclusive (0) 
outcome, and the two orthogonal macrostates cannot be discriminated.}
\label{fig:fig22}
\end{figure}

\end{widetext}

As a first step we consider the generation of a multiphoton quantum field,
obtained by parametric amplification. Let us briefly describe the conceptual
scheme. An entangled pair of two photons in the singlet state $\left\vert
\psi ^{-}\right\rangle _{A,B}$=$2^{-{\frac{1}{2}}}\left( \left\vert
H\right\rangle _{A}\left\vert V\right\rangle _{B}-\left\vert V\right\rangle
_{A}\left\vert H\right\rangle _{B}\right) \ $is produced through a
spontaneous parametric down-conversion (SPDC)\ by crystal 1 pumped by a
pulsed UV pump beam: Fig.\ref{fig:fig22}. There $\left\vert H\right\rangle $
and $\left\vert V\right\rangle $ stands, respectively, for a single photon
with horizontal and vertical polarization $(\overrightarrow{\pi })$ while
the labels $A,B$ refer to particles associated respectively with the spatial
modes $\mathbf{k}_{A}$and $\mathbf{k}_{B}$. The photon
belonging to $\mathbf{k}_{B}$, together with a strong UV pump beam, is fed
into an optical parametric amplifier consisting of a non-linear crystal 2
pumped by the beam $\mathbf{k}_{P}^{\prime }$. The crystal 2 is oriented for
\textquotedblright collinear operation\textquotedblright , i.e., emitting
pairs of amplified photons over the same spatial mode which supports two
orthogonal $\overrightarrow{\pi }$ modes, respectively horizontal and
vertical. Let us analyze the properties of the resultant amplified field.

\subsection{Collinear optical parametric amplifier}

The complete Hamiltonian of the system reads: {\small 
\begin{equation}
\hat{\mathcal{H}}=\hbar \omega \left( \hat{a}_{H}^{\dagger }\hat{a}_{H}+\hat{%
a}_{V}^{\dagger }\hat{a}_{V}\right) +i\hbar \chi \left( \hat{a}_{H}^{\dagger
}\hat{a}_{V}^{\dagger }e^{-2i\omega t}-\hat{a}_{H}\hat{a}_{V}e^{2i\omega
t}\right)  \label{hamcoll1}
\end{equation}%
}\noindent where $\chi $ is proportional to the 2$^{nd}$ order non-linear
susceptibility and to the amplitude of the pump field $E_{0}^{P}$; $\hat{a}%
_{H}^{\dagger }$ and $\hat{a}_{V}^{\dagger }$ are the creation operators
associated to the mode $\mathbf{k}$ respectively, with polarization $\overrightarrow{\pi }
$ horizontal and vertical, ${H,V}$. We assume a classical and undepleted
coherent pump field. Let us consider only the interaction contribution to $%
\hat{\mathcal{H}}$ and a reference system which rotates with angular speed $%
2\omega $. The time-independent interaction Hamiltonian is found to be:
\begin{equation}
\hat{\mathcal{H}}_{I}=i\hbar \chi \left( \hat{a}_{H}^{\dagger }\hat{a}%
_{V}^{\dagger }-\hat{a}_{H}\hat{a}_{V}\right)  \label{ham-coll-int}
\end{equation}%
The Heisenberg evolution equations are:
\begin{equation}
\frac{\partial \hat{a}_{\pi }}{\partial t}=\frac{1}{i\hbar }\left[ \hat{a}%
_{\pi },\hat{\mathcal{H}}_{I}\right] =\chi \hat{a}_{\pi ^{\perp }}^{\dagger }
\label{eq_heisenberg}
\end{equation}%
with general solution:
\begin{equation}
\hat{a}_{\pi }\left( t\right) =\hat{a}_{\pi }\left( 0\right) \cosh \left(
\chi t\right) +\hat{a}_{\pi ^{\perp }}^{\dagger }\left( 0\right) \sinh
\left( \chi t\right)  \label{eq:OPA_coll_Heisen}
\end{equation}%
where $\pi ={H,V}$.

The average number of photons generated in the SPDC\ process can be easily
calculated in the Heisenberg picture formalism by applying the evoluted
operators to the initial vacuum state $|0H,0V\rangle \equiv |0\rangle $.
Heretofore, $|n\pi ,m\pi _{\bot }\rangle $ stands for the Fock-state with $n$
photons with $\pi $ polarization and $m$ photons with the orthogonal $\pi
_{\bot }$ one. For the horizontal polarization we get:
\begin{equation}
\langle 0|\hat{a}_{H}^{\dagger }\left( t\right) \hat{a}_{H}\left( t\right)
|0\rangle =\sinh ^{2}\left( \chi t\right) \equiv \overline{m}
\end{equation}%
The same result holds for the vertical polarization.

\subsection{Output wavefunctions}

\label{teoria_QUIOPA}In any "equatorial" polarization basis $\{\left\vert
\varphi \right\rangle ,\left\vert \varphi ^{\perp }\right\rangle \}$, referred to
a Poincar\'{e} sphere with "poles" $\left\vert H\right\rangle $ and $%
\left\vert V\right\rangle $, the Hamiltonian of the polarization
non-degenerate optical parametric amplifier, described by the expression (%
\ref{ham-coll-int}), can be expressed as: 
\begin{equation}
\hat{\mathcal{H}}_{I}=i\hbar \frac{\chi }{2} e^{-\imath \varphi} \left( (\hat{a}_{\varphi }^{\dagger
})^{2}- e^{\imath 2 \varphi} (\hat{a}_{\varphi \perp }^{\dagger })^{2}\right) +h.c.
\label{OPA_nondeg_pm}
\end{equation}%
where the corresponding field operators are: $\hat{a}_{\varphi }^{\dagger }=%
\frac{1}{\sqrt{2}}\left( \hat{a}_{H}^{\dagger }+e^{i\varphi }\hat{a}%
_{V}^{\dagger }\right) $, $\hat{a}_{\varphi \perp }^{\dagger }=\frac{1}{\sqrt{2}%
}\left( \hat{a}_{H}^{\dagger }-e^{i\varphi }\hat{a}_{V}^{\dagger }\right) $.

The Hamiltonian can then be separated in the two polarization components $\{%
\vec{\pi} _{\varphi },\vec{\pi} _{\varphi ^{\perp }}\}$. The time-dependent field
operators are:
\begin{equation}
\hat{a}_{\varphi }(t)=\hat{a}_{\varphi }(0)\cosh (\chi t) + e^{- \imath \varphi} a_{\varphi }^{\dagger
}(0)\sinh (\chi t)  \label{heis_time_ev}
\end{equation}
Let us now restrict the analysis to the basis $\left\{ \vec{\pi}_{+},\vec{\pi%
}_{-}\right\} $, in which we calculate the output wavefunction in the
Schroedinger picture; the unitary evolution operator is 
\begin{equation}
\hat{U}=\exp \left( -i\frac{\hat{\mathcal{H}}_{I}t}{\hbar }\right) =\hat{U}%
_{+}\hat{U}_{-}
\end{equation}
with 
\begin{equation}
\hat{U}_{\pm }=\exp [\pm \frac{g}{2}\left( \hat{a}_{\pm }^{\dagger 2}-\hat{a}%
_{\pm }^{2}\right) ]
\end{equation}
where $g=\chi t$ is the non linear gain of the process.

The expression of $\hat{U}$ enlightens the decoupling between the two
polarization modes. A simple expression of the operators $\hat{U}_{+}$ and $%
\hat{U}_{-}$ can be obtained adopting the following operatorial relation
\cite{Coll88}: {\small 
\begin{eqnarray}
\hat{U}_{\pm } &=&\exp \left[ \pm \Gamma \left( \frac{\hat{a}_{\pm
}^{\dagger }}{\sqrt{2}}\right) ^{2}\right] \exp \left[ -\ln (\cosh g)\left( 
\hat{a}_{\pm }^{\dagger }\widehat{a}_{\pm }+\frac{1}{2}\right) \right] \times
\notag \\
&\;&\times \exp \left[ \mp \Gamma \left( \frac{\hat{a}_{\pm }}{\sqrt{2}}%
\right) ^{2}\right]
\end{eqnarray}
}with $\Gamma =\tanh g$. Let us consider the injection of a single photon
state with generic equatorial polarization $\vec{\pi}_{\varphi}$ into the
OPA:{\small 
\begin{eqnarray}
\left| \psi _{in}\right\rangle &=& \vert \varphi \rangle = \frac{1}{\sqrt{2}}\left( \left|
H\right\rangle +e^{i\varphi }\left| V\right\rangle \right)  \notag = \\
&=&e^{i\frac{\varphi }{2}}\left( \cos \frac{\varphi }{2}\left|
1+,0-\right\rangle +i\sin \frac{\varphi }{2}\left| 0+,1-\right\rangle \right)
\label{psi_in}
\end{eqnarray}
}The multiphoton output state of the amplifier is found: 
\begin{equation}
\left| \Psi _{out}\right\rangle = \vert \Phi^{\varphi} \rangle =\cos \frac{%
\varphi }{2}\left| \Phi ^{+}\right\rangle +i\sin \frac{\varphi }{2}\left|
\Phi ^{-}\right\rangle  \label{psi_out}
\end{equation}
where: 
\begin{equation}
\left| \Phi ^{\pm }\right\rangle =\hat{U}_{\pm }\hat{U}_{\mp }\left|
1\pm,0\mp\right\rangle =\left( \hat{U}_{\pm }\left| 1\right\rangle _{\pm
}\right) \left( \hat{U}_{\mp }\left| 0\right\rangle _{\mp }\right)
\end{equation}
Simple calculations lead to the expressions: {\small 
\begin{equation}
\left| \Phi ^{\pm }\right\rangle =\frac{1}{C^{2}}\sum_{i,j}\left( -\frac{%
\Gamma }{2}\right) ^{j}\left( \frac{\Gamma }{2}\right) ^{i}\frac{\sqrt{2j!}}{%
j!}\frac{\sqrt{(2i+1)!}}{i!}\left| (2i+1)\pm ,2j\mp \right\rangle
\label{psi+}
\end{equation}
}where $C=\cosh g$. The quantum states $\left| \Phi ^{+}\right\rangle $ and $%
\left| \Phi ^{-}\right\rangle $ are orthogonal, being the unitary evolution
of initial orthogonal states. We observe that $\left| \Phi ^{+}\right\rangle 
$ presents an odd number of $\vec{\pi}_{+}$ polarized photons and an even
number of $\vec{\pi}_{-}$ polarized ones. Conversely for $\left| \Phi
^{-}\right\rangle $.

The average number of photons with polarization $\vec{\pi}_{+}$ can be
estimated in the Heisenberg representation as: 
\begin{equation}
\left\langle \hat{n}_{+}\right\rangle =\left\langle \psi _{in}\right| \hat{a}%
_{+}^{\dagger }(t)\hat{a}_{+}(t)\left| \psi _{in}\right\rangle =\bar{m}%
+\left( 2\bar{m}+1\right) \cos ^{2}\left( \frac{\varphi }{2}\right)
\label{npiu_med}
\end{equation}
where $\bar{m}=\sinh^{2}\left( g\right) $. The phase dependence shows that
for $\varphi =\pi $ the average number of generated photons is equal to the
spontaneous one, while for $\varphi =0$, corresponding to the stimulated
case, an increase of a factor $3$ is observed in the average number of
$\vec{\pi}_{+}$ polarized photons.

Likewise, the average number of photons $\vec{\pi}_{-}$ polarized is given
by: 
\begin{equation}
\left\langle \hat{n}_{-}\right\rangle =\left\langle \psi _{in}\right\vert 
\hat{a}_{-}^{\dagger }(t)\hat{a}_{-}(t)\left\vert \psi _{in}\right\rangle =%
\bar{m}+\left( 2\bar{m}+1\right) \sin ^{2}(\frac{\varphi }{2})
\end{equation}%
Hence, by varying the phase $\varphi $ we observe a fringe pattern which
exhibits a gain dependent visibility: 
\begin{equation}
\emph{V}_{th}^{(1)}=\frac{\left\langle n_{+}\left( \varphi =0\right)
\right\rangle -\left\langle n_{-}\left( \varphi =0\right) \right\rangle }{%
\left\langle n_{+}\left( \varphi =0\right) \right\rangle +\left\langle
n_{-}\left( \varphi =0\right) \right\rangle }=\frac{2\bar{m}+1}{4\bar{m}+1}
\label{visi_th}
\end{equation}%
In the asymptotic limit ($g\rightarrow \infty $) $\emph{V}_{th}^{(1)}=\frac{%
1}{2}$.

%

We now briefly discuss the \emph{phase-covariance} properties of the optical
parametric amplifier when injected by a single-photon state. In this
configuration, this device acts as \emph{optimal phase-covariant cloning
machines}, and due to the unitary evolution of the process, the superposition
character of any generic input state $\vert \varphi \rangle = \frac{1}{\sqrt{%
2}} \left( \vert H \rangle + e^{\imath \varphi} \vert V \rangle \right)$ is
maintained after the amplification: 
\begin{equation}
\vert \Phi^{\varphi} \rangle = \frac{1}{\sqrt{2}} \left( \vert \Phi^{H}
\rangle + e^{\imath \varphi} \vert \Phi^{V} \rangle \right)
\end{equation}
Let us stress that there is freedom in the choice of the macrostates basis
vectors. For example, the $\vert \Phi^{\pm} \rangle$ states can be used in
the expansion of the overall state as in (\ref{psi_out}).
Hence, the equatorial $\vert \Phi^{\varphi} \rangle$ amplified state can be
written as the macroscopic quantum superposition either of the $\left\{
\vert \Phi^{+} \rangle, \vert \Phi^{-} \rangle \right\}$ and the $\left\{
\vert \Phi^{H} \rangle, \vert \Phi^{V} \rangle \right\}$ "basis" states. Furthermore,
due to their resilience to decoherence \cite{DeMa09},  all equatorial
macro-qubits represent a preferred "pointer states basis" \cite{Zure03} for writing $%
\vert \Phi^{\varphi} \rangle$ in the form of a macroscopic quantum
superposition.

\section{Wigner functions of the amplified field}
\label{sec:wigner_no_losses}

In order to investigate the properties of the output field of the QI-OPA device in
more details, we analyze the quasi-probability distribution introduced by
Wigner \cite{Wign32} for the amplified field. The \textit{Wigner function}
is defined as the Fourier transform of the \textit{symmetrically-ordered }%
characteristic function $\chi (\eta )$ of the state described by the general
density matrix $\hat{\rho}$ 
\begin{equation}
\chi (\eta )=Tr\left[ \hat{\rho}\exp \left( \eta \hat{a}^{\dagger }-\eta
^{\ast }\hat{a}\right) \right]  \label{funz_caratt}
\end{equation}%
The associated\textit{\ Wigner function} 
\begin{equation}
W(\alpha )=\frac{1}{\pi ^{2}}\int \exp \left( \eta ^{\ast }\alpha -\eta
\alpha ^{\ast }\right) \chi \left( \eta \right) d^{2}\eta  \label{deffunzwig}
\end{equation}%
exists for any $\hat{\rho}$ but is not always positive definite and,
consequently, can not be considered as a genuine probability distribution.
Since the early decades of quantum mechanics, the emergence of negative
probabilities has been identified as a peculiar feature of quantum physics,
and has been connected to the mathematical segregation of
states which physically live only in combination \cite{Bart44}. In parallel, the non
classicality of a quantum state is expressed by a Glauber's P-representation 
\cite{Glau63,Suda63,Cahi69} which is more singular than a delta function, i.e.
the $P(\alpha)$ proper of coherent states. This means that the system does not possess an
expansion, in terms of the overcomplete semi-classical $|\alpha \rangle $
state basis, that can be interpreted as a probability distribution. However,
the negativity of the Wigner function is not the
only parameter that allows to estimate the non-classicality of a certain
state. For example, the squeezed vacuum state \cite{Wall95} presents a
positive $W$-representation, while its properties cannot be described by
the laws of classical physics. Furthermore, recent papers have shown that
the Wigner function of an EPR state provides direct evidence of its
non-local character \cite{Cohe97,Bana98}, while being completely-positive in
all the phase-space.

The complex variable $\alpha $ in (\ref{deffunzwig}) is the eigenvalue
associated to the non-Hermitian operator $\hat{a}$ which acts upon the
coherent state $|\alpha \rangle $ as follows: $\hat{a}|\alpha \rangle
=\alpha |\alpha \rangle $. It is possible to decompose $\alpha $ in its real
and imaginary part and then to define the quadratures operators $\hat{X}$
and $\hat{Y}$ which allow the representation of the field in the phase space 
$\hat{a}=\frac{\hat{X}+i\hat{Y}}{2}$. The quadratures operators are Hermitian 
operators and thus correspond to
physical observables proportional to the position $\hat{q}$ and the momentum $\hat{p}
$ following the relations: $\hat{X}=\sqrt{\frac{2\omega }{\hbar }}\hat{q}$
and $\hat{Y}=-\sqrt{\frac{2}{\hbar \omega }}\hat{p}.$ The uncertainty
principle leads to: $\Delta \hat{X}\Delta \hat{Y}\geq 1$.


\subsection{Single mode amplifier}
\label{subs:OPA singolo modo}

For the sake of simplicity, let us first consider the Hamiltonian of a
degenerate amplifier acting on a single $k-$mode with polarization $\vec{\pi}_{+}$:
\begin{equation}
\hat{\mathcal{H}}_{I}=i\hbar \frac{\chi }{2}\left( \hat{a}_{+}^{\dagger 2}-%
\hat{a}_{+}^{2}\right)  \label{eq:Ham_single_squeeze}
\end{equation}%
When no seed is injected, the amplifier operates in the regime of
spontaneous emission and the characteristic function reads:
\begin{eqnarray}
\chi _{0}(\eta ,t) &=&\left\langle 0\right\vert \exp [\eta \hat{a}%
_{+}^{\dagger }(t)-\eta ^{\ast }\hat{a}_{+}(t)]\left\vert 0\right\rangle = 
\notag \\
&=&\left\langle 0\right\vert \exp [\eta (t)\hat{a}_{+}^{\dagger }-\eta
^{\ast }(t)\hat{a}_{+}]\left\vert 0\right\rangle
\end{eqnarray}%
with $\hat{a}_{+}(t)=\hat{a}_{+}\cosh (g)+\hat{a}_{+}^{\dagger }\sinh (g)$,
$\eta (t)=\eta \cosh (g)-\eta ^{\ast }\sinh (g)$, and $g=\chi t$. Hereafter,
we explicitly report the dependence of the Wigner function from the interaction 
time $t$. We obtain, using the operatorial relation 
$\exp (\hat{A}+\hat{B})=\exp {\hat{A}}\exp {\hat{B}} \exp (-1/2[\hat{A},\hat{B}])$: 
\begin{equation}
\chi _{0}(\eta ,t)=\exp \left( -\frac{1}{2}\left\vert \eta (t)\right\vert
^{2}\right)
\end{equation}%
The calculation then proceeds as follows. Starting from the definition of the Wigner
function (\ref{deffunzwig}), we perform the two subsequent transformations 
of the integration variables $d^{2}\eta \rightarrow d^{2}\eta (t) \rightarrow xdxd\varphi$,
where $\eta \rightarrow \eta(t)$ has been defined previously and can be expressed as
$\eta (t)=xe^{i\varphi}$. The Wigner function then reads:
%
\begin{equation}
\label{fwigspont}
\begin{aligned}
W_{\vert 0+ \rangle}(\alpha,t )&=\frac{1}{\pi ^{2}}\int e^{x \left[ \left( 
\bar{\alpha}-\bar{\alpha}^{\ast }\right) \cos \varphi +i \left( -\bar{\alpha}%
-\bar{\alpha}^{\ast }\right) \sin \varphi \right] -\frac{1}{2} x^{2}}xdxd\varphi \\
&= \frac{2}{\pi}\int_{0}^{\infty }J_{0}\left( -2\left| \bar{\alpha}\right| x\right) \exp
\left( -\frac{1}{2}x^{2}\right) xdx=\\ &= \frac{2}{\pi }\exp \left[ -2\left|
\bar{\alpha}\right| ^{2}\right]
\end{aligned}
\end{equation}
where $\bar{\alpha}=\alpha \cosh (g)-\alpha ^{\ast }\sinh (g)$, and $J_{0}(x)$ is 
the Bessel function of order 0.
We can now write $\bar{\alpha}=\mathrm{Re}(\alpha )e^{-g}+i\mathrm{Im}(\alpha )e^{g}$ 
as a function of the $X$, $Y$ quadrature operators, defined by the expression
$\alpha =X+iY$. By substituting such variables Eq.(\ref{fwigspont}) becomes 
\begin{equation}
W_{\vert 0+ \rangle}(X,Y,t)=\frac{2}{\pi }\exp \left[ -2\left(
X^{2}e^{-2g}+Y^{2}e^{2g}\right) \right]  \label{wignerspont}
\end{equation}

When we consider the case in which a single photon with polarization $\vec{%
\pi}_{+}$ is injected: $\left| \psi _{in}\right\rangle =\left|
1+\right\rangle$, analogous calculations leads to the characteristic
function 
\begin{eqnarray}
\chi_{1} (\eta ,t) &=&\left\langle 1\right| \exp [\eta (t)\hat{a}%
_{+}^{\dagger }-\eta ^{\ast }(t)\hat{a}_{+}]\left| 1\right\rangle =  \notag
\\
&=&\left( 1-\left| \eta (t)\right| ^{2}\right) \exp \left( -\frac{1}{2}%
\left| \eta (t)\right| ^{2}\right)
\end{eqnarray}
The Wigner function reads: 
\begin{eqnarray}
W_{\vert 1+ \rangle}(X,Y,t) &=&-\frac{2}{\pi }\left[ 1-4\left| \bar{\alpha}%
\right| ^{2}\right] \exp \left[ -2\left| \bar{\alpha}\right| ^{2}\right] = 
\notag \\
&=&-\frac{2}{\pi }\left[ 1-4\left( X^{2}e^{-2g}+Y^{2}e^{2g}\right) \right]
\times  \notag \\
&\;&\times \exp \left[ -2(X^{2}e^{-2g}+Y^{2}e^{2g})\right]
\label{wigner1inj}
\end{eqnarray}

\begin{figure}[tbp]
\includegraphics[width=0.45\textwidth]{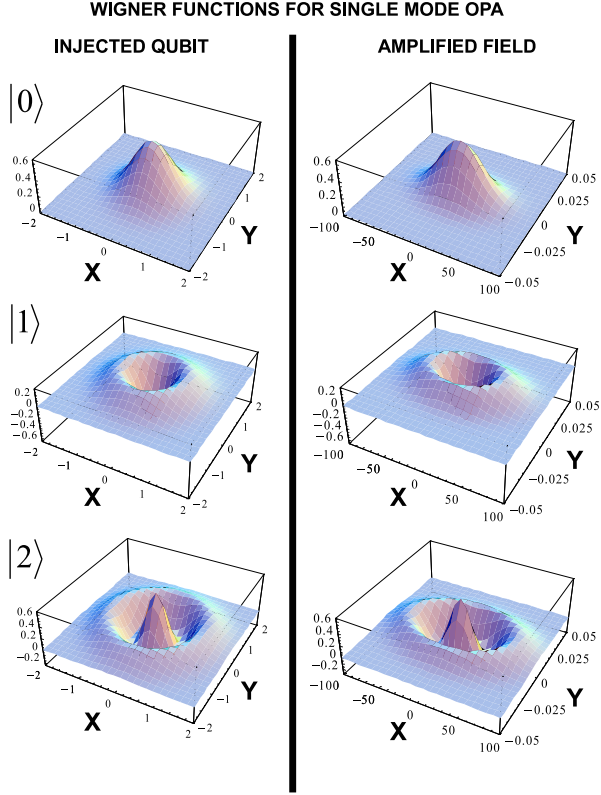}
\caption{Wigner function of the injected states and the corresponding amplified
states, for a single mode degenerate amplifier with gain $g=4$. The axes report 
the quadratures values $X$ and $Y $. $\left| 0\right\rangle $ injection of the 
vacuum state. $\left| 1\right\rangle $ injection of a single photon. 
$\left| 2\right\rangle $ injection of two photons. The amplified field is plotted 
on a different scale with respect to the injected qubit one, due to the high degree 
of squeezing introduced by the amplification process.}
\label{wigner1}
\end{figure}

\begin{figure}[tb]
\begin{center}
\includegraphics[width=0.45\textwidth]{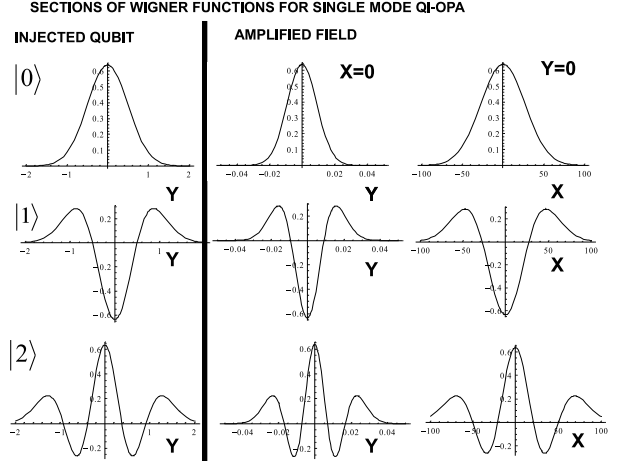}
\end{center}
\caption{Sections of the input states Wigner functions and of the relatives
amplified fields along the directions $X=0$ and $Y=0$. Since the injected
qubit shows a rotational symmetry, we report only the $X=0$ section. $\left|
0\right\rangle $ injection of the vacuum state. $\left| 1\right\rangle $
injection of the single photon state. $\left| 2\right\rangle $ injection of
the two-photon state.}
\label{wigner2}
\end{figure}

As a further example, we consider the injection of the 2-photon state $%
\left| \psi_{in}\right\rangle =\left| 2+ \right\rangle$. We obtain{\small 
\begin{equation}
\chi_{2} (\eta ,t) = \left( 1-2 \left| \eta (t) \right|^{2}+\frac{1}{2} \left|
\eta (t) \right|^{4} \right) \exp \left( -\frac{1}{2}\left| \eta (t)\right| ^{2}\right)
\end{equation}
}and the Wigner function reads:{\small 
\begin{eqnarray}
&&W_{\vert 2+ \rangle}(X,Y,t)=\frac{2}{\pi }\left[ 1-8\left(
X^{2}e^{-2g}+Y^{2}e^{2g}\right) \right.  \notag \\
&\;&\left. +2\left( X^{2}e^{-2g}+Y^{2}e^{2g}\right)^{2} \right] \exp \left[
-2(X^{2}e^{-2g}+Y^{2}e^{2g})\right]  \notag
\end{eqnarray}
}As in the single photon case, the negativity of the Wigner function is
maintained after the amplification process, emphasizing the quantum
properties of the injected field. All these features are shown in
Figs. \ref{wigner1} and \ref{wigner2}, which report respectively the plots of the 
Wigner functions and of their sections $X=0$ and $Y=0$ for QI-OPA amplified
states, with the injection of the $\vert 0 \rangle, \vert 1 \rangle, \vert 2 \rangle$
Fock states.
The analysis of these figures shows that the amplification effect leads 
to the increase of the degree of squeezing for the multi-photon 
output field. The uncertainty on one of the two quadratures is 
decreased while the other one is increased coherently with the \textit{%
Heisenberg uncertainty principle}. Let us note that the quantum character of
the Fock states is underlined by the negativity of the Wigner function in
the central region of the quadratures space.

Finally, this result can be generalized by analogous calculation to the
generic $\vert N+ \rangle$ input state, leading to the Wigner function: 
\begin{equation}
W_{\vert N+ \rangle}(\alpha,t) = \frac{2}{\pi} (-1)^{N} L_{N} \left( 4 \vert 
\overline{\alpha} \vert^{2} \right) e^{-2 \vert \overline{\alpha} \vert^{2}}
\end{equation}
where $L_{N}$ are Laguerre polynomials of order $N$. For all $N$, the
non-classical properties of the injected state are maintained after the
amplification process, as the OPA is described by a unitary evolution
operator.

We conclude this Section on the single-mode amplifier by emphasizing the
connection, shown in \cite{Bisw07}, between the single photon subtracted
squeezed vacuum, and the squeezed single-photon state. It is found that: $%
\hat{a} \hat{S}(\xi) \vert 0 \rangle = \hat{S}(\xi) \vert 1 \rangle$, where $%
\hat{S}(\xi) = \exp \left[\xi \frac{\hat{a}^{\dag \, 2}}{2} -
\xi^{\ast} \frac{\hat{a}^{2}}{2}\right]$, is the single-mode degenerate
squeezing operator, and $\xi = s e^{\imath \theta}$ is the complex squeezing
parameter. This evolution operator is obtained by the single-mode amplifier
interaction Hamiltonian of Eq. (\ref{eq:Ham_single_squeeze}). For small $\xi$%
, this state possess a high value of overlap \cite{Lund04} with the $\left( \vert
\alpha \rangle - \vert - \alpha \rangle \right)$ quantum superposition. This
connection between photon-subtracted squeezed vacuum and squeezed
Fock-states was extended \cite{Bisw07} to the more general p-photon case,
obtaining: 
\begin{equation}
\hat{a}^{p} \hat{S}(\xi) \vert 0 \rangle = \hat{S}(\xi) \vert \psi_{p}
\rangle
\end{equation}
where: 
\begin{equation}
\begin{aligned} \vert \psi_{p} \rangle &= \mathcal{N} \sum_{k=0}^{\left[
\frac{p}{2} \right]} \frac{p! (-1)^{k}}{2^{k} k!\sqrt{(p-2k)!}} \times\\
&\times \left( e^{\imath \theta} \sinh s \, \cosh s\right)^{k} \vert p-2k
\rangle \end{aligned}  \label{eq:photon_subtracted_seed}
\end{equation}
where $\mathcal{N}$ is an opportune normalization constant. Hence, the $p$%
-photon subtracted squeezed vacuum is analogous to the amplified state of a
quantum superposition of odd (or even) Fock-states (\ref%
{eq:photon_subtracted_seed}). For an increasing value of $p$, the overlap
between the $\hat{a}^{p} \hat{S}(\xi) \vert 0 \rangle$ states and the $\left( \vert
\alpha \rangle \pm \vert - \alpha \rangle \right)$ states \cite{Schl91} is
progressively higher, but corresponds to the amplification of a more
sophisticated superposition of Fock-states.

\subsection{Two modes amplifier}

\label{subsec:two_mode_amplifier}

In order to investigate the collinear QI-OPA we have
to analyze the non-degenerate OPA Hamiltonian (\ref{hamcoll1}), i.e. acting
on the both orthogonal polarization modes $\vec{\pi}_{H}$ and $\vec{\pi}_{V}$%
. For a given input state in the amplifier $\left\vert \psi
_{in}\right\rangle $ the characteristic \textit{symmetrically-ordered}
function can be written as: 
\begin{equation}
\chi _{\psi }(\eta ,\xi ,t)=\left\langle \psi _{in}\right\vert e^{(\eta \hat{%
a}_{H}^{\dagger }(t)-\eta ^{\ast }\hat{a}_{H}(t))}e^{(\xi \hat{a}%
_{V}^{\dagger }(t)-\xi ^{\ast }\hat{a}_{V}(t))}\left\vert \psi
_{in}\right\rangle  \label{funz_caratt2}
\end{equation}%
where the time dependent operators solve the Heisenberg equation of motion 
(\ref{heis_time_ev}) and are expressed in the basis $\left\{ \vec{\pi}_{H}, 
\vec{\pi}_{V} \right\}$. 

It is useful to rewrite the expression (\ref{funz_caratt2}) by using the $%
\left\{ \vec{\pi}_{+},\vec{\pi}_{-}\right\} $ polarization basis. We obtain: 
\begin{eqnarray}
\chi_{\psi} (\eta ,\xi ,t) &=&\left\langle \psi _{in}\right| \exp \left(
\eta (t)\frac{\hat{a}_{+}^{\dagger }}{\sqrt{2}}-\eta ^{\ast }(t)\frac{\hat{a}%
_{+}}{\sqrt{2}}\right) \times  \notag \\
&\;&\times \exp \left( \xi (t)\frac{\hat{a}_{-}^{\dagger }}{\sqrt{2}}-\xi
^{\ast }(t)\frac{\hat{a}_{-}}{\sqrt{2}}\right) \left| \psi _{in}\right\rangle
\end{eqnarray}
where 
\begin{eqnarray}
\label{eq:trans_wigner_char_1}
\eta (t) &=&(\eta +\xi )C-(\eta ^{\ast }+\xi ^{\ast })S \\
\label{eq:trans_wigner_char_2}
\xi (t) &=&(\eta -\xi )C+(\eta ^{\ast }-\xi ^{\ast })S 
\end{eqnarray}
and $C=\cosh (g)$, $S=\sinh (g)$.

Let us consider the case of single photon injection on the polarization mode 
$\vec{\pi}_{+}$, the input state wavefunction can be written as $\left| \psi
_{in}\right\rangle =\left| 1+,0-\right\rangle $. Hence the characteristic
function is 
\begin{equation}
\chi_{1,0} (\eta ,\xi ,t)=\left( 1-\frac{\left| \eta (t)\right| ^{2}}{2}%
\right) \exp [-\frac{1}{4}(\left| \eta (t)\right| ^{2}+\left| \xi (t)\right|
^{2})]
\end{equation}

In this case the Wigner function is the quadri-dimensional Fourier transform
of the characteristic function given by \cite{DeMa98a}: 
\begin{eqnarray}
&&W_{\vert 1+,0- \rangle}(\alpha ,\beta ,t)=\frac{1}{\pi ^{4}}\int d^{2}\eta
d^{2}\xi e^{(\eta ^{\ast }\alpha -\eta \alpha ^{\ast })} e^{(\xi ^{\ast
}\beta -\xi \beta ^{\ast })}  \notag \\
&&\exp [-\frac{1}{4}(\left| \eta (t)\right| ^{2}+\left| \xi (t)\right|
^{2})]\left( 1-\frac{\left| \eta (t)\right| ^{2}}{2}\right) =  \notag \\
&=&-\left( \frac{2}{\pi }\right) ^{2}\left[ \exp \left( -\left| \Delta
\right| ^{2}\right) \right] \left( 1-\left| \Delta _{A}+\Delta _{B}\right|
^{2}\right)  \label{funz_wig_opa_nondeg}
\end{eqnarray}
where we have used: 
\begin{eqnarray}
\left| \Delta \right| ^{2} &=&\frac{1}{2}\left[ \left| \gamma _{A+}\right|
^{2}+\left| \gamma _{A-}\right| ^{2}+\left| \gamma _{B+}\right| ^{2}+\left|
\gamma _{B-}\right| ^{2}\right] \\
\Delta _{K} &=&\frac{1}{\sqrt{2}}\left( \gamma _{K+}-i\gamma _{K-}\right) 
\end{eqnarray}
with $K = A,B$.

In this case the squeezing variables are $\gamma _{A+}$ and $\gamma _{B-}$
while $\gamma _{B+}$ and $\gamma _{A-}$ are the respective conjugated
variables: 
\begin{eqnarray}
\gamma_{A +} = (\alpha + \beta^{\ast}) e^{-g}; \; \gamma_{B +} = (\alpha^{\ast} + \beta ) e^{-g} \\
\gamma_{A -} = \imath (\alpha - \beta^{\ast}) e^{g}; \; \gamma_{B -} = \imath (\beta - \alpha^{\ast}) e^{g}
\end{eqnarray}
The integration of (\ref{funz_wig_opa_nondeg}) is analogous to the one reported
in section \ref{subs:OPA singolo modo}.

\begin{figure}[!tb]
\begin{center}
\includegraphics[width=0.45\textwidth]{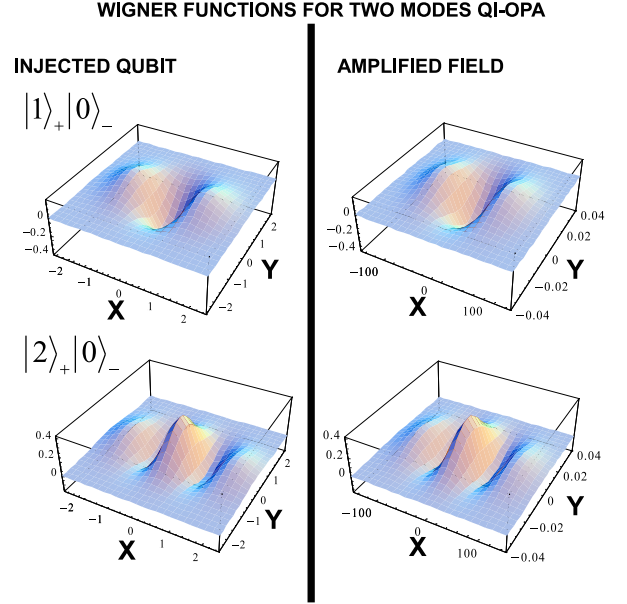}
\end{center}
\caption{Wigner function of the injected states and the corresponding amplified
states, for a non-degenerate amplifier with gain $g=4$. The axes
scales on different graphs are different due to the high degree of squeezing
introduced by the amplifier. $\left| 1+,0-\right\rangle$
injection of a single photon $\vec{\protect\pi}_{+}$ polarized. $\left|
2+,0-\right\rangle$ injection of a two-photon state $\vec{\protect\pi}_{+}$
polarized. }
\label{wigner3}
\end{figure}

\begin{figure}[h]
\begin{center}
\includegraphics[scale=0.45]{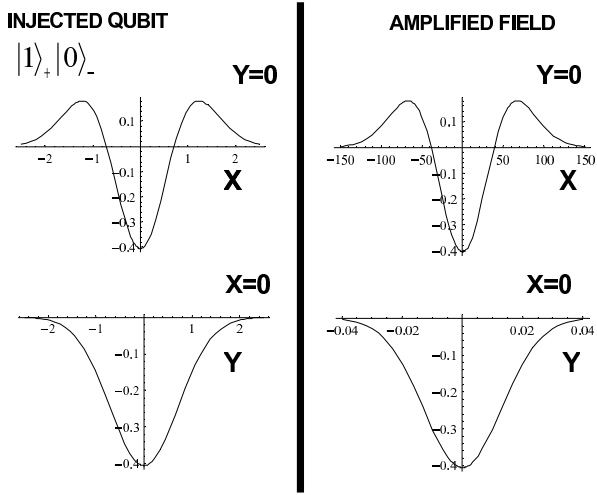}
\end{center}
\caption{Sections of the Wigner function for the injection of the single
photon state and for the corresponding amplified state obtained with a
non-collinear amplifier with a gain value $g=4$.}
\label{wigner4}
\end{figure}

If $\alpha =\left| \alpha \right| \exp i\varphi _{\alpha }$ and $\beta
=\left| \beta \right| \exp i\varphi _{\beta }$ have a well-defined phase
relation, $\varphi _{\alpha }=-\varphi _{\beta }=\varphi $, then for every $%
\varphi $ value it is possible to represent the Wigner function in a
three-dimensional graph, reported in Fig.\ref{wigner3}, that is, the 
projection of the total function onto a certain subspace. The quadrature 
variables in this subspace are: $X=(\alpha +\beta ^{\ast })$ e 
$Y=(\beta -\alpha ^{\ast })$. Furthermore, in Fig.\ref{wigner4} we report
the $X=0$ and $Y=0$ sections of the Wigner function for the single
photon amplified state in comparison with the injected seed. We again note
the resilience of the negative region, centered in the origin of the
phase-space, and the presence of the degree of squeezing induced by the
amplifier.

When the input state is the state of two photons with $\vec{\pi}_{+}$
polarization $\left| \psi _{in}\right\rangle =\left| 2+,0-\right\rangle $
the characteristic function is: 
\begin{equation}
\chi_{2,0} (\eta ,\xi ,t)=e^{-\frac{1}{4}(\left| \eta (t)\right| ^{2}+\left|
\xi (t)\right| ^{2})} \left( 1-\left| \eta (t)\right| ^{2}+\frac{\left| \eta
(t)\right| ^{4}}{8}\right)
\end{equation}
and the Wigner function is: 
\begin{eqnarray}
&&W_{\vert 2+,0- \rangle}(\alpha ,\beta ,t) =\left( \frac{2}{\pi }\right)
^{2}\left[ \exp \left( -\left| \Delta \right| ^{2}\right) \right] \times 
\notag \\
&\times&\left( 1-2\left| \Delta_{A}+\Delta _{B}\right| ^{2} + \frac{1}{2}%
\left| \Delta _{A}+\Delta _{B}\right| ^{4}\right)
\end{eqnarray}
In Fig.\ref{wigner3} we report the plots of the Wigner function of both the single- 
and double-photon amplified states compared to the original W-representations
of the injected seed.

These Wigner functions can also be evaluated by employing the results obtained
in the single mode amplifier case. In this context the characteristic
function factorizes into two parts $\chi _{+}(\eta ,t)\chi _{-}(\eta ,t)$,
that refer to polarizations $\vec{\pi}_{+}$ and $\vec{\pi}_{-}$
respectively, as a consequence of the independence of the two oscillators. 
Analogously, the Wigner function factorizes and when
the state $\left| \psi _{in}\right\rangle =\left| 1+,0-\right\rangle $ is
injected, it reads: 
\begin{eqnarray}
&&W_{\vert 1+,0- \rangle}(\tilde{\alpha},\tilde{\beta},t) =W_{+}(\tilde{%
\alpha})W_{-}(\tilde{\beta})= - \left( \frac{2}{\pi }\right) ^{2}  \notag \\
&&\left[ 1-4\left| \bar{\alpha}\right| ^{2}\right] \exp \left[ -2\left| \bar{%
\alpha}\right| ^{2}\right] \exp \left[ -2\left| \bar{\beta}\right| ^{2}%
\right]
\end{eqnarray}
where $\bar{\alpha}(t)=\tilde{\alpha}\cosh (g)-\tilde{\alpha}^{\ast }\sinh
(g)$ and $\bar{\beta}(t)=\tilde{\beta}\cosh (g)-\tilde{\beta}^{\ast }\sinh
(g)$. The variables $\left( \tilde{\alpha},\tilde{\beta}\right) $ are
related to $\left( \alpha ,\beta \right) $ through a simple rotation into
the quadri-dimensional phase space: 
\begin{eqnarray}
\cosh \left( g\right) \alpha -\sinh \left( g\right) \beta ^{\ast } &=&\cosh
\left( g\right) \tilde{\alpha}-\sinh \left( g\right) \tilde{\alpha}^{\ast }
\\
\cosh \left( g\right) \beta -\sinh \left( g\right) \alpha ^{\ast } &=&\cosh
\left( g\right) \tilde{\beta}-\sinh \left( g\right) \tilde{\beta}^{\ast }
\end{eqnarray}
We note that these rotations take the form of the Bogolioubov
transformations which express the time evolution of the field operators in
the collinear Optical Parametric Amplifier.

\subsection{Measurement of the quadratures with a double homodyne}
\label{subsec:homodyne}

Quadratures measurement can be obtained by homodyne technique \cite{Wall95}
largely adopted in the context of quantum optics. Quantum fields showing
quadratures entanglement have been realized \cite%
{Ou92,Zhan00,Silb01,Bowe03,Giac92,Scho02} and adopted in several quantum
information protocols. 
We now discuss the realization of a double homodyne experiment in
order to investigate the Wigner distributions in the QIOPA case. We define the
quadrature operators in a general form, introducing the phase dependence
from the variable $\theta $: 
\begin{equation}
\hat{X}_{\theta }=\frac{\hat{a}e^{-i\theta }+\hat{a}^{\dagger }e^{i\theta }}{%
2};\hat{Y}_{\theta }=\frac{-i\hat{a}e^{-i\theta }+i\hat{a}^{\dagger
}e^{i\theta }}{2}
\end{equation}

Let us briefly review the scheme of an homodyne experiment. The impinging field under 
investigation $\hat{E}_{1}(\mathbf{r},t)$ is combined into a beam splitter
with a \emph{local oscillator} $\hat{E}_{L.O}(\mathbf{r},t)$, usually prepared 
in a coherent state $\vert \beta \rangle$ with $\beta = \vert \beta \vert 
e^{\imath \theta}$. When $\beta$ is larger than the amplitude of the field
$\hat{E}_{1}$, the difference between the photon-counts of the two output modes
of the beam-splitter is proportional to:
\begin{equation}
\langle \hat{c}^{\dagger }\hat{c}\rangle -\langle \hat{d}^{\dagger }\hat{d}%
\rangle \approx \sqrt{2\tau \left( 1-\tau \right) }\langle \hat{Y}_{\theta
}\rangle  \label{misuraomodina}
\end{equation}
where $\tau$ is the BS transmittivity. By varying the local oscillator phase
by $\pi /2$ it is possible to select the measured quadrature.

\begin{figure}[h]
\begin{center}
\includegraphics[scale=0.35]{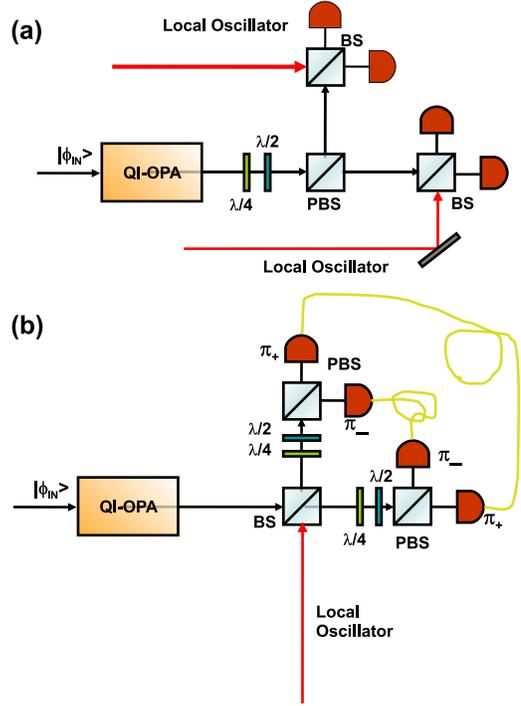}
\end{center}
\caption{Two equivalent schemes for the double homodyne measurements.
In both schemes, the choice of the basis for the analysis depends from the
waveplates $\protect\lambda /4+\protect\lambda /2$ settings.
(a) A PBS divides the two polarization components, each component is 
combined with a coherent radiation (with same polarization) on a BS. 
At the output of the PBS the field is revealed by two photodiodes. 
The signals are analyzed following the theory of the homodyne
technique. (b) A BS combines the field with a local
oscillator. The two output field impinge on a PBS which divides the two
polarization components. The choice of the analysis basis depends on the
waveplates setting $\protect\lambda /4+\protect\lambda /2$. Signals relative
to the different polarization components are detected through two
photodiodes and the results are pair-correlated according to homodyne
theory. 
}
\label{doppiaomodina}
\end{figure}

In the context of the QI-OPA we need to generalize the homodyne measurement
for the two polarization modes. A BS is inserted in the experimental scheme, 
following two different
solutions. In the first case (Fig.\ref{doppiaomodina}-(a)) at the exit of the
QI-OPA a couple of waveplates $\lambda /4+\lambda /2$ and a PBS divide the
two orthogonal polarizations which are combined with two equally polarized
local oscillators. 

In the second case (Fig.\ref{doppiaomodina}-(b)) the field at the exit of the QI-OPA 
is combined with a local oscillator onto a BS; the two output modes are analyzed in
polarization through a couple of waveplates $\lambda /4+\lambda /2$ (same
setting on both modes) and a PBS and finally detected by a pair of
photodiodes. The local oscillator polarization must be intermediate with respect
to the analysis basis, for example if the analysis basis is the linear $\pm
45^{\circ }$ rotated one, the polarization of the local oscillator can be
horizontal or vertical. The intensity measurements on both arms must be
correlated, coupling equal polarizations in order to obtain the result of (%
\ref{misuraomodina}). The results obtained in the two configurations are
equal, and the two schemes are completely equivalent for the characterization
of QI-OPA amplified states.

\subsection{Optical parametric amplification of N$>1$ Fock states}

Let us now consider the two-photon input state $\left| \psi _{in}\right\rangle
=\left| 2+,0-\right\rangle$ the characteristic function reads:{\small 
\begin{equation}
\chi_{2,0} (\eta ,\xi ,t)=\exp [-\frac{1}{4}(\left| \eta (t)\right|
^{2}+\left| \xi (t)\right| ^{2})]\left( 1-\left| \eta (t)\right| ^{2}+\frac{%
\left| \eta (t)\right| ^{4}}{8}\right)  \label{Characteristic2,0ph}
\end{equation}
}and the Wigner function becomes: 
\begin{equation}
\begin{aligned} &W_{\vert 2+,0- \rangle}(\alpha ,\beta ,t)=\left(
\frac{2}{\pi }\right) ^{2}e^{-\left| \Delta \right| ^{2}}\\ &\left(
1-2\left| \Delta _{A}+\Delta _{B}\right| ^{2}+\frac{1}{2}\left| \Delta
_{A}+\Delta _{B}\right| ^{4}\right) \end{aligned}
\label{WignerFunction2,0ph}
\end{equation}
A three dimensional plot of this function is obtained (see Fig. 4).

When both polarization modes are one photon injected $\left| \psi
_{in}\right\rangle =\left| 1+,1-\right\rangle$, the characteristic function
reads:

{\small 
\begin{equation}
\chi_{1,1} (\eta ,\xi ,t)=\left( 1-\frac{\left| \eta (t)\right| ^{2}}{2}%
\right) \left( 1-\frac{\left| \xi (t)\right| ^{2}}{2}\right) e^{-\frac{1}{4}%
(\left| \eta (t)\right| ^{2}+\left| \xi (t)\right| ^{2})}
\label{Characteristic1,1ph}
\end{equation}
}and Wigner function is: {\small 
\begin{eqnarray}
W_{\vert 1+,1- \rangle}(\alpha ,\beta ,t) &=&\left( \frac{2}{\pi }\right) ^{2}\left[
1-2\left| \Delta \right| ^{2}+\left| \Delta _{A}+\Delta _{B}\right|
^{2}\times \right.  \notag \\
&\times &\left. \left( 1-\left| \Delta _{A}+\Delta _{B}\right| ^{2}\right) 
\right] e^{-\left| \Delta \right| ^{2}}  \label{WignerFunction1,1ph}
\end{eqnarray}
}

As in the single mode case, we can generalize the results obtained for a
generic Fock-state as input $\vert \psi_{in} \rangle = \vert N+,M- \rangle$.
The Wigner functions of the amplified field with these generalized seeds
read: 
\begin{equation}  
\label{eq:wigner_NM_no_losses}
\begin{aligned} 
W_{\vert N+,M- \rangle}(\alpha, \beta, t) &= \left(
\frac{2}{\pi} \right)^{2} (-1)^{M+N} L_{N}\left(\vert \Delta_{A} + 
\Delta_{B} \vert^{2} \right) \\ &L_{M}\left(\vert \Delta_{B} - 
\Delta_{A} \vert^{2} \right) e^{-2 \vert \Delta \vert^{2}} 
\end{aligned}
\end{equation}
The Wigner function approach given in this Section to the QI-OPA device
allows to stress the quantum properties of the amplified field. The
negativity of the Wigner function can be deduced by the explicit general
expression of Eq.(\ref{eq:wigner_NM_no_losses}). Indeed, in analogy with the
un-amplified Fock-states $\vert N+, M- \rangle$ \cite{Schl01}, the Wigner
function $W_{\vert N+, M- \rangle}(\alpha, \beta, t)$ is the product of two
Laguerre polynomials \cite{Abra65} $L_{N}$ and $L_{M}$, which are negative
in several region. Furthermore, the amplified states possess a high degree
of squeezing in the field $\left\{ \overline{\alpha}, \overline{\beta}
\right\}$ quadratures, at variance with the injected states.

\section{Fock-Space analysis of macroscopic quantum superpositions over a
lossy channel}
\label{sec:fock_space}

In the present Section we analyze how the peculiar quantum interference
properties of the amplified single-photon states are smoothed and cancelled
when a decoherence process, i.e. a "system-enviroment" interaction, is
affecting their time evolution. More specifically, in the specific case of
optical fields, the main decoherence process can be identified with the
presence of lossy elements, as for example photo-detectors. Such
process is mathematically described by an artificial beam-splitter (BS)
scattering model, since this optical element expresses the coupling between
the transmission channel and a different spatial mode. The same analysis is
carried out also on a different class of MQS based on coherent $|\alpha
\rangle $ states, to emphasize analogies and differences.

\subsection{The lossy channel model}

As said, losses are analyzed through the effect of a generic linear
beam-splitter (BS)\ with transmittivity $T$ and reflectivity $R=1-T$, acting
on a generic quantum state associated with a single mode beam: Fig.\ref%
{fig:fig1} \cite{Loud00,Durk04,Durk04a,Leon93}. The action of the lossy channel on a
generic density matrix $\hat{\rho}$ is obtained by the application of the BS
unitary transformation and by the evaluation of the partial trace on the BS\
reflected mode (R-trace). The linear map describing the interaction is
expressed by the following expansion \cite{Durk04,Durk04a}: 
\begin{equation}
\mathcal{L}[\hat{\rho}]=\sum_{p}\hat{M}_{p}\,\hat{\rho}\,\hat{M}_{p}^{\dag }
\end{equation}%
where the $\hat{M}_{p}$ operators are: 
\begin{equation}
\hat{M}_{p}=R^{p/2}T^{(\hat{a}^{\dag }\hat{a})/2}\frac{\hat{a}^{p}}{\sqrt{p!}%
}
\end{equation}%
To evaluate the average values of operators, the lossy channel can be expressed in
the Heisenberg picture, exploiting the BS unitary transformations and
performing the average on the initial state $\hat{\rho}$.

\begin{figure}[t]
\centering
\includegraphics[width=0.45\textwidth]{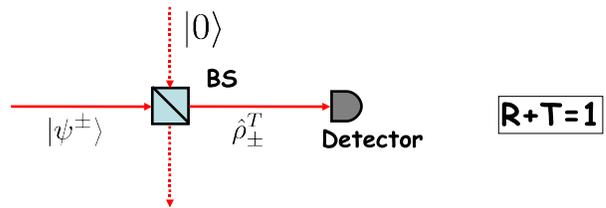}
\caption{Schematization of the decoherence model by a linear beam-splitter
of transmittivity T.}
\label{fig:fig1}
\end{figure}

\subsection{De-coherence on the quantum superposition of coherent $\vert 
\protect\alpha \rangle$ states}

\begin{figure}[th]
\centering
\includegraphics[width=0.45\textwidth]{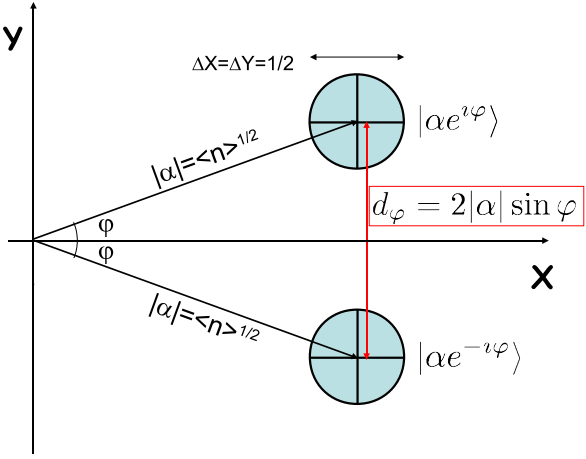}
\caption{Schematic representation in the bidimensional phase-space $\left\{
X,Y \right\}$ of the coherent states $\vert \protect\alpha e^{\imath \protect%
\varphi} \rangle$ which are the components of the MQS treated in this
section. The circles represent the two-dimensional projection of the
Gaussian functions which represent the Wigner function of these states. The
distance between the center of the two Gaussian is $d_{\protect\varphi}^{2}
= 4 \vert \protect\alpha \vert^{2} \sin^{2} \protect\varphi$.}
\label{fig:coherent_phase_space_blob}
\end{figure}

In this Section we investigate the evolution over a lossy channel of the
quantum superpositions of coherent states (CSS) \cite{Schl91}: 
\begin{equation}
|\Psi _{\pm }^{\varphi }\rangle {}=\mathcal{N}_{\pm }^{\varphi }\frac{1}{%
\sqrt{2}}\left( |\alpha e^{\imath \varphi }\rangle {}\pm |\alpha e^{-\imath
\varphi }\rangle \right)  \label{eq:superposition_coherent}
\end{equation}
with $\alpha $ real and $\mathcal{N}_{\varphi }^{\pm }=\left( 1\pm 
e^{-2|\alpha |^{2}\sin ^{2}\varphi}\cos \left[ |\alpha |^{2}\sin 2\varphi \right] 
\right) ^{-\frac{1}{2}}$ is an appropriate normalization factor.
The two states with opposite relative phases $|\Psi _{+}^{\varphi }\rangle
{} $ and $|\Psi _{-}^{\varphi }\rangle {}$ are orthogonal when $|\alpha
|^{2}\sin ^{2}\varphi >1$ (Fig.\ref{fig:coherent_phase_space_blob}).
In such case the two components $|\alpha
e^{\imath \varphi }\rangle {}$ and $|\alpha e^{-\imath \varphi }\rangle $
are distinguishable. This class of Macroscopic Quantum Superposition
presents several peculiar properties, such as squeezing and sub-poissonian
statistics, which can not be explained by the characteristics of the
coherent $\vert \alpha \rangle$ states.

We now proceed with the analysis of the density matrix after the propagation
over the lossy channel. In the following we assume $|\alpha |^{2}\sin
^{2}\varphi >1$, hence $\mathcal{N}_{\varphi }^{\pm }\sim 1 $. The density
matrix of the quantum state after the beam-splitter transformation is: 
\begin{equation}
\begin{aligned} \hat{\rho'}^{\pm \, \varphi} &= \frac{1}{2} \left( \vert
\beta e^{\imath \varphi} \rangle_{C} \, _{C}\langle \beta e^{\imath \varphi}
\vert \otimes \vert \imath \gamma e^{\imath \varphi} \rangle_{D} \,
_{D}\langle \imath \gamma e^{\imath \varphi} \vert + \right. \\ &+ \left.
\vert \beta e^{- \imath \varphi} \rangle_{C} \, _{C}\langle \beta e^{-
\imath \varphi} \vert \otimes \vert \imath \gamma e^{- \imath \varphi}
\rangle_{D} \, _{D}\langle \imath \gamma e^{- \imath \varphi} \vert \right.
+ \\ &\pm \left. \vert \beta e^{\imath \varphi} \rangle_{C} \, _{C}\langle
\beta e^{- \imath \varphi} \vert \otimes \vert \imath \gamma e^{\imath
\varphi} \rangle_{D} \, _{D}\langle \imath \gamma e^{- \imath \varphi} \vert
+ \right.\\ &\pm \left. \vert \beta e^{- \imath \varphi} \rangle_{C} \,
_{C}\langle \beta e^{\imath \varphi} \vert \otimes \vert \imath \gamma e^{-
\imath \varphi} \rangle_{D} \, _{D}\langle \imath \gamma e^{\imath \varphi}
\vert \right) \end{aligned}
\end{equation}
with $\beta =\sqrt{T}\alpha $ and $\gamma =\sqrt{R}\alpha $. The output
state over the transmitted field is obtained by tracing the density matrix $%
\hat{\rho ^{\prime }}^{\pm \,\varphi }$ over the mode $\hat{d}$. The final
expression for the density matrix after losses reads: 
\begin{equation}  \label{eq:cat_losses_density_matrix}
\begin{aligned} \hat{\rho}_{C}^{\pm} &= \frac{1}{2}\left( |\beta e^{\imath
\varphi} \rangle _{{}}\,_{{}}\langle \beta e^{\imath \varphi}|+|\beta e^{-
\imath \varphi}\rangle _{{}}\,_{{}}\langle \beta e^{-\imath
\varphi}\vert+\right.\\ &\left.\pm e^{-2R|\alpha |^{2} \sin^{2}\varphi}
e^{\imath R \vert \alpha \vert^{2} \sin 2\varphi} |\beta e^{\imath
\varphi}\rangle _{{}}\,_{{}}\langle \beta e^{- \imath \varphi}|+ \right.\\
&\left.\pm e^{-2R|\alpha|^{2} \sin^{2}\varphi} e^{- \imath R \vert \alpha
\vert^{2} \sin 2\varphi} |\beta e^{-\imath
\varphi}\rangle_{{}}\,_{{}}\langle \beta e^{\imath \varphi}|\right)
\end{aligned}
\end{equation}
Let us analyze first the case $\varphi = \frac{\pi}{2}$. The distribution in
the Fock space exhibits only elements with an even number of photons for $%
|\Psi _{+}\rangle $ or an odd number of photons for $|\Psi _{-}\rangle $.
This peculiar \textit{comb} structure is indeed very fragile under the
effect of losses since the R-trace operation must be carried out in the
space of the \textit{non-orthogonal} coherent-states. In fig.\ref%
{fig:distribution_coherent_losses} are reported the photon number
distribution for different values of $R$ with an initial average number of
photon equal to $\langle n\rangle =16$ and $\varphi =\frac{\pi }{2} $. 
\begin{figure}[t]
\centering
\includegraphics[width=0.5\textwidth]{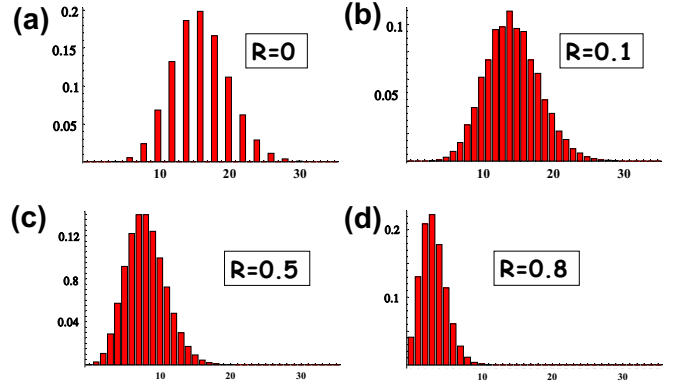}
\caption{(a)-(d): Plot of the distribution of the number of photons in the $%
|\Psi_{+}^{\frac{\protect\pi}{2}}\rangle $ state for $\protect\alpha =4$,
corresponding to an average number of photons $\langle n\rangle =16$, for
reflectivities $R=0$ (fig.2-a), $R=0.1$ (fig.2-b), $R=0.5$ (fig.2-c) and $%
R=0.8$ (fig.2-d)}
\label{fig:distribution_coherent_losses}
\end{figure}
We observe that for a reflectivity $R=0.1$, corresponding to about $\sim 1.5$
photon lost in average, the distribution is similar to the Poisson
distribution associated to the coherent states. Such graphical analysis
points out how the ortoghonality between $|\Psi _{+}\rangle $ and $|\Psi
_{-}\rangle $ quickly decrease as soon as $R$ differs from $0$, since the
phase relation between the components $|\alpha \rangle $ and $|-\alpha
\rangle $ becomes undefined.

\subsection{De-coherence in a lossy channel for equatorial amplified qubits}

Analogously to the previous Section, in which the effect of losses where
analyzed for the MQS of coherent state, we begin our treatment of QIOPA
amplified states with the evaluation of the density matrix after the
propagation over a lossy channel. 

Before the lossy process, the density matrix of the state $\hat{\rho}^{\varphi} =
\vert \Phi^{\varphi} \rangle \, \langle \Phi^{\varphi} \vert$ has the form:
\begin{equation}
\label{eq:density_matrix_unperturbed_equatorial}
\hat{\rho}^{\varphi} = \sum_{i,j,k,q=0}^{\infty} \gamma_{ij} \gamma^{\ast}_{kq} 
\vert (2i+1) \varphi, (2j) \varphi_{\bot} \rangle \, \langle (2k+1) \varphi, (2q) 
\varphi_{\bot} \vert
\end{equation}
We note from this expression that only elements with an odd number of photons
in the $\vec{\pi}_{\varphi}$ and an even number in the $\vec{\pi}_{\varphi_{\bot}}$
polarization are present. Furthermore, in Fig.\ref{fig:fig3}-(a) we note that the photon number
distribution presents a strong unbalancement due to the quantum injection of the 
$\vec{\pi}_{\varphi}$ single photon. Indeed, the QIOPA seeded by a photon with
equatorial polarization acts as a \emph{phase-covariant optimal cloning machine}, 
and is stimulated to generate an output field containing more photons in the
polarization of the injected seed. 
Let us now analyze the effects of the transmission in a lossy channel for
the equatorial amplified qubits by plotting the photon number distributions. 
The output density matrix after the transmission over the lossy channel
is the sum of four terms:
\begin{widetext}

\begin{equation}
\begin{aligned}
\hat{\rho}^{\varphi}_{T} &= \sum_{i,j,k,q=0}^{\infty} \left( \hat{\rho}^{\varphi}_{T} 
\right)_{ijkq} \vert (2i+1) \varphi, (2j) \varphi_{\bot} \rangle \, \langle (2k+1) \varphi, (2q) 
\varphi_{\bot} \vert + \left( \hat{\rho}^{\varphi}_{T} \right)_{ijkq}
\vert (2i) \varphi, (2j) \varphi_{\bot} \rangle \, \langle (2k) \varphi, (2q) 
\varphi_{\bot} \vert +\\
&+ \left( \hat{\rho}^{\varphi}_{T} \right)_{ijkq}
\vert (2i+1) \varphi, (2j+1) \varphi_{\bot} \rangle \, \langle (2k+1) \varphi, (2q+1) 
\varphi_{\bot} \vert + \left( \hat{\rho}^{\varphi}_{T} \right)_{ijkq}
\vert (2i) \varphi, (2j+1) \varphi_{\bot} \rangle \, \langle (2k) \varphi, (2q+1) 
\varphi_{\bot} \vert
\end{aligned}
\end{equation}

\end{widetext}
The details on the calculation and on the expressions of the coefficients are 
reported in appendix \ref{app:equatorial_Fock}.

Let us now analyze this result. When the original state propagates through a 
lossy channel, the first effect at low values of $R$ is the cancellation of the 
peculiar \emph{comb} structure (Fig.\ref{fig:fig3}-(a)) given by the presence 
in the density matrix (\ref{eq:density_matrix_unperturbed_equatorial}) only of 
terms with a specific parity $\vert (2i+1) \varphi, (2j) \varphi_{\bot} \rangle \, \langle (2k+1) 
\varphi, (2q) \varphi_{\bot} \vert$, similarly to the coherent state MQS's previously
studied. However, at progressively higher values of $R$, the distributions
in the Fock space remain unbalanced in the polarization of the injected photon
(Fig.\ref{fig:fig3}-(a)). The resilience of this unbalancement  allows to distinguish 
the orthogonal macro-qubits $\left \{ \vert \Phi^{\varphi} \rangle , \vert \Phi^{\varphi_{\bot}} 
\rangle \right\}$ even after the propagation over the lossy channel, by exploiting
this property with a suitable detection scheme, such as the OF device reported in \cite{DeMa08}. 
All these considerations will be discussed and quantified later in the paper in Section 
\ref{sec:Bures_Fock} by analyzing the distinguishability of such states as a function of the
lossy channel efficiency $T$.

\subsection{De-coherence in a lossy channel for amplified $\vec{\protect\pi}%
_{H}, \vec{\protect\pi}_{V}$ qubits}

\begin{widetext}

\begin{figure}[ht!]
\includegraphics[width=.9\textwidth]{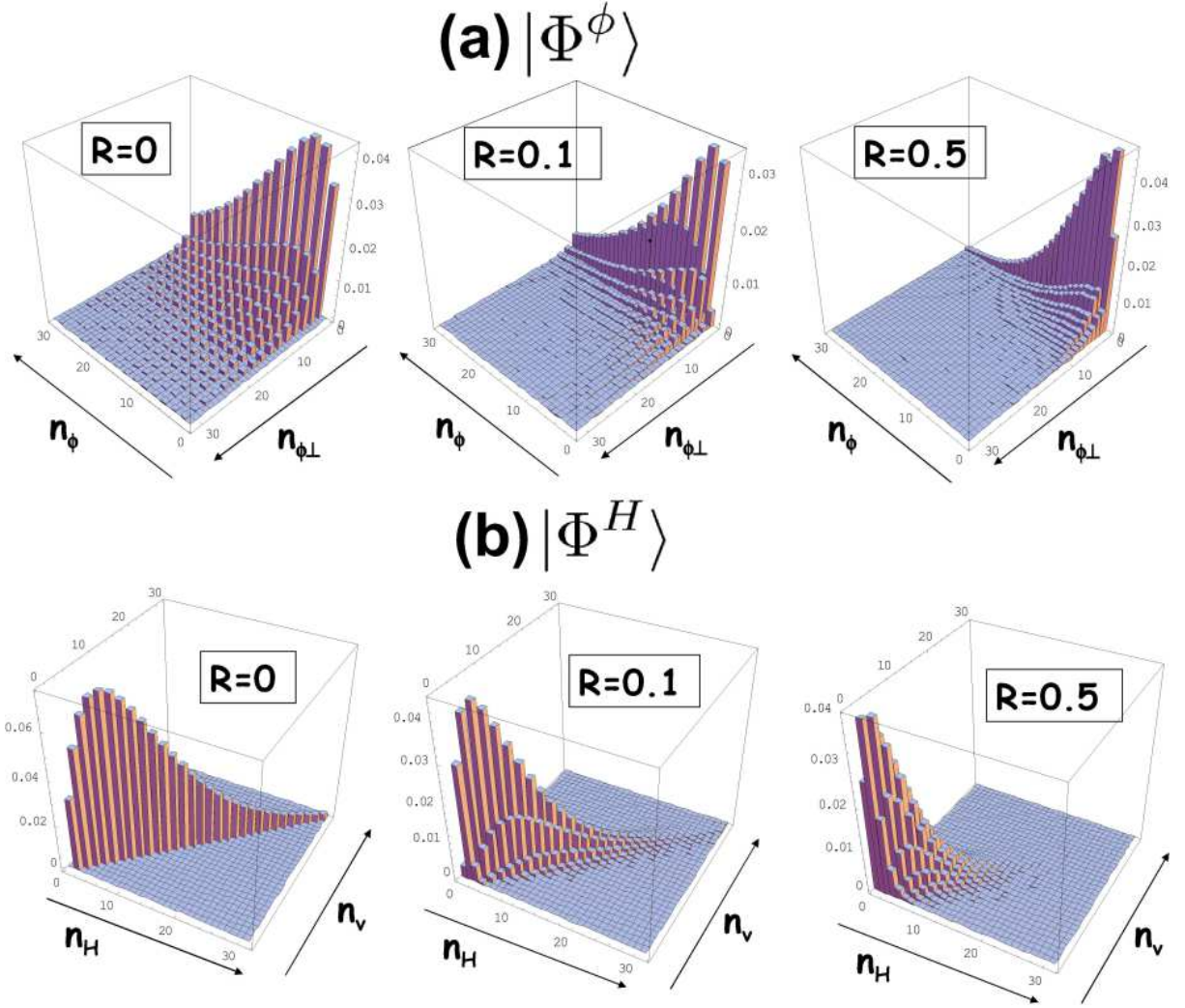}
\caption{(a) Probability distribution in the Fock space $(n_{\protect\phi%
},n_{\protect\phi_{\bot}})$ for the amplified
$|\Phi^{\protect\phi}\rangle$ state of a generic equatorial qubit
for different values of the
transmittivity. (b) Probability distribution in the Fock space $%
(n_{H},n_{V}) $ for the amplified $|\Phi^{H}\rangle$ state for
different
values of the transmittivity. All distributions refer to a gain value of $%
g=1.5$, corresponding to an average number of photons $\langle n
\rangle \approx 19$. } \label{fig:fig3}
\end{figure}

\end{widetext}

For the sake of completeness, in this Section we shall analyze the evolution of $%
\vert \Phi^{H} \rangle$ and $\vert \Phi^{V} \rangle$ amplified states. As a first remark,
we note that the collinear Optical Parametric Amplifier is not an \emph{optimal cloner} for
states with $\pi_{H}$ and $\pi_{V}$ polarization, and the output states do
not possess the same peculiar properties obtained with an equatorial
injected qubit. The density matrix of the $\vert \Phi^{H} \rangle$
amplified state is:
\begin{equation}
\begin{aligned} \hat{\rho}^{H} &= \vert \Phi^{H} \rangle \, \langle \Phi^{H}
\vert = \frac{1}{C^{4}} \sum_{n,m=0}^{\infty} \Gamma^{n+m} \sqrt{n+1}
\sqrt{m+1} \\ &\vert (n+1)H, nV \rangle \, \langle (m+1)H,mV \vert
\end{aligned}
\end{equation}
In Fig.\ref{fig:fig3}-(b) we plotted the photon number distribution of this 
state (R=0). We note that the $\vec{\pi}_{H}$ amplified state does not
possess the same unbalancement of the equatorial macro-qubits
$\vert \Phi^{\varphi} \rangle$ analyzed in the previous section.

After the propagation over the lossy channel, the density matrix reads:
\begin{equation}
\begin{aligned} &\hat{\rho}^{H}_{T} = \sum_{i=1}^{\infty} \sum_{j=0}^{i-1}
\sum_{k=0}^{\infty} \left( \hat{\rho}^{H}_{T} \right)_{ijk} 
\vert iH,jV \rangle \, \langle kH, (k+j-i)V \vert +\\ &+
\sum_{i=0}^{\infty} \sum_{j=i}^{\infty} \sum_{k=0}^{\infty} \left(
\hat{\rho}_{T}^{H} \right)_{ijk} \vert iH,jV
\rangle \, \langle kH, (k+j-i)V \vert \end{aligned}
\end{equation}
where details on the calculation and on the coefficients are reported in
appendix \ref{app:HV_Fock}.
The effect of the propagation over the lossy channel is shown in Fig.\ref%
{fig:fig3}-(b). The original distribution for $R=0$ is pseudo-diagonal,
corresponding to the presence only of terms $\vert (n+1)H, nV \rangle$. Here
the difference of one photon between the two polarization is due to the
injection of the seed. For values of $R$ different from 0, the distribution
is no longer pseudo-diagonal and this characteristic becomes progressively
smoothed. Furthermore, the absence of the unbalancement in the photon
number distribution typical of the equatorial macro-qubits does not allow to
exploit this feature to discriminate among the orthogonal states 
$\left\{ \vert \Phi^{H} \rangle, \vert \Phi^{V} \rangle \right\}$. We then expect
that these couple of states possess a lower resilience to losses than
the equatorial $\vert \Phi^{\varphi} \rangle$ macro-states analyzed in the
previous section.


\section{Wigner function representation of coherent states MQS in presence
of de-coherence}

For the sake of clarity, we briefly review previous results on the Wigner
functions associated to coherent superposition states after the propagation
over a lossy channel. We start from the general definition of the Wigner
function for mixed states \cite{Wign32}: 
\begin{equation}  \label{eq:Wigner_def}
W_{\hat{\rho}}(X,Y) = \frac{1}{\pi} \int_{-\infty}^{\infty} d\xi e^{2 \imath Y
\xi} _{}\langle X-\xi \vert \hat{\rho} \vert X+\xi \rangle_{}
\end{equation}
Considering the density matrix of the CSS after losses (\ref%
{eq:cat_losses_density_matrix}) we obtain: 
\begin{equation}
\begin{aligned} W_{\op{\rho}_{C}^{\varphi \, \pm}}(X,Y) &=
\frac{\mathcal{N}^{\varphi \, 2}_{\pm}}{2} \left( W_{\ket{\beta e^{\imath
\varphi}}{}}(X,Y) + W_{\ket{\beta e^{-\imath \varphi}}{}}(X,Y) + \right.\\
&\left. \pm W^{int}_{\op{\rho}_{C}^{\varphi \, \pm}}(X,Y) \right)
\end{aligned}
\end{equation}
In the last expression, the first two components are analogous to the
diagonal ones of the unperturbed case \cite{Schl91}, and can be written as : 
\begin{equation}
W_{|\beta e^{\pm \imath \varphi }\rangle {}}(X,Y)=\frac{1}{\pi }e^{-\left( X-%
\sqrt{T}\overline{x}_{\varphi }\right) ^{2}}e^{-\left( Y \mp \sqrt{T}%
\overline{Y}_{\varphi }\right) ^{2}}
\end{equation}
where $\overline{X}_{\varphi}^{2} = 2 \vert \alpha \vert^{2} \cos^{2} \varphi$ and
$\overline{Y}_{\varphi}^{2} = 2 \vert \alpha \vert^{2} \sin^{2} \varphi$.
Hence losses reduce the average value of the quadratures $\hat{X}$ and $\hat{Y}
$. 

The interference contribution reads: 
\begin{equation}
\label{eq:Wigner_cat_losses_interf}
\begin{aligned} W^{int}_{\op{\rho}_{C}^{\varphi \, \pm}}&(X,Y) =
\frac{2}{\pi} e^{-Y^{2}} e^{- \left( X - \sqrt{T} \overline{X}_{\varphi}
\right)^{2}} e^{- R \overline{Y}_{\varphi}^{2}} \times \\ &\times \cos
\left[ 2 \sqrt{2} \alpha \sqrt{T} \sin \varphi \left( X - \frac{\alpha
(2T-1)}{\sqrt{2T}} \cos \varphi \right) \right] \end{aligned}
\end{equation}
which is strongly reduced in amplitude by a factor
proportional to $e^{-R\overline{Y}_{\varphi }^{2}}$. 

\begin{figure}[th]
\centering
\includegraphics[width=0.50\textwidth]{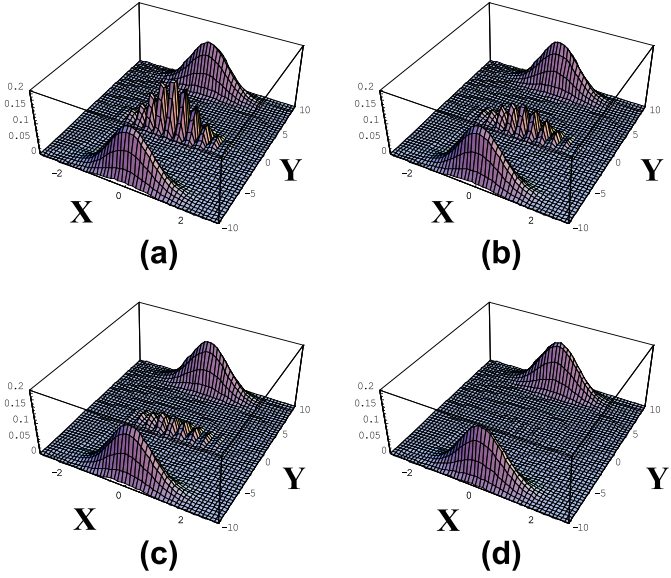}
\caption{Plots of the Wigner function for $\protect\alpha =6$ and $\protect%
\varphi =\frac{\protect\pi }{2}$ for different values of the riflectivity
(a) $R=0$, (b) $R=0.01$, (c) $R=0.02$ and (d) $R=0.1$.}
\label{fig:wigner_coherent_phi_losses}
\end{figure}

In Fig.\ref{fig:wigner_coherent_phi_losses} are plotted the Wigner functions
associated to different values of $R$, for the same initial conditions $%
\varphi =\frac{\pi }{2}$ and $\alpha =6$. 
As expected, by increasing the degree of losses the central peak is
progressively attenuated up to a complete deletion of the quantum features
associated to the negativity of the Wigner functions. We observe that the 
damping factor $e^{-2R|\alpha|^{2}\sin ^{2}\varphi }$ of the coherence terms 
derives from the exponential decrease of the non-diagonal terms of the density 
matrix (\ref{eq:cat_losses_density_matrix}).

More specifically, we now focus on the $\varphi = \frac{\pi}{2}$ case, i.e.
the $\vert \alpha \rangle \pm \vert - \alpha \rangle$ state. In Fig. \ref%
{fig:Coherent_negative_section} we report the plots of the $Y=0$ section of
the Wigner function for different values of the reflectivity $R$, which
corresponds to the interference pattern.

\begin{figure}[ht]
\includegraphics[width=0.5\textwidth]{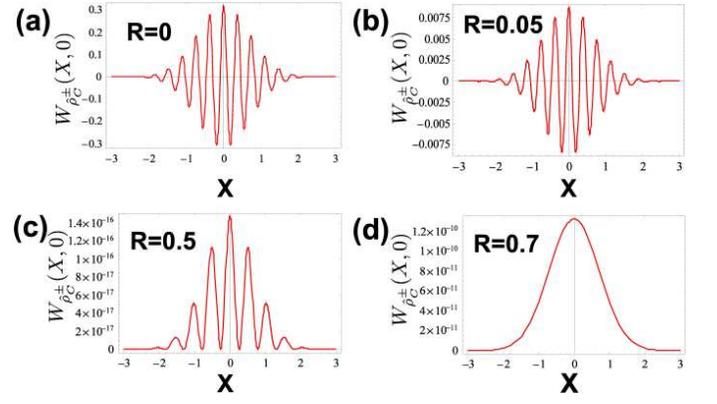}
\caption{$Y=0$ sections of the Wigner function for the quantum superposition
of coherent states, for $\protect\alpha = 6$ and $\protect\varphi = \frac{%
\protect\pi}{2}$ and different values of $R$. (a) $R=0$: Unperturbed
interference pattern. (b) $R=0.05$: Exponentially damped interference
pattern. (c) $R=0.5$: Transition to a completely positive Wigner function,
as the interference pattern is shifted towards positive values. (d) $R=0.7$:
The interference pattern in the central region is deleted by decoherence.}
\label{fig:Coherent_negative_section}
\end{figure}

At low R, the amplitude of the oscillation is exponentially
damped. This exponential factor $e^{- 2 R \vert \alpha \vert^{2} \sin^{2} \varphi}$
is responsible for the fast decrease in the negativity of the Wigner function,
but the alternance of positive and negative regions is maintained.
However, when $R$ approaches the $0.5$ value, the interference pattern is
progressively shifted towards positive values in all the X-axis range, and
at $R=0.5$ it ceases to be non-positive. This evolution depends on the presence of
the $W_{|\beta e^{\pm \imath \varphi }\rangle {}}(X,Y)$ diagonal terms, which are
not exponentially damped in amplitude and for $R \sim 0.5$ become comparable
to the interference term. This transition is graphically shown in Fig.\ref%
{fig:Negativity_coherent}, where the negativity is plotted as a function of $%
R$. This quantity has been evaluated by calculating the value of the Wigner
function in the first minimum of the cosine term of $W^{int}_{\hat{\rho}%
_{C}^{\varphi \, \pm}}(X,Y)$. We note the transition from negative to
positive at $R=0.5$, where for higher values this point ceases to be the
minimum of the complete Wigner function in the $\left\{X,Y\right\}$ plane.
\begin{figure}[ht]
\centering
\includegraphics[width=0.4\textwidth]{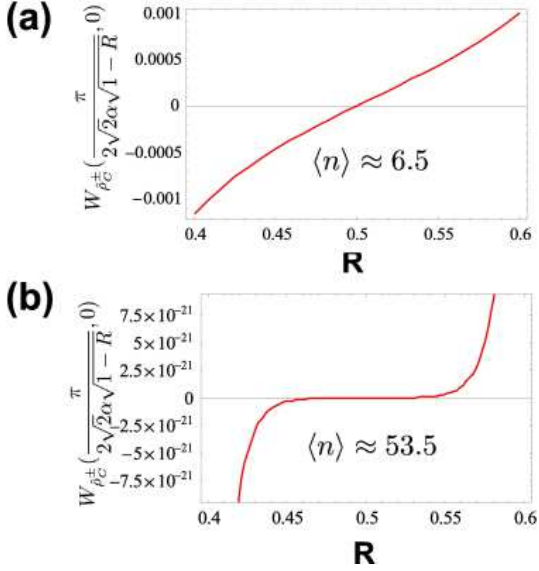}
\caption{Transition from non-positive to completely positive Wigner function
as a function of the reflectivity $R$, for a number of average photons of $%
\langle n \rangle \approx 6.5$ (a) and $\langle n \rangle \approx 53.5$ (b).
The negativity has been evaluated as described in the text. We note that, while the
interval of $R$ in which the Wigner function is non positive is independent of
the number of photons, the amount of negativity decreases exponentially with 
$\langle n \rangle$ due to the exponential factor $e^{-2 R \vert \alpha \vert^{2} 
\sin^{2} \varphi}$ in the interference term $W^{int}_{\hat{\rho}%
_{C}^{\varphi \, \pm}}(X,Y)$.}
\label{fig:Negativity_coherent}
\end{figure}
This property can be derived explicitly by the complete form of the
Wigner function. For $\varphi =\frac{\pi }{2}$, $Y_{0}=0$ and $X_{0}=\frac{%
\pi }{2\sqrt{2}\sqrt{1-R}\alpha }$, corresponding to the minimum of the
cosine $\cos \left[ 2\sqrt{2}\alpha \sqrt{T}\sin \varphi \left( X-\frac{%
\alpha (2T-1)}{\sqrt{2T}}\cos \varphi \right) \right] $, we get: {\small 
\begin{equation}
\begin{aligned} W_{\hat{\rho}_{C}^{\frac{\pi}{2} \, +}}(X_{0}, 0) &=
\frac{\mathcal{N}^{\frac{\pi}{2}}_{+}}{\pi} e^{-2(1-R) \vert \alpha
\vert^{2}}\left( e^{-2 \vert \alpha \vert^{2}(1-R)} - e^{-2 \vert \alpha
\vert^{2}R}\right) = \\ &= \left\{ \begin{array}{lrl} < 0 & \mathrm{if} & R<
\frac{1}{2} \\ > 0 & \mathrm{if} & R> \frac{1}{2} \end{array} \right.
\end{aligned}
\end{equation}%
} thus obtaining the desired result.

As a general statement, it is well known that the negativity of the
"quasi-probability" Wigner function of a state $\rho $ is a sufficient,
albeit non necessary, condition for the "quantumness" of $\rho $%
\cite{Schl01}.

\section{Wigner function representation of the
phase-covariant quantum cloning process in presence of de-coherence}

Combining the approach of Section III and the lossy channel method
introduced in Section IV, we derive the analytical expressions for the
Wigner function of the QI-OPA amplified states in presence of losses. The
calculation will be performed in the Heisenberg picture, starting from the
evaluation of the characteristic function of the BS-transmitted field. This
analysis will be performed for both the single-mode degenerate amplifier,
i.e. a single-mode squeezing Hamiltonian, and for the two-mode optical
parametric amplifier, in which the polarization degree of freedom plays an
important role as stressed in the Fock-space analysis previuosly performed.
Finally, we shall focus our attention on the negativity of the Wigner
function, evidence of the non-classical properties of this class of states.

\subsection{Single-mode degenerate amplifier}

\begin{figure}[tbp]
\centering
\includegraphics[width=0.5\textwidth]{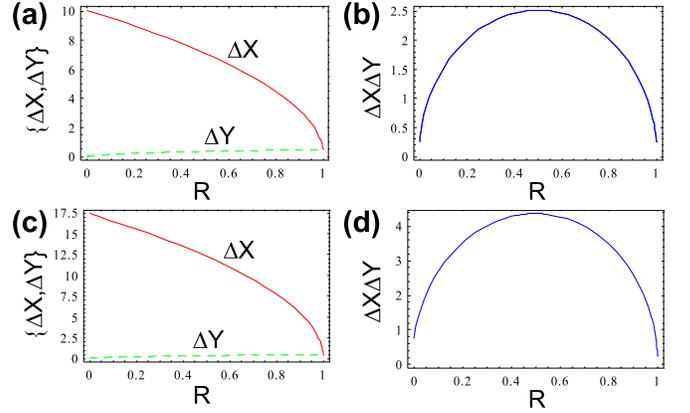}
\caption{(\textbf{a})-(\textbf{b}) Uncertainty relations for the quadrature operators in the
spontaneous emission case for a single mode OPA. \textbf{Left figure}:
Fluctuation on the $X$ quadrature (red straight line) and on the $Y$ quadrature
(green dashedline) as a function of the parameter $R$ of the lossy channel. 
\textbf{Right figure}: Uncertainty relation $\Delta X \Delta Y$ as a
function of the parameter $R$. (\textbf{c})-(\textbf{d}) Uncertainty relations for the
quadrature operators in the single-photon amplified case for a single mode
OPA. \textbf{Left figure}: Fluctuation on the $X$ quadrature (red straight line) and
on the $Y$ quadrature (green dashed line) as a function of the parameter $R$ of the
lossy channel. \textbf{Right figure}: Uncertainty relation $\Delta X \Delta
Y $ as a function of the parameter $R$. All figures refer to a gain value of 
$g=3$.}
\label{fig:uncertainty_relations}
\end{figure}

Let us first analyze the case of a single mode-degenerate amplifier. We
begin our analysis by evaluating the characteristic function in presence of
a lossy channel. The operators describing the output field can be written in
the Heisenberg picture in the form: 
\begin{equation}  \label{eq:IO_field}
\hat{c}^{\dag}(t) = \sqrt{T} \hat{a}^{\dag}(t) + \imath \sqrt{R} \hat{b}%
^{\dag}
\end{equation}
where $\hat{a}^{\dag}(t) = \hat{a}^{\dag} \cosh g + \hat{a} \sinh g$ is the
time evolution of the field operator in the amplifier, and $R = 1 - T$.
Hence, the characteristic function for a generic input state $\vert N
\rangle_{A} \vert 0 \rangle_{B}$ can be calculated as: 
\begin{equation}
\chi_{N}(\eta, R, t) = \langle N,0 \vert e^{\eta \hat{c}^{\dag}(t) -
\eta^{\ast} \hat{c}(t)} \vert N,0 \rangle
\end{equation}
Inserting the explicit expression (\ref{eq:IO_field}) for the output field
operators, we obtain the expression: 
\begin{equation}  \label{eq:char_func_1_mode_losses}
\begin{split}
\chi_{N}(\eta, R, t) &= e^{-\frac{1}{2} T \vert \eta(t) \vert^{2}} \, e^{- 
\frac{1}{2} R \vert \eta \vert^{2}} \times \\
&\times \left[ \sum_{n=0}^{N} 
\begin{pmatrix}
N \\ 
n%
\end{pmatrix}
\frac{(-1)^{n}}{n!} \left( T \vert \eta (t)\vert^{2} \right)^{n} \right]
\end{split}%
\end{equation}
In this expression, the transformation $\left( \eta, \eta^{\ast} \right)
\rightarrow \left( \eta(t),\eta^{\ast}(t) \right)$ has the form $\eta(t) =
\eta C - \eta^{\ast} S$, equivalent to the ideal case previously analyzed.
The Wigner function of the output field is then obtained from its definition
(\ref{deffunzwig}) as the two-dimensional Fourier transform of the characteristic function.
Inserting the explicit expression of $\chi_{N} (\eta, R, t)$, and by
changing the variables with the transformation $\eta \rightarrow \eta(t)$
with unitary Jacobian we obtain: 
\begin{equation}
\begin{split}
W_{\vert N \rangle} (\alpha, R, t) &= \frac{1}{\pi^{2}} \sum_{n=0}^{N}
c_{n}^{N} \int d^{2} \eta(t) \, \vert \eta(t) \vert^{2n} \, e^{ - \varepsilon \vert
\eta(t) \vert^{2}} \\
&\times \exp\left[ - \kappa \left( \eta^{2}(t) + \eta^{\ast \, 2}(t) \right) %
\right] \\
&\times \exp\left[ - \overline{\alpha}^{\ast} \, \eta (t) + \overline{\alpha}
\, \eta^{\ast}(t) \right]
\end{split}%
\end{equation}
where: 
\begin{eqnarray}
c_{n}^{N} &=& 
\begin{pmatrix}
N \\ 
n%
\end{pmatrix}
\frac{(-1)^{n}}{n!} T^{n} \\
\label{eq:epsilon_wigner}
\varepsilon &=& \frac{1}{2} \left( 1 + 2 R S^{2} \right) \\
\label{eq:kappa_wigner}
\kappa &=& \frac{1}{2} R C S \\
\overline{\alpha} &=& \alpha C - \alpha^{\ast} S
\end{eqnarray}
In order to evaluate the integral in the previous expression, we recall the
following identity \cite{Bisw07,Puri01}: 
\begin{equation}
\label{eq:fond_int_rel}
\begin{split}
&\frac{1}{\pi} \int d^{2} \alpha \, e^{- \tau \vert \alpha \vert^{2}}\exp %
\left[ - \mu \alpha^{2} - \nu \alpha^{\ast \, 2} - z^{\ast} \alpha + z
\alpha^{\ast} \right] = \\
&= \frac{1}{\sqrt{\tau^{2} - 4 \mu \nu}} \exp \left[ - \frac{\mu z^{2} + \nu
z^{\ast \, 2} + \tau \vert z \vert^{2}}{\tau^{2} - 4 \mu \nu} \right]
\end{split}%
\end{equation}
From the latter equation, we can derive the following useful identities: 
\begin{widetext}
\begin{equation}
\label{eq:integral_relations_wigner}
\begin{split}
I_{n} \left( \mu, \nu, \tau, z \right) &= \frac{1}{\pi} \int d^{2} \alpha \, \vert \alpha \vert^{2n} e^{- \tau \vert \alpha \vert^{2}} \exp\left[ - \mu \alpha^{2} - \nu \alpha^{\ast \, 2} - z^{\ast} \alpha + z \alpha^{\ast} \right] = \\
&= \frac{1}{\pi} (-1)^{n} \frac{\partial^{2n}}{\partial z^{n} \partial z^{\ast \, n}} \int d^{2} \alpha e^{- \tau \vert \alpha \vert^{2}} \exp\left[ - \mu \alpha^{2} - \nu \alpha^{\ast \, 2} - z^{\ast} \alpha + z \alpha^{\ast} \right] = \\
&= (-1)^{n} \frac{\partial^{2n}}{\partial z^{n} \partial z^{\ast \, n}} \left\{ \frac{1}{\sqrt{\tau^{2} - 4 \mu \nu}} \exp \left[ - \frac{\mu z^{2} + \nu z^{\ast \, 2} + \tau \vert z \vert^{2}}{\tau^{2} - 4 \mu \nu} \right] \right\}
\end{split} 
\end{equation}
\end{widetext}
These expressions can then be used to evaluate the Fourier transform of the
characteristic function (\ref{eq:char_func_1_mode_losses}), leading to the
Wigner function: 
\begin{equation}
W_{\vert N \rangle} (\alpha, R, t) = \frac{1}{\pi} \sum_{n=0}^{N} c_{n}^{N}
I_{n} \left( \kappa, \kappa, \varepsilon, \overline{\alpha} \right)
\end{equation}
As a first case, we analyze the evolution of the Wigner function of a
squeezed vacuum state, corresponding to the value $N=0$ in the previous
expression. Analogously to the unperturbed case, the quadrature variables
for the single-mode OPA are defined by $\alpha = X + \imath Y$. The Wigner
function for the squeezed vacuum then reads: 
\begin{equation}
\begin{split}
&W_{\vert 0 \rangle}(X,Y,R,t) = \frac{2}{\pi} \frac{1}{\sqrt{1 + 4 (1-R) R
S^{2}}} \\
&\times \exp\left[-2 \frac{\left( X^{2} e^{-2g} + Y^{2} e^{2g} \right) + 2 R
S \left( X^{2} e^{-g} + Y^{2} e^{g} \right)}{1 + 4 (1-R) R S^{2}}\right]
\end{split}%
\end{equation}
From this quantity, we can explicitly calculate how the degree of squeezing
changes with an increasing value of the parameter R. We obtain for the
fluctuations of the two field quadrature operators $\left\{\hat{X}, \hat{Y}
\right\}$ (Fig.\ref{fig:uncertainty_relations}-(\textbf{a})): 
\begin{eqnarray}
\Delta X &=& \frac{1}{2} \sqrt{T e^{2g} + (1 - T)} \\
\Delta Y &=& \frac{1}{2} \sqrt{T e^{-2g} + (1 - T)}
\end{eqnarray}
This two operators do not satisfy anymore the minimum uncertainty relation,
as in the unperturbed case, after the propagation over the lossy channel:
Fig.\ref{fig:uncertainty_relations}-(\textbf{b}). This is due to the additional
poissonian noise belonging to the photon loss process. 

\begin{figure}[tbp]
\centering
\includegraphics[width=0.5\textwidth]{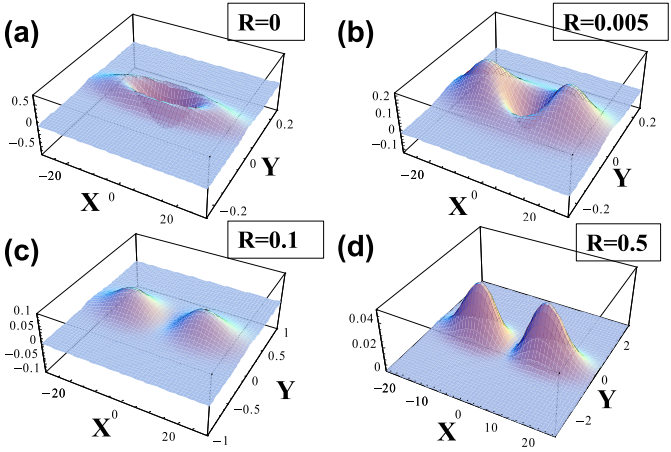}
\caption{Wigner function of a single-photon amplified state in a single-mode
degenerate OPA for $g=3$. (a) (R=0) Unperturbed case. (b) (R=0.005) For
small reflectivity, the Wigner function remains negative in the central
region. (c) (R=0.1) The Wigner function progressively evolve in a positive
function in all the phase-space. (d) (R=0.5) Transition from a non-positive
to a completely positive Wigner function.}
\label{fig:single_mode_1_photon_amplified_losses}
\end{figure}

\begin{figure}[tbp]
\centering
\includegraphics[width=0.5\textwidth]{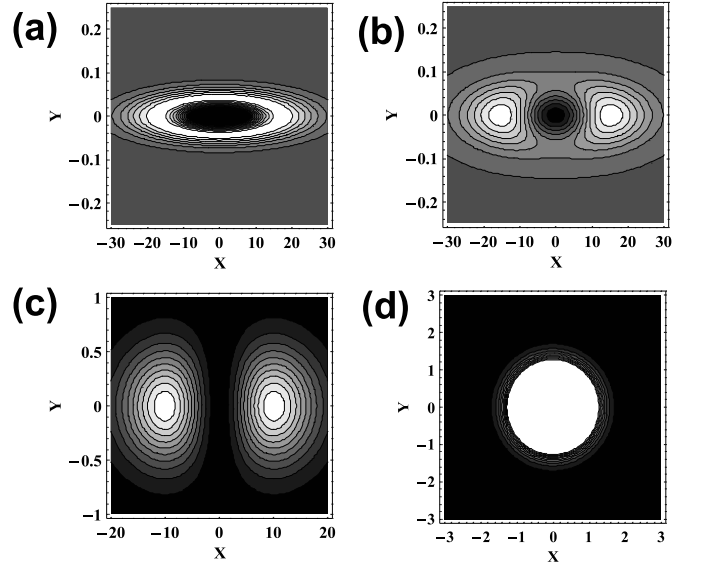}
\caption{Contour plots of the Wigner function of a single-photon amplified
state in a single-mode degenerate OPA for $g=3$. (a) (R=0) Unperturbed case.
(b) (R=0.005) For small reflectivity, the Wigner distribution begin to
evolve in a double-peaked function with a negative central region,
analogously to the CSS case. (c) (R=0.5) Transition from the regime with a
negative central region and a completely positive distribution (d) (R=1)
Evolution of the Wigner function into a vacuum state when all photons are
lost.}
\label{fig:contour_wigner_losses}
\end{figure}

A similar behaviour is obtained for the single-photon squeezed state. The
Wigner function for $N=1$ reads: 
\begin{equation}
\begin{split}
&W_{\vert 1 \rangle}(X,Y,R,t) = \frac{2}{\pi} \frac{1}{\sqrt{1 + 4 (1-R) R
S^{2}}} \, P_{\vert 1 \rangle}(X,Y,R,t) \\
&\times \exp\left[-2 \frac{\left( X^{2} e^{-2g} + Y^{2} e^{2g} \right) + 2 R
S \left( X^{2} e^{-g} + Y^{2} e^{g} \right)}{1 + 4 (1-R) R S^{2}}\right]
\end{split}%
\end{equation}
where: 
\begin{equation}
\begin{split}
&P_{\vert 1 \rangle}(X,Y,R,t) = 1- \frac{4 (1-R)}{1 + 4 (1-R) R S^{2}} \left[
\frac{1}{2} \left( 1 + 2 R S^{2} \right) + \right. \\
&+ \left( X^{2} e^{-2g} + Y^{2} e^{2g} \right) + 2 \left(1 + 2 R S^{2}
\right) \times \\
&\times \left. \frac{\left( X^{2} e^{-2g} + Y^{2} e^{2g} \right) + 2 R S
\left( X^{2} e^{-g} + Y^{2} e^{g} \right)}{1 + 4 (1-R) R S^{2}} \right]
\end{split}%
\end{equation}
The region where the Wigner function is negative becomes smaller when the
parameter $R$ of the lossy channel is increased. In Fig.\ref%
{fig:single_mode_1_photon_amplified_losses} we report the plots of $W_{\vert
1 \rangle}(X,Y,R,t)$ for different values of the reflectivity $R$. As a
first effect, the negative region is deleted for a
reflectivity $R=1/2$: Fig.\ref{fig:single_mode_1_photon_amplified_losses}-(d).
Then, the form of the distribution remain unchanged until the reflectivity
becomes close to 1 and all the photons present in the states are lost: $R \langle n
\rangle \simeq \langle n \rangle$. As a further analysis, let us consider
the value at $X=0$ and $Y=0$, in which the Wigner function has the maximum
negativity. We obtain that $W_{\vert 1 \rangle} (0,0,R,t) < 0$ for $R \leq 
\frac{1}{2}$, showing that the negativity is maintained in that range of the
lossy channel efficiency. This property can be analyzed by the
two-dimensional contour plots of fig.\ref{fig:contour_wigner_losses}.

As a further analysis, starting from the Wigner function, we can evaluate
the fluctuations in the quadrature operators after losses
(Fig.\ref{fig:uncertainty_relations}-(\textbf{c}): 
\begin{eqnarray}
\Delta X &=& \frac{1}{2} \sqrt{3 T e^{2g} + (1 - T)} \\
\Delta Y &=& \frac{1}{2} \sqrt{3 T e^{-2g} + (1 - T)}
\end{eqnarray}
As in the squeezed vacuum case, poissonian noise added by the lossy channel
increases the fluctuations $\Delta X \Delta Y$ for any non-zero value of the
losses parameter $R$: Fig.\ref{fig:uncertainty_relations}-(\textbf{d}).

\subsection{Two-mode collinear amplifier}

The previous calculation, performed in the case of a single-mode degenerate
OPA, can be again used to analyze the collinear QI-OPA case. The
characteristic function with a generic input two-mode state $\vert \Psi_{in}
\rangle$ can be evaluated as: 
\begin{equation}
\begin{split}
\chi_{N,M}(\eta,\xi,R,t) &= \langle \Psi_{in} \vert e^{\eta \hat{c}%
^{\dag}_{H}(t) - \eta^{\ast} \hat{c}_{H}(t)} \otimes \\
& \otimes e^{\xi \hat{c}^{\dag}_{V}(t) - \xi^{\ast} \hat{c}_{V}(t)} \vert
\Psi_{in} \rangle
\end{split}%
\end{equation}
Analogously to the single-mode OPA case, the time evolution of the field
operators, due to the amplification process and to the propagation in the
lossy channel, takes the form: 
\begin{eqnarray}
\label{eq:Heisen_OPA_loss_1}
\hat{c}^{\dag}_{H}(t) &=& \sqrt{1-R} \, \hat{a}^{\dag}_{H}(t) + \imath \sqrt{%
R} \, \hat{b}_{H}^{\dag} \\
\label{eq:Heisen_OPA_loss_2}
\hat{c}^{\dag}_{V}(t) &=& \sqrt{1-R} \, \hat{a}^{\dag}_{V}(t) + \imath \sqrt{%
R} \, \hat{b}_{V}^{\dag}
\end{eqnarray}
where the expressions of the equation of motion for $\hat{a}_{H,V}^{\dag}(t)$
are (\ref{eq:OPA_coll_Heisen}). We now proceed by writing the characteristic
function in the $\left\{ \vec{\pi}_{+}, \vec{\pi}_{-} \right\}$ basis and by
substituting the relations (\ref{eq:Heisen_OPA_loss_1}-\ref{eq:Heisen_OPA_loss_2}) 
for the time evolution in the Heisenberg picture of
the field operators. The characteristic function then reads: 
\begin{equation}
\begin{split}
&\chi_{N,M}(\eta,\xi,R,t) = \langle \Psi_{in} \vert e^{\frac{\eta(t) \sqrt{%
1-R}}{\sqrt{2}} \hat{a}^{\dag}_{+} - \frac{\eta^{\ast}(t) \sqrt{1-R}}{\sqrt{2%
}} \hat{a}_{+}} \\
&\otimes e^{\frac{\imath \sqrt{R} (\eta + \xi)}{\sqrt{2}} \hat{b}^{\dag}_{+}
- \frac{- \imath \sqrt{R} (\eta^{\ast} + \xi^{\ast})}{\sqrt{2}} \hat{b}_{+}}
\otimes e^{\frac{\xi(t) \sqrt{1-R}}{\sqrt{2}} \hat{a}^{\dag}_{-} - \frac{%
\xi^{\ast}(t) \sqrt{1-R}}{\sqrt{2}} \hat{a}_{-}} \\
&\otimes e^{\frac{\imath \sqrt{R} (\xi - \eta)}{\sqrt{2}} \hat{b}^{\dag}_{-}
- \frac{- \imath \sqrt{R} (\xi^{\ast} - \eta^{\ast})}{\sqrt{2}} \hat{b}_{-}}
\vert \Psi_{in} \rangle
\end{split}%
\end{equation}
The transformation between $\left(\eta, \eta^{\ast}, \xi, \xi^{\ast} \right)
\rightarrow \left(\eta(t),\eta^{\ast}(t), \xi(t), \xi^{\ast}(t) \right)$ is
the same (\ref{eq:trans_wigner_char_1}-\ref{eq:trans_wigner_char_2}) of the
unperturbed case. The evaluation of the average on an input injected state $%
\vert N+,M- \rangle$ in the QI-OPA and on the vacuum-injected port $b$ of
the beam-splitter leads to the following result: 
\begin{equation}
\begin{split}
&\chi_{N,M}(\eta,\xi,R,t) = e^{- \frac{1}{4} R \left( \vert \eta + \xi
\vert^{2} + \vert \xi - \eta \vert^{2}\right)} \\
&e^{-\frac{1}{4}(1-R) \left( \vert \eta(t) \vert^{2} + \vert \xi(t)
\vert^{2}\right)} \left\{ \sum_{n=0}^{N} \sum_{m=0}^{M} c_{n}^{N} c_{m}^{M}
\right. \\
& \left. \left( \frac{\vert \eta(t) \vert^{2}}{2} \right)^{n} \left( \frac{%
\vert \xi(t) \vert^{2}}{2} \right)^{m} \right\}
\end{split}%
\end{equation}
The Wigner function is then calculated as the 4-dimensional Fourier
transform of the characteristic function.
Analogously to the single-mode case, we proceed with the calculation by
evaluating the Fourier integral that, after changing the integration
variables with the transformation $\left(\eta, \eta^{\ast}, \xi, \xi^{\ast}
\right) \rightarrow \left(\eta(t), \eta^{\ast}(t), \xi(t), \xi^{\ast}(t)
\right)$, can be written as: 
\begin{widetext}
\begin{equation}
\begin{split}
W_{\vert N+,M- \rangle}(\alpha, \beta, R, t) &= \frac{1}{\pi^{4}} \int d^{2}\eta(t) \int d^{2}\xi(t) \vert J \vert \exp \left[ -\frac{1}{2} \varepsilon \left( \vert \eta(t) \vert^{2} + \vert \xi(t) \vert^{2} \right) \right] \exp \left[ -\frac{1}{2} \kappa \left( \eta^{2}(t) + \eta^{\ast \, 2}(t) - \xi^{2}(t) - \xi^{\ast \, 2} (t) \right) \right] \\
&\left( \sum_{n=0}^{N} c_{n}^{N} \frac{\vert \eta(t) \vert^{2n}}{2^{n}} \right)
\left( \sum_{m=0}^{M} c_{m}^{M} \frac{\vert \xi(t) \vert^{2m}}{2^{m}} \right)
\exp \left[ \frac{1}{\sqrt{2}} \left( \eta^{\ast}(t) \overline{\alpha} - \eta(t) \overline{\alpha}^{\ast} \right) \right] \exp \left[\frac{1}{\sqrt{2}}\left( \xi^{\ast}(t) \overline{\beta} - \xi(t) \overline{\beta}^{\ast} \right) \right]
\end{split}
\end{equation}
\end{widetext}
where the parameters $\varepsilon, \kappa$ have been defined in 
(\ref{eq:epsilon_wigner}-\ref{eq:kappa_wigner}), and the transformation between $%
\alpha \rightarrow \overline{\alpha}$ and $\beta \rightarrow \overline{\beta}
$ is: 
\begin{eqnarray}
\overline{\alpha} &=& \frac{1}{\sqrt{2}} \left[ (\alpha + \beta) C -
(\alpha^{\ast} + \beta^{\ast}) S \right] \\
\overline{\beta} &=& \frac{1}{\sqrt{2}} \left[ (\beta - \alpha) C +
(\beta^{\ast} - \alpha^{\ast}) S \right]
\end{eqnarray}
The derived integral relations (\ref{eq:integral_relations_wigner}) lead to
the final result: 
\begin{equation}
\begin{split}
W_{\vert N+,M- \rangle}(\alpha, \beta, R, t) &= \frac{1}{\pi^{2}} \left(
\sum_{n=0}^{N} c_{n}^{N} I_{n}(\kappa, \kappa, \varepsilon, \overline{\alpha}%
) \right) \\
& \times \left( \sum_{m=0}^{M} c_{m}^{M} I_{m} (-\kappa, -\kappa,
\varepsilon, \overline{\beta}) \right)
\end{split}%
\end{equation}

Let us analyze the spontaneous emission case, when $N=M=0$. Adopting the
same definition of the phase space used in the unperturbed case in Section %
\ref{subsec:two_mode_amplifier}, we obtain: 
\begin{equation}
\begin{split}
W&_{\vert 0+,0- \rangle}(X,Y,R,t) = \left( \frac{2}{\pi} \right)^{2} \frac{1%
}{1 + 4 S^{2} R (1-R)} \\
&\times \exp \left[- \frac{\left( 1 + 2 R S^{2} + 2 R C S \right)
X^{2} e^{-2g}}{1 + 4 S^{2} R (1-R)} \right] \\
&\times \exp \left[ - \frac{\left( 1 + 2 R S^{2} - 2 R C S \right) Y^{2}
e^{2g}}{1 + 4 S^{2} R (1-R)} \right]
\end{split}%
\end{equation}
As in the single mode case, the degree of squeezing in the quadrature
operators is progressively decreased by the poissonian noise introduced by
the lossy channel.

\begin{figure}[tbp]
\centering
\includegraphics[width=0.5\textwidth]{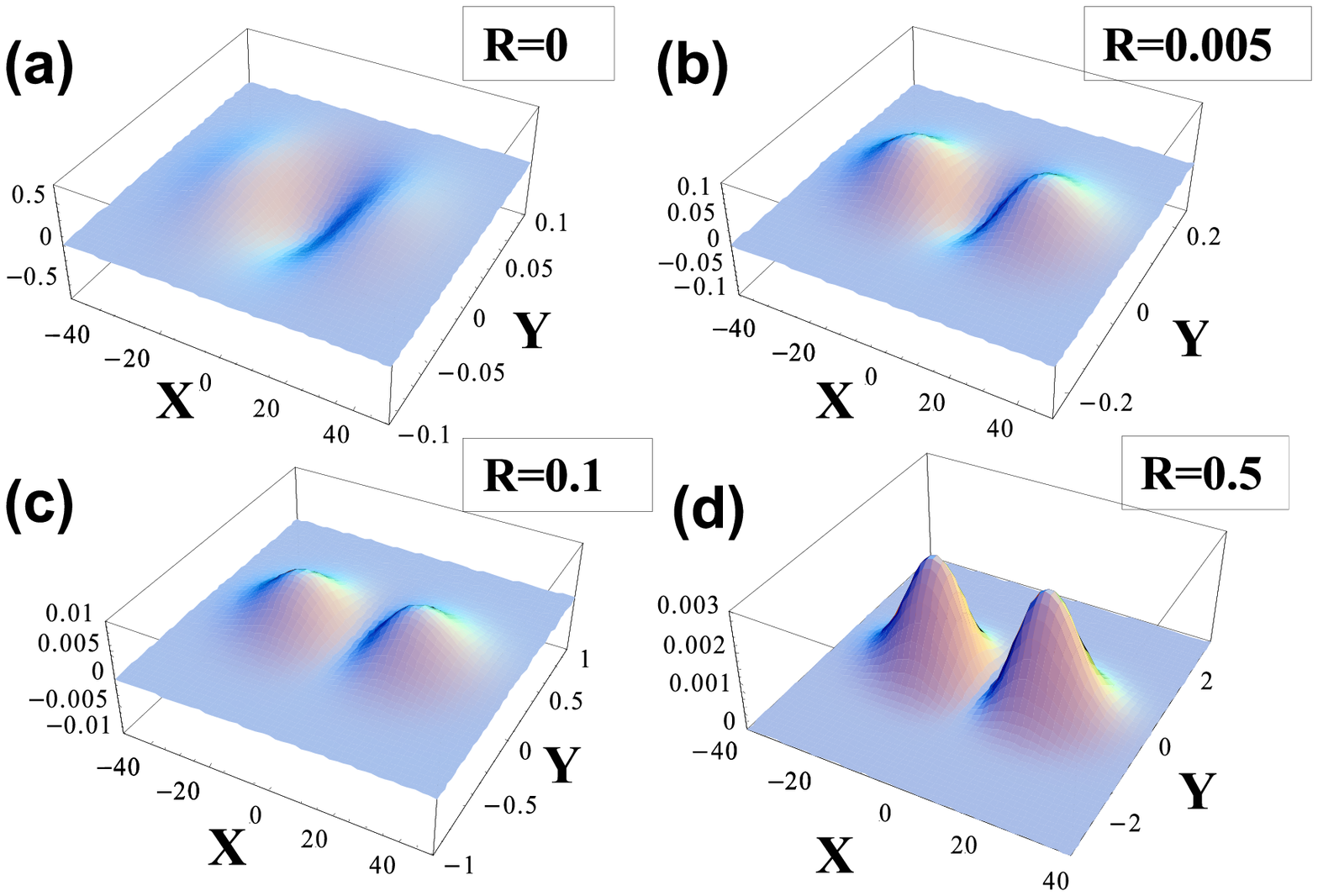}
\caption{Wigner function of a single-photon $\vert 1+ \rangle$ amplified
state in a two-mode degenerate OPA for $g=3$. (\textbf{a}) (R=0) Unperturbed case.
(\textbf{b}) (R=0.005) For small reflectivity, the Wigner function remains negative
in the central region. (\textbf{c}) (R=0.1) The Wigner function progressively evolve
in a positive function in all the phase-space. (\textbf{d}) (R=0.5) Transition from a
non-positive to a completely positive Wigner function.}
\label{fig:two_mode_1_photon_amplified_losses.eps}
\end{figure}

Furthermore, let us analyze the case of the single photon amplified states,
i.e. $N=1,M=0$ and $N=0,M=1$. The Wigner functions after losses for this
quantum states are: 
\begin{eqnarray}
W_{\vert 1+,0- \rangle}(X,Y,R,t) &=& W_{\vert 0+,0- \rangle} \, P_{\vert
1+,0- \rangle}(X,R,t) \\
W_{\vert 0+,1- \rangle}(X,Y,R,t) &=& W_{\vert 0+,0- \rangle} \, P_{\vert
0+,1- \rangle}(Y,R,t)
\end{eqnarray}
where $P_{\vert 1+,0- \rangle}(X,R,t)$ and $P_{\vert 0+,1- \rangle}(Y,R,t)$
are the following $2^{nd}$ order polynomials: 
\begin{equation}
\begin{split}  \label{eq:polyn_1p0m}
P_{\vert 1+,0- \rangle}&(X,R,t) = 1 + \frac{4 (1-R)}{1 + 4 S^{2} R (1-R)}
\times \\
&\times \left[ - \frac{1}{2} \left( 1 + X^{2} e^{-2g} \right) - R S^{2} +
\left( 1 + 2 R S^{2} \right) \times \right. \\
&\times \left.\frac{1 + 2 R S^{2} + 2 R C S X^{2} e^{-2g}}{1 + 4 R (1-R)
S^{2}} \right]
\end{split}%
\end{equation}
\begin{equation}
\begin{split}
P_{\vert 0+,1- \rangle}&(Y,R,t) = 1 + \frac{4 (1-R)}{1 + 4 S^{2} R (1-R)}
\times \\
&\times \left[ - \frac{1}{2} \left( 1 + Y^{2} e^{2g} \right) - R S^{2} +
\left( 1 + 2 R S^{2} \right) \times \right. \\
&\times \left.\frac{1 + 2 R S^{2} - 2 R C S Y^{2} e^{2g}}{1 + 4 R (1-R) S^{2}%
} \right]
\end{split}%
\end{equation}
In Fig.\ref{fig:two_mode_1_photon_amplified_losses.eps} we report the Wigner
function $W_{\vert 1+,0- \rangle}(X,Y,R,t)$ for different values of the
reflectivity $R$. The evolution of this distribution is similar to the
single-mode OPA case analyzed previously.

\subsection{Resilience of quantum properties after decoherence}

The Wigner functions calculated in the previous section allows to obtain a
complete overview of the phase-space properties of the QI-OPA amplified states
after the propagation over a lossy channel. In particular, we focus on the
negativity of the W-representation in the specific case of a single-photon
injected qubit. As said, the presence of a negative region in
the phase-space domain is one possible parameter to recognize the
non-classical properties of a generic quantum state.

Let us now consider the expression (\ref{eq:polyn_1p0m}) for the polynomial $%
P_{\vert 1+,0- \rangle}(X,R,t)$. In the lossless regime considered in Fig.%
\ref{wigner3}, the Wigner function takes its minimum value in the origin of
the phase-space (X=0,Y=0). In presence of losses, the Wigner function
remains negative in the origin for $R<\frac{1}{2}$. 
This behaviour is shown in the plots of the $Y=0$ sections of $%
W_{\vert 1+, 0- \rangle}(X,Y,R,t)$ reported in Fig.\ref%
{fig:QIOPA_negative_section}, where for the value of $R=\frac{1}{2}$ ceases
to be negative. This is also shown in Fig.\ref%
{fig:Negativity_QIOPA}, that reports the value of $W_{\vert 1+,0-
\rangle}(0,0,R,t)$ as a function of the reflectivity $R$ of the
beam-splitter that models the lossy channel. Finally, this behaviour, as for
the quantum superposition of coherent states in the previous Section, can be
analytically obtained by calculating $P_{\vert 1+,0- \rangle}(X,R,t)$ in $%
X=0 $. We obtain: 
\begin{equation}
P_{\vert 1+,0- \rangle}(X,R,t) = \frac{2R - 1}{1 + 4 R (1-R) S^{2}} \left\{ 
\begin{array}{lrl}
< 0 & \mathrm{if} & R< \frac{1}{2} \\ 
> 0 & \mathrm{if} & R> \frac{1}{2}%
\end{array}
\right.
\end{equation}

\begin{figure}[tbp]
\centering
\includegraphics[width=0.5\textwidth]{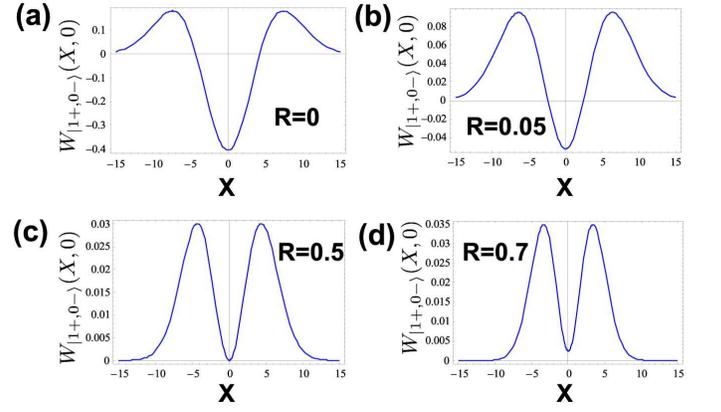}
\caption{$Y=0$ section of the Wigner function for QIOPA single photon
amplified states for $\langle n \rangle \approx 36$. (a) $R=0$: Unperturbed
Wigner function, negative in the origin. (b) $R=0.05$: The negativity in the
origin progressively decreases due to the coupling with the enviroment. (c) $%
R=0.5$: Transition from a non-positive to a completely positive Wigner
function. (d) $R=0.7$: The Wigner function is positive in all the $\left\{
X,Y \right\}$ plane.}
\label{fig:QIOPA_negative_section}
\end{figure}

\begin{figure}[tbp]
\centering
\includegraphics[width=0.4\textwidth]{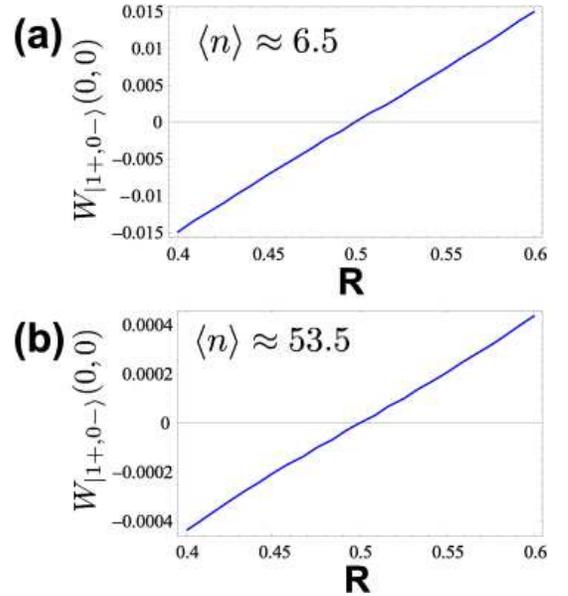}
\caption{Transition from a non-positive distribution to a completely
positive Wigner function for QI-OPA single-photon amplified states in
presence of decoherence, as a function of the beam-splitter reflectivity $R$%
. Average photon number of (a) $\langle n \rangle \approx 6.5$ and (b) $%
\langle n \rangle \approx 53.5$. The negativity is evaluated as the value of
the Wigner function in the origin.}
\label{fig:Negativity_QIOPA}
\end{figure}

\noindent This analysis explicitly shows the persistence of quantum
properties in the QI-OPA single-photon amplified state even in presence of
losses. However, the negativity of the Wigner function is not the only
parameter that reveals the quantum behaviour in a given system. Hence, more
detailed analysis have to be performed including different approaches in
order to investigate the $R>\frac{1}{2}$ regime.

\begin{figure}[tbp]
\centering
\includegraphics[width=0.4\textwidth]{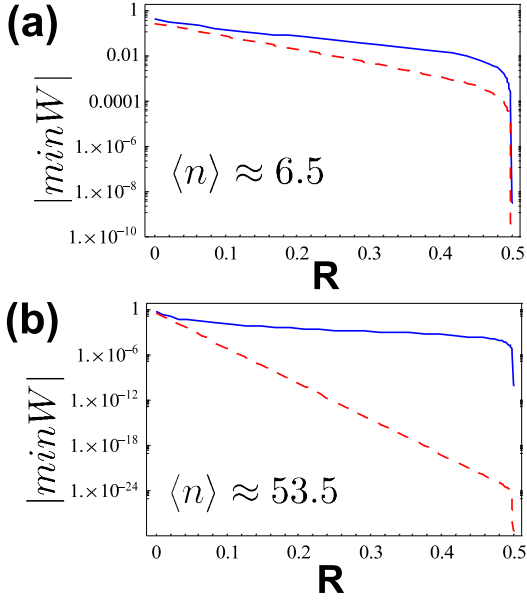}
\caption{Absolute value of the negativity, evaluated as the minimum of the
Wigner function, in the reflectivity range $0 \leq R \leq \frac{1}{2}$. The
blue straight lines correspond to QI-OPA amplified states, while red dashed lines to the $%
\vert \protect\alpha \rangle \pm \vert - \protect\alpha \rangle$ quantum
superposition. (a) $\langle n \rangle \approx 6.5$ and (b) $\langle n \rangle 
\approx 53.5$.}
\label{fig:Negativity_comparison}
\end{figure}

We conclude our analysis by comparing the behaviour of the two different
macroscopic quantum superpositions analyzed in this paper. In Fig.
\ref{fig:Negativity_comparison} we compare the decrease of the Wigner function
negativity for the QI-OPA single-photon amplified states and for the
superposition of coherent states, for two different values of the average
number of photons. We observe that the QI-OPA solution possesses a higher
resilience to losses, i.e. a slower decrease of the negative part, with
respect to the $\vert \alpha \rangle \pm \vert - \alpha \rangle$ states.
However, the Wigner function for both quantum superposition ceases to be
negative at $R=\frac{1}{2}$.

\section{Persistence of coherence in macroscopic quantum superpositions:
Fock-space analysis}
\label{sec:Bures_Fock}

In this Section we perform a complementary analysis in the Fock-space of the
macroscopic quantum superposition generated by the quantum cloning of
single-photon states, by applying two criteria \cite{DeMa09} based on the
concept of distance in Hilbert spaces, more specifically the Bures metric.
This approach allows to quantify from a different point of view the different 
resilience to losses that this quantum superposition posses in contrast to 
the fragility of the $\vert
\alpha \rangle$ states MQS.

First, we introduce the coherence criteria and discuss their interpretation.
Then we apply this last approach to the two different MQS under investigation
showing analogies and differences. Furthermore, with an opportune POVM
technique, based on the O-Filter device introduced in \cite{Naga07,DeMa08},
the properties in the Fock-space of QIOPA amplified states can be exploited
to obtain a higher discrimination in the measurement stage, at the cost of a
lower events rate.

\subsection{Criteria for Macroscopic Quantum Superpositions}

\textbf{Metrics in Hilbert spaces}. In order to distinguish between two
different quantum states, we need to define a metric distance in the Hilbert
space. An useful parameter to characterize quantitatively the overlap of two
quantum states is the fidelity between two generic density matrices $%
\widehat{\rho }$ and $\widehat{\sigma }$, defined as: $\mathcal{F}(\widehat{%
\rho },\widehat{\sigma })=\mathrm{Tr}^{2}\left( \sqrt{\widehat{\rho }^{\frac{%
1}{2}}\widehat{\sigma }\widehat{\rho }^{\frac{1}{2}}}\right) $ \cite{Jozs94}%
. This parameter reduces to $\mathcal{F}(|\psi \rangle ,|\varphi \rangle
)=\left| \langle \psi |\varphi \rangle \right| ^{2}$ for pure states, and is
an extension of the scalar product between quantum states to the density
matrix formalism. We have $0\leq \mathcal{F}\leq 1$, where $\mathcal{F}=1$
for identical states, and $\mathcal{F}=0$ for orthogonal states. This
quantity is not a metric, but can be adopted to define two different useful
metrics, which are the angle distance $D_{A}(\widehat{\rho },\widehat{\sigma 
})=\arccos \mathcal{F}(\widehat{\rho },\widehat{\sigma })$ \cite{Niel00},
and the Bures distance \cite{Bure69,Hubn92,Hubn93}: 
\begin{equation}  \label{eq:Bures_distance_definition}
D(\widehat{\rho },\widehat{\sigma })=\sqrt{1-\sqrt{\mathcal{F}(\widehat{\rho 
},\widehat{\sigma })}}
\end{equation}
Furthermore, the fidelity can also be used to calculate a lower and an upper
bound for the trace distance, defined as $D_{T}(\widehat{\rho },\widehat{%
\sigma })=\frac{1}{2}\mathrm{Tr}\left| \widehat{\rho }-\widehat{\sigma }%
\right| $ and related to the fidelity by \cite{Niel00}: $1-\sqrt{\mathcal{F}
(\widehat{\rho },\widehat{\sigma })}\leq D(\widehat{\rho },\widehat{\sigma }%
) \leq \sqrt{1-\mathcal{F}(\widehat{\rho },\widehat{\sigma })}$. In this
Section we will adopt the Bures distance as a metric in the quantum state
space, as it is connected to the probability of obtaining an inconclusive
result with a suitable Positive Operator Valued Measurement (POVM) \cite%
{Hutt96,Pere95}, which is $\sqrt{F\left(|\phi \rangle ,|\psi \rangle \right) 
}=|\langle \psi |\phi \rangle |$ for pure states .

\textbf{Distinguishability, MQS\ Visibility}. Let us characterize two
macroscopic states $|\phi _{1}\rangle $ and $|\phi _{2}\rangle $ and the
corresponding MQS's: $|\phi ^{\pm }\rangle =\frac{\mathcal{N}_{\pm }}{\sqrt{2%
}}\left( |\phi _{1}\rangle \pm |\phi _{2}\rangle \right) $ by adopting two
criteria. \textbf{I)} The \textit{distinguishability} between $|\phi
_{1}\rangle $ and $|\phi _{2}\rangle $ can be quantified as $D\left( |\phi
_{1}\rangle ,|\phi _{2}\rangle \right) $. \textbf{II)} The ''\textit{%
Visibility''}, i.e. ''degree of orthogonality'' of the MQS's $|\phi ^{\pm
}\rangle $ is expressed again by: $D\left( |\phi ^{+}\rangle ,|\phi
^{-}\rangle \right) $. Indeed, the value of the MQS\ visibility depends
exclusively on the relative phase of the component states:$\ |\phi
_{1}\rangle $ and $|\phi _{2}\rangle $. Assume two orthogonal superpositions 
$|\phi ^{\pm }\rangle $: $D\left( |\phi ^{+}\rangle ,|\phi ^{-}\rangle
\right) =1$. In presence of losses the relative phase between $\ |\phi
_{1}\rangle $ and $|\phi _{2}\rangle $ progressively randomizes and the
superpositions $|\phi ^{+}\rangle $ and $|\phi ^{-}\rangle $ approach an
identical fully mixed state leading to: $D\left( |\phi ^{+}\rangle ,|\phi
^{-}\rangle \right) =0$. The aim of this Section is to study the evolution
in a lossy channel of the phase decoherence acting on two macroscopic states 
$|\phi _{1}\rangle $ and $|\phi _{2}\rangle \ $and on the corresponding
superpositions $|\phi ^{\pm }\rangle $ and the effect on the size of the
corresponding $D\left( |\phi _{1}\rangle ,|\phi _{2}\rangle \right) $ and $%
D\left( |\phi ^{+}\rangle ,|\phi ^{-}\rangle \right) $.

\subsection{Bures distance for $\vert \protect\alpha \rangle$ states MQS}

In order to quantify the loss of coherence in the macroscopic superposition
of coherent states, we estimate their relative Bures distance. For the basis 
$\alpha$ states we have: 
\begin{equation}
D(|\alpha e^{\imath \varphi }\rangle ,|\alpha e^{-\imath \varphi }\rangle )=%
\sqrt{1-e^{-2T|\alpha |^{2}\sin ^{2}\varphi }}
\end{equation}
The distinguishability between these two states keeps close to $1$ up to
high values of the beamsplitter reflectivity, when almost all photons are
reflected.

Let us now estimate the distance between the superposition states. We
consider the following condition $|\langle \beta e^{\imath \varphi }|\beta
e^{-\imath \varphi }\rangle| =e^{-2T|\alpha |^{2}\sin ^{2}\varphi
} \approx 0$, which corresponds to $T|\alpha |^{2}\sin ^{2}\varphi >1$. Hence,
except for very small $T$, the coherent states $\vert \beta e^{\pm \imath \varphi} 
\rangle$ after the propagation over a lossy channel remain almost orthogonal. In
such situation we can associate to $\left\{ |\beta e^{\imath \varphi }
\rangle {},|\beta e^{-\imath \varphi }\rangle {}\right\} $ the two orthogonal 
states of a qubit $\left\{ |0\rangle {},|1\rangle {}\right\} $. 
Let us introduce the parameters $\chi =2|\alpha |^{2}R\sin ^{2}\varphi $ and 
$\psi =|\alpha |^{2}R\sin 2\varphi $. The density matrices after 
losses can be represented as $2\times 2$ matrices associated to 
the qubit state: 
\begin{equation}
\hat{\rho}_{C}^{\pm \,\varphi }=\frac{1}{2} 
\begin{pmatrix}
1 & \pm e^{-\chi }e^{\imath \psi } \\ 
\pm e^{-\chi }e^{-\imath \psi } & 1%
\end{pmatrix}%
\end{equation}
To estimate the fidelity $\mathcal{F}\left( \hat{\rho}_{C}^{+\,\varphi },%
\hat{\rho}_{C}^{-\,\varphi }\right) $ we need to calculate: 
%
\begin{equation}
\left( \hat{\rho}_{C}^{+\,\varphi }\right) ^{\frac{1}{2}}\hat{\rho}%
_{C}^{-\,\varphi }\left( \hat{\rho}_{C}^{+\,\varphi }\right) ^{\frac{1}{2}}=%
\frac{1}{4} 
\begin{pmatrix}
1-e^{-2\chi } & 0 \\ 
0 & 1-e^{-2\chi }%
\end{pmatrix}%
\end{equation}
Hence we get: 
\begin{equation}
\mathcal{F}\left( \hat{\rho}_{C}^{+\,\varphi },\hat{\rho}_{C}^{-\,\varphi
}\right) =1-e^{-2\chi }=1-e^{-4|\alpha |^{2}R\sin ^{2}\varphi }
\end{equation}
From the definition (\ref{eq:Bures_distance_definition}), we found: 
\begin{equation}  \label{eq:universal_coherent}
D\left( \hat{\rho}_{C}^{\varphi \,+},\hat{\rho}_{C}^{\varphi \,-}\right) =%
\sqrt{1-\sqrt{1-e^{-4R|\alpha |^{2}\sin ^{2}\varphi }}}
\end{equation}
This curve represents all the coherent states MQS's of the form 
(\ref{eq:superposition_coherent}), for any value of $\alpha$ and $\varphi$.
The distance depends exclusively from the average number of reflected
photons, $R|\alpha |^{2}$, multiplied by a scale factor $\xi _{\varphi
}^{2}=\sin ^{2}\varphi $. This term is proportional to the phase-space
distance (Fig.\ref{fig:coherent_phase_space_blob}) between the two 
components $|\alpha e^{\imath \varphi }\rangle {}$%
and $|\alpha e^{-\imath \varphi }\rangle {}$, equal to $d_{\varphi
}^{2}=4 |\alpha |^{2}\sin ^{2}\varphi $. In fig.\ref%
{fig:bures_distance_coherent} are reported the distances for different
values of $\varphi $ for $\alpha =4$. 
\begin{figure}[th]
\centering
\includegraphics[width=0.50\textwidth]{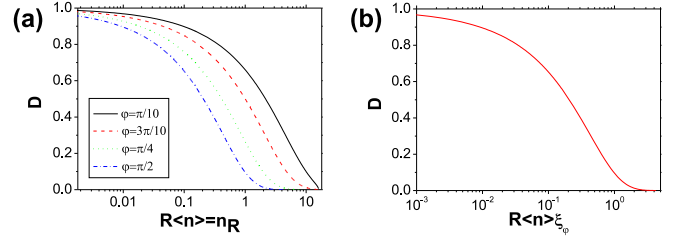}
\caption{(a) Analytical results of the Bures distance for the superposition
of coherent states, with $\protect\alpha =4$ and for different values of the
phase, as function of the average number of reflected photons. (b) Universal
curve of the Bures distance for the superposition of coherent states for any 
$\alpha$ and $\varphi$, obtained by plotting eq.(\ref{eq:universal_coherent}) 
as a function of the parameter $x = R\langle n\rangle \protect\xi _{\protect\varphi }$, 
with $\protect \xi _{\protect\varphi }=\sin ^{2}\protect\varphi $.}
\label{fig:bures_distance_coherent}
\end{figure}
For $\varphi =\frac{\pi }{2}$, the coherent states exhibit opposite phases.
Such condition represents the limit situation, in which the loss of
coherence is higher since $\sin ^{2}\varphi =1$. We observe, as is shown by
fig.\ref{fig:bures_distance_coherent}, that the value of $D$ is reduced down
to $\sim 0.1$ once $R|\alpha |^{2}=R\langle n\rangle =1$. Hence, the loss 
on the average of a single photon cancels most of the coherence in the quantum 
superposition state.

\subsection{Bures distance for QI-OPA amplified states}

\begin{figure}[t]
\centering
\includegraphics[scale=.38]{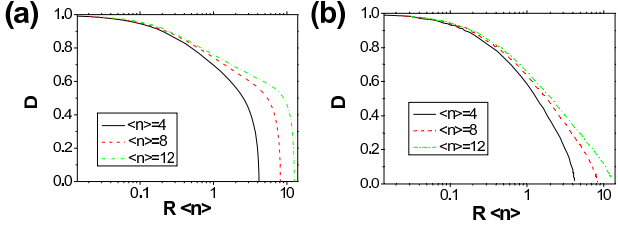}
\caption{(\textbf{a}) Numerical evaluation of the distance $D(x)$ between
two orthogonal equatorial macro-qubits $|\Phi^{\protect\phi,\protect\phi%
_{\bot}}\rangle$ as function of the average lost particle $x=R<n>$. Black straight
line refers to $g=0.8$ and hence to $\langle n \rangle \approx 4$, red dashed line
to $g=1.1$ and $\langle n \rangle \approx 8$, green dash-dotted line to $g=1.3$ and $%
\langle n \rangle \approx 12$. (\textbf{b}) Numerical evaluation of the
distance $D(x)$ between two orthogonal linear macro-qubits $%
|\Phi^{H,V}\rangle$ as function of the average lost particle $x=R<n>$. Black straight
line refers to $g=0.8$ and hence to $\langle n \rangle \approx 4$, red dashed line
to $g=1.1$ and $\langle n \rangle \approx 8$, green dash-dotted line to $g=1.3$ and $%
\langle n \rangle \approx 12$.}
\label{fig:fig4}
\end{figure}

As a following step, we have evaluated numerically the \textit{%
distinguishability of }$\left\{ |\Phi ^{+,-}\rangle \right\} $ through the
distance $D(|\Phi ^{+}\rangle ,|\Phi ^{-}\rangle )$ between the multiphoton
states generated by QI-OPA. It is found that this property of $\left\{
|\Phi^{+,-}\rangle \right\} $ coincides with the MQS\ Visibility of $|\Psi
^{\pm}\rangle $, in virtue of the phase-covariance of the process: $%
D(|\Psi^{+}\rangle ,|\Psi ^{-}\rangle )$ = $D(|\Phi ^{R}\rangle ,|\Phi
^{L}\rangle )$= $D(|\Phi ^{+}\rangle ,|\Phi ^{-}\rangle )$. The \textit{%
visibility} of the MQS $\left\{ |\Psi ^{+,-}\rangle \right\} $ has been
evaluated numerically analyzing the Bures distance as a function of the
average lost photons: $x\equiv R<n>$. This calculation have been performed
by taking the complete expression of the density matrix, reported in Section
IV and Appendices A-B, and by performing an approximate calculation of the 
fidelity through numerical algebraic matrix routines. This algorithm has been 
tested by evaluating numerically the Bures distance between the quantum superposition
of coherent states $\vert \alpha \rangle \pm \vert - \alpha \rangle$. The
comparison with the analytical result of Eq.(\ref{eq:universal_coherent})
gave a high confidence level for the approximate results. The results for
different values of the gain for equatorial macroqubits are reported in Fig.%
\ref{fig:fig4} -(\textbf{a}).

Note that for small values of $x$ the decay of $D(x)$ is far slower than for
the coherent state case shown in Fig.\ref{fig:bures_distance_coherent}- (%
\textbf{b}). Furthermore, after a common inflection point at $D\sim 0.6$ the 
\textit{slope} of all functions $D(x)$ corresponding to different values of $%
<n>$ increases fast towards the \textit{infinite} value, for increasing $\
x\rightarrow <n>$ and: $R\rightarrow 1$. The latter property can be
demonstrated with a perturbative approach on the density matrix. We find, in
the low $T$ and high gain limit of $\frac{\partial D(\hat{\rho}%
^{T}_{\varphi}, \hat{\rho}^{T}_{\varphi_{\bot}})}{\partial T}$, the slope: 
$ \lim_{g \rightarrow \infty} \lim_{T \rightarrow 0} \frac{\partial 
D(\hat{\rho}^{T}_{\varphi}, \hat{\rho}^{T}_{\varphi_{\bot}})}{\partial T} 
= \lim_{g \rightarrow \infty} (1 + 4 C^{2} + 2 C^{2} \Gamma 
(1 + 2 \Gamma^{2})^{\frac{1}{2}}) = \infty$. 
For total particle loss, $R$ = $1$ and $x$ = $<n>$ is: $D(x)$ = $0$. All
this means that the MQS\ Visibility can be significant even if the average
number $x$ of lost particles is very close to the initial total number $<n>$,
i.e. for $R\sim 1$. This behavior is opposite to the case of \textit{%
coherent states} where the function $D(R|\alpha |^{2})$ approaches the zero
value with a \textit{slope} equal to \textit{zero}: Fig.\ref%
{fig:bures_distance_coherent}-(b).

For sake of completeness, we then performed the same calculation for the
linear macroqubits $\vert \Phi^{H,V} \rangle$, and the results are reported
in Fig.\ref{fig:fig4}-(b). For this injected qubit, not lying in the
equatorial plane of the Bloch sphere, the amplification process does not
correspond to an "\emph{optimal cloning machine}". For this reason, the flow
of noise from the enviroment in the amplification stage is not the minimum
optimal value, and hence the output states possess a faster decoherence
rate. Indeed, the output distributions, as shown in Fig.\ref{fig:fig3}-(b),
do not possess the strong unbalancement in polarization of the equatorial
macroqubits $\vert \Phi^{\varphi} \rangle$ that is responsible of their
resilience structure.

\subsection{Distinguishability enhancement through an Orthogonality-Filter}

\begin{figure}[ht]
\centering
\includegraphics[width=0.5\textwidth]{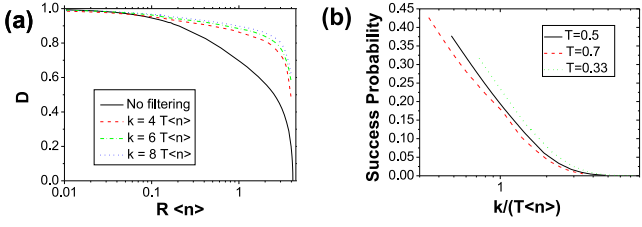}
\caption{(a) Numerical evaluation of the Bures distance between two orthogonal
equatorial macro-qubits after discrimination with the O-Filter for
different values of the threshold $k$ ($g=0.8$). The filtering probabilities
for the three cases are respectively $P(k=4T\langle n \rangle) = 1.6*10^{-3}$, 
$P(k=6T\langle n \rangle) = 7.14*10^{-5}$ and $P(k=8T\langle n \rangle) = 3.06*10^{-6}$.
(b) Success probability for the O-Filter for $g=0.8$ and different values of the
transmittivity as a function of the parameter $\frac{k}{T \langle n \rangle}$. The latter
expresses the threshold $k$ with respect to the average number of the incident
photons in the detection apparatus. We note that the curves for different transmittivities
can be almost superimposed.}
\label{fig:fig5}
\end{figure}

As a further investigation, we consider the case of a more sophisticated
measurement scheme based on an electronic device named O-Filter (OF). The
demonstration of microscopic-macroscopic entanglement by adopting the
O-Filter based measurement strategy was reported in \cite{DeMa08}. The POVM like technique \cite{Pere95}
implied by this device locally selects, after an intensity measurement the
events for which the difference between the photon numbers associated with
two orthogonal polarizations $|n_{\pi}-m_{\pi_{\bot}}|>k$, i.e. larger than
an adjustable threshold $k$ \cite{Naga07}, where $n_{\pi}$ is the number of
photon polarized $\pi$ and $m_{\pi_{\bot}}$ the number of photon polarized $%
\pi_{\bot}$. By this method a sharper discrimination between the output
states $|\Phi ^{\varphi }\rangle $ and $|\Phi ^{\varphi _{\bot }}\rangle $ can
be achieved. The action of the OF, sketched in Fig.\ref{fig:fig22} can be
formalized through the POVM elements: 
\begin{eqnarray}
\hat{F}^{(+1)}_{\pi,\pi_{\bot}}(k)&=& \sum_{n=k}^{\infty} \sum_{m=0}^{n-k}%
\hat{\Pi}_{n,m} \\
\hat{F}^{(-1)}_{\pi,\pi_{\bot}}(k) &=& \sum_{m=k}^{\infty} \sum_{n=0}^{m-k} 
\hat{\Pi}_{n,m} \\
\hat{F}^{(0)}_{\pi,\pi_{\bot}}(k) &=& \hat{I} - \hat{F}^{(+1)}_{\pi} - \hat{F%
}^{(-1)}_{\pi}
\end{eqnarray}
where the $\hat{\Pi}_{n,m} = |n \pi,m \pi_{\bot}\rangle \, \langle n \pi,m
\pi_{\bot}|$ are Fock-state Von-Neumann projectors that describe the
performed intensity measurement. The average of the couple of operators 
$\langle \hat{F}^{(-1)}_{\pi,\pi_{\bot}}(k) \rangle + \langle 
\hat{F}^{(+1)}_{\pi,\pi_{\bot}}(k) \rangle$ defines the success probability
of the O-Filter, i.e. the rate of events leading to one of the conclusive outcomes
($\pm 1$).
To calculate the action of the O-Filter in the Bures distance, 
we projected the density matrix of the states over the joint subset
corresponding to the ($\pm1$) outcomes, neglecting only the terms leading to
the inconclusive (0) result. Then, the same numerical analysis of the
previous section has been performed. In Fig.\ref{fig:fig5}-(a) the results for $%
g=0.8$ and different values of $k$\ are reported. Note the increase of the
value of $D(x)$, i.e. of the MQS\ Visibility, by increasing $k$. 
Of course, the increase in the MQS Visibility through the O-Filter device 
is achieved at the cost of a lower success probability (Fig.\ref{fig:fig5}-(a)).
According to the graphical analysis of Section \ref{sec:fock_space}
on the photon number distributions of the equatorial amplified macro-states, 
the O-Filter device improves the MQS Visibility since it exploits the peculiar
unbalancement in polarization of the equatorial amplified macro-states in a 
Fock-space analysis. In the selected regions, the $\left\{ \vert 
\Phi^{\varphi} \rangle, \vert \Phi^{\varphi_{\bot}} \rangle \right \}$ states 
can be discriminated with a higher fidelity.
As a further analysis, we reported in fig.\ref{fig:fig5}-(b) the trend of the success
probability as a function of the threshold $k$ for different values of the transmittivity.
Interestingly enough, we note that the success probability depends only on the ratio between
the threshold $k$ and the number of transmitted photons $T \langle n \rangle$, since the three
curves for different $T$ are almost superimposed. This is a consequence of the property
of the photon number distribution of the equatorial macro-qubits, since its form is almost
left unaltered after the propagation over a lossy channel.

\section{Wigner function for non-degenerate quantum injected optical parametric 
amplifier in the EPR configuration}
\label{sec:non_coll_wigner_unpert}

\begin{figure}[ht!]
\centering
\includegraphics[width=0.5\textwidth]{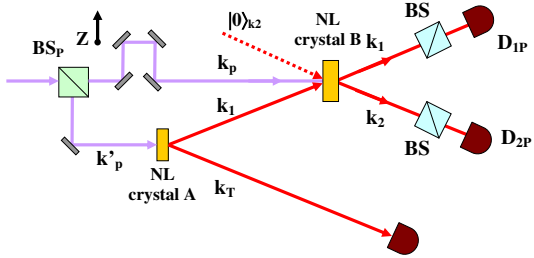}
\caption{Scheme for the non-collinear optical parametric 
amplification of a single-photon state. An EPR pair is generated in a
first non-linear crystal. The photon on mode $\mathbf{k}_{T}$ acts as a
trigger to conditionally prepare the photon on mode $\mathbf{k}_{1}$. The latter
is superimposed with a strong $\mathbf{k}_{P}$ pump beam, and is amplified
in a second non-linear crystal oriented in the same optical configuration
of the EPR source. The beam-splitters on modes $\mathbf{k}_{1}$ and $\mathbf{k}_{2}$ 
in the figure represent the lossy transmission channel.}
\label{fig:scheme_non_coll}
\end{figure}

In this section we derive the Wigner function associated to the non-degenerate 
quantum injected optical parametric amplifier, working in a non-collinear EPR 
configuration. The scheme for this device is reported in 
fig.\ref{fig:scheme_non_coll}.

In the stimulated regime, 
this amplifier acts as an $N \rightarrow M$ \emph{universal optimal 
quantum cloning machine} \cite{Pell03,DeMa04}. 
The interaction Hamiltonian of this device can be written in the form:
\begin{equation}
\label{eq:Ham_noncoll}
\hat{\mathcal{H}}_{epr} = \imath \hbar \chi \left( \hat{a}_{1\pi}^{\dag} 
\hat{a}_{2\pi_{\bot}}^{\dag} - \hat{a}_{1\pi_{\bot} }^{\dag} \hat{a}_{2\pi}^{\dag} 
\right) + \mathrm{h.c.}
\end{equation}
where $\pi$ stands for any polarization state, and $i=1,2$ for the two spatial modes.
The time evolution of the field operators can be directly derived by the Heisenberg
equations, obtaining:
\begin{eqnarray}
\label{eq:non_coll_evol_1}
\hat{a}_{1\pi(2 \pi_{\bot})}^{\dag}(t) &=& \hat{a}_{1\pi(2 \pi_{\bot})}^{\dag} C + 
\hat{a}_{2 \pi_{\bot}(1 \pi)} S \\
\label{eq:non_coll_evol_4}
\hat{a}_{1\pi_{\bot}(2 \pi)}^{\dag}(t) &=& \hat{a}_{1\pi_{\bot}(2 \pi)}^{\dag} C - 
\hat{a}_{2\pi(1 \pi_{\bot})} S
\end{eqnarray}
In the following paragraphs we shall proceed with the calculation of the Wigner 
function for this amplifier in the stimulated regime, i.e. when a generic 
Fock-state is injected on the spatial mode $\mathbf{k}_{1}$.

\subsection{Wigner function for the non-collinear QIOPA in absence of decoherence}

In this section we derive the Wigner function for the QIOPA in a non collinear
configuration.
Without loss of generality, let us restrict our attention to the equatorial polarization basis 
$\vec{\pi}_{\varphi} = \frac{1}{\sqrt{2}} \left( \vec{\pi}_{H} + e^{\imath \varphi}
\vec{\pi}_{V} \right)$ and
$\vec{\pi}_{\varphi_{\bot}} = \frac{1}{\sqrt{2}} \left( - e^{- \imath \varphi} 
\vec{\pi}_{H} + \vec{\pi}_{V} \right)$.
As for the collinear case, the injected state over spatial mode $\mathbf{k}_{1}$ 
is the generic Fock-state $\vert \psi_{in} \rangle_{12} = \vert N \varphi, M \varphi_{\bot} 
\rangle_{1} \otimes \vert 0\varphi, 0\varphi_{\bot} \rangle_{2}$. The characteristic
function is then evaluated starting from the definition:
{\small
\begin{equation}
\begin{aligned}
\chi_{N,M} \left\{ \eta, \xi,t \right\} &= \, _{12}\langle \psi_{in} \vert e^{\eta_{1} \hat{a}_{1H}^{\dag}(t) 
- \eta_{1}^{\ast} \hat{a}_{1H}(t)} e^{\eta_{2} \hat{a}_{2V}^{\dag}(t) - \eta_{2}^{\ast} 
\hat{a}_{2V}(t)} \\
&e^{\xi_{1} \hat{a}_{1V}^{\dag}(t) - \xi_{1}^{\ast} \hat{a}_{1V}(t)} 
e^{\xi_{2} \hat{a}_{2H}^{\dag}(t) - \xi_{2}^{\ast} \hat{a}_{2H}(t)} \vert \psi_{in} 
\rangle_{12}
\end{aligned}
\end{equation}
}
Let us apply the transformation $\left\{ \eta, \xi \right\} \rightarrow \left\{ \overline{\eta}(t)
, \overline{\xi}(t) \right\}$, where:
\begin{eqnarray}
\label{eq:non_coll_int_var_trans_basis_1}
\overline{\eta}_{1} = \frac{1}{\sqrt{2}} \left( \eta_{1} + e^{- \imath \varphi} \xi_{1} \right) &;&
\overline{\xi}_{1} = \frac{1}{\sqrt{2}} \left( - e^{\imath \varphi} \eta_{1} + \xi_{1} \right) \\
\label{eq:non_coll_int_var_trans_basis_4}
\overline{\eta}_{2} = \frac{1}{\sqrt{2}} \left( \eta_{2} - e^{\imath \varphi} \xi_{2} \right) &;&
\overline{\xi}_{2} = \frac{1}{\sqrt{2}} \left( e^{- \imath \varphi} \eta_{2} + \xi_{2} \right)
\end{eqnarray}
and:
\begin{eqnarray}
\overline{\eta}_{1(2)}(t) &=& \overline{\eta}_{1(2)} C - \overline{\eta}_{2(1)}^{\ast} S \\
\overline{\xi}_{1(2)}(t) &=& \overline{\xi}_{1(2)} C + \overline{\xi}_{2(1)}^{\ast} S
\end{eqnarray}
With analogous calculation to the one performed in Section \ref{sec:wigner_no_losses} 
we obtain:
\begin{equation}
\begin{aligned}
\chi_{N,M} \left\{ \eta, \xi,t \right\} &= \exp\left[- \frac{1}{2} \sum_{j=1}^{2} \left( \vert 
\overline{\eta}_{j}(t) \vert^{2} + \vert \overline{\xi}_{j}(t) \vert^{2} \right) \right] \\
& \times \left( \sum_{n=0}^{N} \frac{N! (-1)^{n}}{(N-n)! n!^{2}} \vert \overline{\eta}_{1}(t) 
\vert^{2n} \right) \\
& \times \left( \sum_{m=0}^{M} \frac{M! (-1)^{m}}{(M-m)! m!^{2}} \vert 
\overline{\xi}_{1}(t) \vert^{2m} \right)
\end{aligned}
\end{equation}

The Wigner function of the amplified field can be then expressed as the 8-dimensional
Fourier transform of the characteristic function, according to its definition.
%
%
By inserting the explicit expression of the characteristic function, by
separating the integrals on each couple of complex conjugate variables, and 
by exploiting the integral relations already used in sec.\ref{sec:wigner_no_losses}
we find:
\begin{equation}
\begin{aligned}
W&_{\vert N \varphi, M \varphi_{\bot} \rangle} \left\{ \alpha, \beta,t \right\} = 
\left( \frac{2}{\pi} \right)^{4} (-1)^{N+M} e^{- 2 \vert \Delta \vert^{2}} \\
&L_{N} \left( \vert \Delta_{A} \left\{ \alpha \right\} + e^{- \imath \varphi} 
\Delta_{B} \left\{ \beta \right\}\vert^{2} \right) \\
&L_{M} \left( \vert - \Delta_{A} \left\{ \alpha \right\} e^{\imath \varphi} +
\Delta_{B} \left\{ \beta \right\}\vert^{2} \right) 
\end{aligned}
\end{equation}
The $\Delta_{A}\left\{ \alpha \right\}$, $\Delta_{B}\left\{ \beta \right\}$ and
$\vert \Delta \vert^{2}$ variables are defined in Appendix \ref{app:non_coll_no_losses}, 
where all the details of the calculation of the Wigner function are reported.

Let us we consider the injection of a single photon with polarization
state $\vert \varphi \rangle_{1} = \frac{1}{\sqrt{2}} \left( \vert H \rangle + 
e^{\imath \varphi} \vert V \rangle \right) $. The Wigner function reads \cite{DeMa98}:
{\small
\begin{equation}
\label{eq:non_coll_wigner_1_ideal}
W_{\vert 1 \varphi, 0 \varphi_{\bot} \rangle} \left\{ \alpha, \beta,t \right\} =
\frac{16}{\pi^{4}} e^{-2 \vert \Delta \vert^{2}} (\vert \Delta_{A} \left\{ 
\alpha \right\} + e^{- \imath \varphi} \Delta_{B} \left\{ \beta \right\}\vert^{2} - 1)
\end{equation}
}where $\vert \Delta_{A} \left\{ \alpha \right\} + e^{- \imath \varphi} 
\Delta_{B} \left\{ \beta \right\}\vert^{2}$ represents the interference term 
due to the quantum superposition form of the input state. Indeed this interference
term defines a phase-space region where the Wigner function is negative,
showing the broadcasting of the quantum properties of the injected single photon
state to the amplified field. 
For more details on the amplification of a single photon, refer to \cite{DeMa98}.

\subsection{Wigner function for the non-collinear EPR optical parametric
amplifier in presence of decoherence}
\label{sec:non_coll_wigner_losses}

In this section we are interested in the Wigner function 
for the non-degenerate optical parametric amplifier in presence of losses.
The lossy channels on the two spatial modes $\left\{ \mathbf{k}_{1}, \mathbf{k}_{2} \right\}$ 
are again simulated by a beam-splitter model, with input-output relations:
\begin{equation}
\label{eq:bs_non_coll}
\hat{c}^{\dag}_{i} = \sqrt{T} \hat{a}_{i}(t) + \imath \sqrt{R} \hat{b}_{i}
\end{equation}
where a vacuum state is injected in the $\hat{b}_{i}$ ports of the 
beam-splitters. We considered the case of two transmission channels 
for modes $\mathbf{k}_{1}$ and $\mathbf{k}_{2}$ with equal efficiencies.

The characteristic function is then evaluated as:

\begin{widetext}

\begin{equation}
\chi_{N,M} \left\{ \eta, \xi, R,t \right\} = \, _{AB}\langle \psi_{in} \vert 
e^{\overline{\eta}_{1} \hat{a}_{1\varphi}^{\dag}(t) - \overline{\eta}_{1}^{\ast} \hat{a}_{1\varphi}(t)} 
e^{\overline{\eta}_{2} \hat{a}_{2\varphi_{\bot}}^{\dag}(t) - \overline{\eta}_{2}^{\ast} 
\hat{a}_{2\varphi_{\bot}}(t)}
e^{\overline{\xi}_{1} \hat{a}_{1\varphi_{\bot}}^{\dag}(t) - \overline{\xi}_{1}^{\ast} 
\hat{a}_{1\varphi_{\bot}}(t)} e^{\overline{\xi}_{2} \hat{a}_{2\varphi}^{\dag}(t) - 
\overline{\xi}_{2}^{\ast} \hat{a}_{2\varphi}(t)} \vert \psi_{in} 
\rangle_{AB}
\end{equation}
The transformation to the $\left\{ \overline{\eta}, \overline{\xi} \right\}$ 
variables is the same as in the unperturbed case, and has been defined previously
in Eqq.(\ref{eq:non_coll_int_var_trans_basis_1}-\ref{eq:non_coll_int_var_trans_basis_4}).
Following the approach of Section \ref{sec:non_coll_wigner_unpert}, we insert 
the beam-splitter relations (\ref{eq:bs_non_coll}) and evaluate the 
averages after writing the exponential operators in normally ordered form.
We obtain:
{\small
\begin{equation}
\begin{aligned}
\chi_{N,M} &\left\{ \eta, \xi, R,t \right\} = \exp \left[ - \frac{1}{2} \left(
1 + 2 R S^{2}\right) \sum_{j=1}^{2} \left( \vert \overline{\eta}_{j}(t) 
\vert^{2} + \vert \overline{\xi}_{j}(t) \vert^{2} \right) \right] 
\exp\left[ - R CS \left( \overline{\eta}_{1}(t) 
\overline{\eta}_{2}(t) + \overline{\eta}_{1}^{\ast}(t) 
\overline{\eta}_{2}^{\ast}(t) \right) \right] \\ 
&\times \exp \left[ R CS \left( \overline{\xi}_{1}(t) 
\overline{\xi}_{2}(t) + \overline{\xi}_{1}^{\ast}(t) \overline{\xi}_{2}^{\ast}(t)
\right)\right] \left( \sum_{n=0}^{N} \frac{N! (-T)^{n}}{(N-n)! n!^{2}} 
\vert \overline{\eta}_{1}(t) \vert^{2n}  \right) \left( \sum_{m=0}^{M} 
\frac{M! (-T)^{m}}{(M-m)! m!^{2}} \vert \overline{\xi}_{1}(t) \vert^{2m} \right)
\end{aligned}
\end{equation}
}
The Wigner function of the field can then be evaluated as the 8-dimensional Fourier 
transform:
\begin{equation}
\begin{aligned}
W_{\vert N \varphi, M \varphi_{\bot} \rangle} \left\{ \alpha, \beta, R,t\right\} &= 
\frac{1}{\pi^{8}} \int \int \int \int \left( \prod_{j=1}^{2} d^{2} 
\overline{\eta}_{j} d^{2} \overline{\xi}_{j} \right) \chi_{N,M} \left\{ \eta, \xi,R,t \right\}
\exp\left\{ \sum_{j=1}^{2} \left[ \overline{\eta}^{\ast}_{j}(t) \overline{\alpha}_{j}(t) - \overline{\eta}_{j}(t) \overline{\alpha}^{\ast}_{j}(t) \right] \right\} \\
&\times \exp \left\{ \sum_{j=1}^{2} \left[ \overline{\xi}^{\ast}_{j}(t) \overline{\beta}_{j}(t) - \overline{\xi}_{j}(t) \overline{\beta}^{\ast}_{j}(t) \right] \right\}
\end{aligned}
\end{equation}

\end{widetext}
After the explicit insertion of the characteristic function, the Wigner function can 
be written as a product of two integrals:
\begin{equation}
W_{\vert N \varphi, M \varphi_{\bot} \rangle} \left\{ \alpha, \beta, R,t \right\} = 
I^{+}_{N}\left\{ \alpha, R,t \right\} I_{M}^{-}\left\{ \beta, R,t \right\}
\end{equation}
where the $I_{N}^{+}\left\{ \alpha, R,t \right\}$ and $I_{M}^{-}\left\{ \alpha, R,t \right\}$
are:
{\small
\begin{eqnarray}
I^{+}_{N}\left\{ \alpha,R,t \right\} &=& \frac{1}{\pi^{2}} \sum_{n=0}^{N} d_{n}^{N}(T) J_{n,0} 
\left( \varepsilon^{'}, \mu; \overline{\alpha}_{1}(t), \overline{\alpha}_{2}(t) \right)\\
I^{-}_{M}\left\{ \beta,R,t \right\} &=& \frac{1}{\pi^{2}} \sum_{m=0}^{M} d_{m}^{M}(T) J_{m,0} 
\left( \varepsilon^{'}, - \mu; \overline{\beta}_{1}(t), \overline{\beta}_{2}(t) \right)
\end{eqnarray}
}The $J_{n,0}$ integrals are explicited in Appendix A in Eq.(\ref{eq:jnm_def}).
The parameters $\varepsilon^{'}, \mu, d_{n}^{N}(T)$ for the non-collinear optical parametric
amplifier are:
\begin{eqnarray}
d_{n}^{N}(T) &=& \begin{pmatrix} N \\ n \end{pmatrix} \frac{(- T)^{n}}{n!} \\
\varepsilon^{'} &=& \frac{1}{2} \left( 1 + 2 R S^{2} \right) \\
\mu &=& R C S
\end{eqnarray}

\begin{figure}[tbp]
\centering
\includegraphics[width=0.4\textwidth]{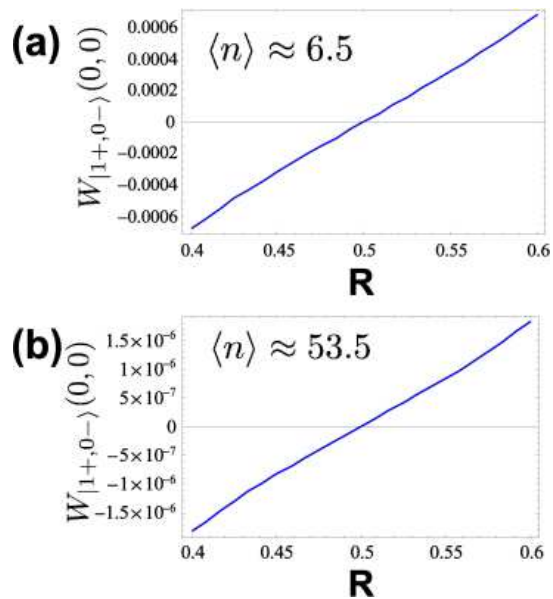}
\caption{Transition from a non-positive distribution to a completely
positive Wigner function of the single-photon amplified states, obtained by the non-collinear 
QI-OPA in presence of decoherence, as a function of the beam-splitter reflectivity $R$. 
Average photon number of (a) $\langle n \rangle \approx 6.5$ and (b) $%
\langle n \rangle \approx 53.5$. The negativity is evaluated as the value of
the Wigner function in the origin.}
\label{fig:Negativity_non_collinear}
\end{figure}

Let us again focus on the single-photon case, corresponding to $N=1$ and $M=0$.
With an analogous procedure to the collinear case, we analyze the persistence of negative
regions in the Wigner function after the decoherence process. Analyzing the form of
Eq.(\ref{eq:non_coll_wigner_1_ideal}), we note that the minimum occurs when $\vert 
\Delta_{A} \left\{ \alpha \right\} + e^{- \imath \varphi} \Delta_{B} \left\{ \beta
\right\}\vert^{2} = 0$ and $\vert \Delta \vert^{2} = 0$. This point corresponds to the
origin of the 8-dimensional phase space given by the $\left\{ \overline{\alpha}(t), 
\overline{\beta}(t) \right\}$ variables. The evolution of the Wigner function in this
point is explicitly evaluated as:
\begin{equation}
W_{\vert 1 \varphi, 0 \varphi_{\bot} \rangle} \left\{ 0, 0, R,t\right\} = \frac{16}{\pi^{4}} \frac{2R - 1}{\left[ 1 + 
4 R (1-R) S^{2} \right]^{3}}
\end{equation}
The value in the origin of the Wigner function is reported in Fig.
\ref{fig:Negativity_non_collinear}. We note an analogous trend with respect to 
the collinear  case, while the absolute value of the amount of negativity is 
smaller. This is due to the \emph{universal cloning} feature of the non-collinear 
QI-OPA, in which the cloning fidelity is smaller than the one in \emph{phase-covariant} 
case.

We conclude this Section on the non-collinear QI-OPA by stressing that a full 
phase-space characterization of the quantum states generated by this device
would require a couple of double homodyne measurement setups, one for each
spatial mode, as the one discussed in Section \ref{subsec:homodyne}.

\section{Conclusion and perspectives}

The quantum properties of the QI-OPA generated macroscopic system, have been
investigated in phase-space by a Wigner quasi-probability function analysis
when this class of states is transmitted over a lossy channel, i.e. in
presence of a decohering \emph{system - enviroment} interaction which
represents the main limitation in the implementation of quantum
information tasks. We first considered the ideal case, in absence
of losses, showing the presence of peculiar quantum properties such as
squeezing and a non-positive W-representation. Then, after a brief review of
the properties of the coherent states MQS, the resilience to losses of
QI-OPA amplified states in a lossy configuration was investigated, allowing
to observe the persistence of the non-positivity of the Wigner function in a
certain range of the \emph{system-enviroment} interaction parameter $R$.
This behaviour was analyzed in close comparison with the $|\alpha \rangle $
states MQS, which possesses a non-positive W-representation in the same
interval of the interaction parameter $R$. Moreover, the more resilient
structure of the QI-OPA amplified states was emphasized by their slower
decoherence rate, represented by both the slower decrease in the negative
part of the Wigner function and by the behavior of the Bures distance
between orthogonal macrostates, the latter evaluated in the Fock-Space.
Since the negativity of the W-representation is a sufficent but not a 
necessary condition for the non-classicality of any physical system,
future investigations could be aimed to the analysis of the decoherence regime in
which the Wigner function is completely positive, analyzing the presence
of quantum properties from a different point of view.

We acknowledge useful discussions with Wolfgang Schleich.
Work supported by PRIN 2005 of MIUR and INNESCO 2006 of CNISM.

\appendix

\section{Density matrix of the equatorial amplified states after the propagation
over a lossy channel}
\label{app:equatorial_Fock}

In this appendix we perform the derivation of the density matrix of the equatorial
amplified states after the propagation over a lossy channel. The applied model
is the same BS-scattering process used throughout the whole paper.

The starting point is the expression of the unperturbed state $|\Phi
^{\varphi }\rangle {}$: 
\begin{equation}
\begin{aligned} |\Phi^{\varphi}\rangle{} &= \frac{1}{C^{2}}
\sum_{i,j=0}^{\infty} \left( e^{- \imath \varphi} \frac{\Gamma}{2}
\right)^{i} \left( - e^{\imath \varphi} \frac{\Gamma}{2} \right)^{j} \\
&\times \frac{\sqrt{(2i+1)!} \, \sqrt{2j!}}{i! \, j!}
|(2i+1)\varphi,(2j)\varphi_{\bot}\rangle{} \end{aligned}
\end{equation}
where $C=\cosh g$ and $\Gamma =\tanh g$, with $g$ gain of the amplifier, with
the state written in the polarization basis $\left\{ \vec{\pi}_{\varphi },%
\vec{\pi}_{\varphi _{\bot }}\right\} $. The state can be written in terms of
the creation operators: 
\begin{equation}
|\Phi ^{\varphi }\rangle {}=\frac{1}{C^{2}}\sum_{i,j=0}^{\infty }\left(
e^{-\imath \varphi }\frac{\Gamma }{2}\right) ^{i}\left( -e^{\imath \varphi }%
\frac{\Gamma }{2}\right) ^{j}\frac{\left( \hat{a}_{\varphi }^{\dag }\right)
^{2i+1}\,\left( \hat{a}_{\varphi _{\bot }}^{\dag }\right) ^{2j}}{i!\,j!}%
|0\rangle {}
\end{equation}
We then apply the time evolution independent operators $\hat{U}%
_{BS}^{(\varphi )}$ and $\hat{U}_{BS}^{(\varphi _{\bot })}$, which describe
the interaction in the beam-splitter: $\hat{U}_{BS} = e^{\imath \frac{\theta}{2} 
\left( \hat{a}^{\dag} \hat{b} + \hat{a} \hat{b}^{\dag}\right)}$.
As a reasonable assumption, we consider the transmittivity $T_{\varphi}$ and 
$T_{\varphi_{\bot}}$ to be equal for all polarization states. 
We can then write the evolved state in the Schrodinger picture 
$\vert \Phi^{^{\prime}\varphi} \rangle$ as: 
\begin{equation}
\begin{aligned} |\Phi'^{\varphi}\rangle{} &= \hat{U}_{BS}^{(\varphi)}
\hat{U}_{BS}^{(\varphi_{\bot})} |\Phi^{\varphi}\rangle{} = \frac{1}{C^{2}}
\sum_{i,j=0}^{\infty} \left( e^{- \imath \varphi} \frac{\Gamma}{2}
\right)^{i} \left( - e^{\imath \varphi} \frac{\Gamma}{2} \right)^{j} \\
&\times \frac{\left( \sqrt{T} \op{c}_{\varphi}^{\dag} + \imath \sqrt{R}
\op{d}_{\varphi}^{\dag} \right)^{2i+1}\, \left( \sqrt{T}
\op{c}_{\varphi_{\bot}}^{\dag} + \imath \sqrt{R}
\op{d}_{\varphi_{\bot}}^{\dag} \right)^{2j}}{i! \, j!} |0\rangle{}
\end{aligned}
\end{equation}
We can now exploit the binomial expansion $(a+b)^{n}=%
\sum_{i=0}^{n}a^{i}b^{n-i} 
\begin{pmatrix}
n \\ 
i%
\end{pmatrix}
$, as the operators for the two output mode of the beam-splitter $\hat{c},%
\hat{d}$ commute and can be treated as algebraic variables. 
Subsequently applying the creation operators to the vacuum 
we then obtain the final expression of the state after the propagation in
the beam-splitter: 
\begin{equation}
\begin{aligned} |\Phi'^{\varphi}\rangle{} &= \frac{1}{C^{2}}
\sum_{i,j=0}^{\infty} \left( e^{- \imath \varphi} \frac{\Gamma}{2}
\right)^{i} \left( - e^{\imath \varphi} \frac{\Gamma}{2} \right)^{j} \\
&\times \sum_{m=0}^{2i+1} \sum_{n=0}^{2j} \left[ \begin{pmatrix} 2i+1 \\ m
\end{pmatrix} \begin{pmatrix} 2j \\ n \end{pmatrix} \right]^{\frac{1}{2}}
\frac{\sqrt{(2i+1)!} \, \sqrt{2j!}}{i! \, j!} \\ &\times \left( \sqrt{T}
\right)^{n+m} \left( \imath \sqrt{R} \right)^{2i+2j+1-n-m} \\ &\times
|m,n\rangle_{C} \otimes |2i+1-m,2j-n\rangle_{D} \end{aligned}
\end{equation}
The density matrix of the quantum state is then
obtained as $\hat{\rho ^{\prime }}^{\varphi }=|\Phi ^{\prime }{}^{\varphi
}\rangle \langle \Phi ^{\prime }{}^{\varphi }|$. Let us call: 
\begin{equation}
\begin{aligned} \gamma_{ij,mn} &= \frac{1}{C^{2}} \left( e^{- \imath
\varphi} \frac{\Gamma}{2} \right)^{i} \left( - e^{\imath \varphi}
\frac{\Gamma}{2} \right)^{j} \left[ \begin{pmatrix} 2i+1 \\ m \end{pmatrix}
\begin{pmatrix} 2j \\ n \end{pmatrix} \right]^{\frac{1}{2}} \\ &\times
\frac{\sqrt{(2i+1)!} \, \sqrt{2j!}}{i! \, j!} \left( \sqrt{T} \right)^{n+m}
\left( \imath \sqrt{R} \right)^{2i+2j+1-n-m} \end{aligned}
\end{equation}
Tracing the density matrix $\hat{\rho ^{\prime }}^{\varphi }$ on the
reflected $\hat{d}$ BS-mode in order to consider our ignorance
over the number of reflected photons we obtain: 
\begin{equation}
\begin{aligned} \hat{\rho}_{T}^{\varphi} &= \mathrm{Tr}_{D} \left(
\hat{\rho'}^{\varphi} \right) = \sum_{x,y=0}^{\infty} \, _{D}\langle
x,y|\Phi'^{\varphi}\rangle \, \langle \Phi'^{\varphi}|x,y\rangle_{D} = \\ &=
\sum_{x,y=0}^{\infty} \sum_{i,j,k,l=0}^{\infty} \sum_{m=0}^{2i+1}
\sum_{n=0}^{2j} \sum_{p=0}^{2k+1} \sum_{q=0}^{2l} \gamma_{ij,mn}
\gamma^{\ast}_{kl,pq} \vert m,n \rangle_{} \, _{}\langle p,q \vert \\
&\otimes \, _{D}\langle x,y|2i+1-m,2j-n\rangle_{D} \, _{D}\langle
2k+1-p,2l-q|x,y\rangle_{D} \end{aligned}
\end{equation}
The evaluation of the scalar products between Fock-states lead to the
desired result. The expression of the density matrix on the transmitted $%
\hat{c}$ spatial BS-mode is then written in the form: 
\begin{equation}
\hat{\rho}_{T}^{\varphi }=\sum_{i,j,k,q=0}^{\infty }\,_{{}}\langle i,j|\hat{%
\rho}_{T}^{\varphi }|k,q\rangle _{{}}\,|i,j\rangle _{{}}\,_{{}}\langle k,q|
\end{equation}
\begin{widetext}
We now report the expressions of the density matrix coefficients, which
depend on the parity. 
For $i,j,k,q$ even we obtain: 
\begin{equation}
\label{eq:matrix_elements_even_even}
\begin{aligned} 
_{}\langle i,j \vert \hat{\rho}^{\varphi}_{T} \vert k,q
\rangle_{} &= \frac{1}{C^{4}} \left( \frac{\Gamma}{2} \right)^{\frac{i+k}{2}} \left( - \frac{\Gamma}{2} \right)^{\frac{j+q}{2}} \left( e^{\imath \varphi}\right)^{\frac{j+k-i-q}{2}}
R \left( \sqrt{T} \right)^{i+j+k+q} \frac{\sqrt{i!j!k!q!}}{\frac{i}{2}! \frac{j}{2}! \frac{k}{2}! \frac{q}{2}!} 
(i+1)(k+1) \\
&\left[\frac{1}{1 - R^{2} \Gamma^{2}} \right]^{2+\frac{i+j+k+q}{2}} \, _{2}F_{1} \left( - \frac{i}{2}, - \frac{k}{2};
\frac{3}{2}; R^{2} \Gamma^{2} \right) \, _{2}F_{1} \left( - \frac{j}{2}, - \frac{q}{2}; \frac{1}{2}; R^{2} \Gamma^{2} \right)
\end{aligned}
\end{equation}
For $i,k$ odd and $j,q$ even, we obtain: 
\begin{equation}
\begin{aligned} 
_{}\langle i,j \vert \hat{\rho}^{\varphi}_{T} \vert k,q
\rangle_{} &= \frac{1}{C^{4}} \left( \frac{\Gamma}{2} \right)^{\frac{i+k}{2}-1} \left( - \frac{\Gamma}{2} \right)^{\frac{j+q}{2}} \left( e^{\imath \varphi}\right)^{\frac{j+k-i-q}{2}}
\left( \sqrt{T} \right)^{i+j+k+q} \frac{\sqrt{i!j!k!q!}}{\frac{i-1}{2}! \frac{j}{2}! \frac{k-1}{2}! \frac{q}{2}!} \\
&\left[\frac{1}{1 - R^{2} \Gamma^{2}} \right]^{2+\frac{i+j+k+q}{2}}
\, _{2}F_{1} \left( - \frac{1+i}{2}, - \frac{1+k}{2};
\frac{1}{2}; R^{2} \Gamma^{2} \right) \, _{2}F_{1} \left( - \frac{j}{2},
- \frac{q}{2}; \frac{1}{2}; R^{2} \Gamma^{2} \right)
\end{aligned}
\end{equation}
For $i,k$ even and $j,q$ odd, we obtain: 
\begin{equation}
\begin{aligned}
_{}\langle i,j \vert \hat{\rho}^{\varphi}_{T} \vert k,q
\rangle_{} &= \frac{1}{C^{4}} \left( \frac{\Gamma}{2} \right)^{\frac{i+k}{2}} \left( - \frac{\Gamma}{2} \right)^{\frac{j+q}{2}-1} \left( e^{\imath \varphi}\right)^{\frac{j+k-i-q}{2}}
R^{2} \Gamma^{2} \left( \sqrt{T} \right)^{i+j+k+q} (i+1)(k+1) \frac{\sqrt{i!j!k!q!}}{\frac{i}{2}! \frac{j-1}{2}! \frac{k}{2}! \frac{q-1}{2}!} \\
&\left[\frac{1}{1 - R^{2} \Gamma^{2}} \right]^{2+\frac{i+j+k+q}{2}} \, _{2}F_{1} \left( - \frac{i}{2}, - \frac{k}{2};
\frac{3}{2}; R^{2} \Gamma^{2} \right) \,_{2}F_{1} \left( \frac{1-j}{2}, \frac{1-q}{2}; \frac{3}{2}; R^{2} \Gamma^{2} \right)
\end{aligned}
\end{equation}
Finally, for $i,j,k,q$ odd we obtain: 
\begin{equation}
\label{eq:matrix_elements_odd_odd}
\begin{aligned}
_{}\langle i,j \vert \hat{\rho}^{\varphi}_{T} \vert k,q
\rangle_{} &= \frac{1}{C^{4}} \left( \frac{\Gamma}{2} \right)^{\frac{i+k}{2}-1} \left( - \frac{\Gamma}{2} \right)^{\frac{j+q}{2}-1} \left( e^{\imath \varphi}\right)^{\frac{j+k-i-q}{2}} R \Gamma^{2} \left( \sqrt{T} \right)^{i+j+k+q} \frac{\sqrt{i!j!k!q!}}{\frac{i-1}{2}! \frac{j-1}{2}! \frac{k-1}{2}! \frac{q-1}{2}!} \\
&\left[\frac{1}{1 - R^{2} \Gamma^{2}} \right]^{2+\frac{i+j+k+q}{2}} \, _{2}F_{1} 
\left( - \frac{1+i}{2}, - \frac{1+k}{2}; \frac{1}{2}; R^{2} \Gamma^{2} \right) \,
_{2}F_{1} \left( \frac{1-j}{2}, \frac{1-q}{2}; \frac{3}{2}; R^{2} \Gamma^{2} \right)
\end{aligned}
\end{equation}
\end{widetext}
In these expressions, $_{2}F_{1}\left( \alpha,\beta;\gamma;z \right) $ are
Hyper-geometric functions \cite{Slat66}.

\section{Density matrix of the $\vec{\pi}_{H}, \vec{\pi}_{V}$ amplified states after 
the propagation over a lossy channel}

\label{app:HV_Fock}

The procedure for the evaluation of the density matrix of the state after
losses is the same applied in the previous section. Let us analyze the $%
\vert \Phi^{H} \rangle$ state, whose density matrix after the amplification
process reads: 
\begin{equation}
\begin{aligned} \hat{\rho}^{H} &= \vert \Phi^{H} \rangle \, \langle \Phi^{H}
\vert = \frac{1}{C^{4}} \sum_{n,m=0}^{\infty} \Gamma^{n+m} \sqrt{n+1}
\sqrt{m+1} \\ &\vert (n+1)H, nV \rangle \, \langle (m+1)H,mV \vert
\end{aligned}
\end{equation}
After the insertion of the BS unitary transformation, the joint
density matrix of the transmitted and reflected modes is: 
\begin{equation}
\begin{aligned} \hat{\rho}^{' H} &= \frac{1}{C^{4}}�\sum_{m,n=0}^{\infty}
\sum_{i=0}^{n+1} \sum_{j=0}^{n} \sum_{k=0}^{m+1} \sum_{q=0}^{m} \Gamma^{n+m}
\sqrt{n+1} \sqrt{m+1} \\ & (-1)^{2m+1-k-q} \left[ \begin{pmatrix} n+1 \\ i
\end{pmatrix} \begin{pmatrix} n \\ j \end{pmatrix} \begin{pmatrix} m+1 \\ k
\end{pmatrix} \begin{pmatrix} m \\ q \end{pmatrix} \right]^{\frac{1}{2}} \\
& \left( \sqrt{T} \right)^{i+j+k+q} \left( \sqrt{R} \right)^{n+m+2-i-j-k-q}
\\ &\vert iH, jV \rangle_{C} \langle kH, qV \vert \otimes \\ &\otimes \vert
(n+1-i)H, (n-j)V \rangle_{D} \langle (m+1-k)H, (m-q)V \vert \end{aligned}
\end{equation}
After tracing over the reflected mode, we finally obtain: 
\begin{equation}
\begin{aligned} &\hat{\rho}^{H}_{T} = \sum_{i=1}^{\infty} \sum_{j=0}^{i-1}
\sum_{k=0}^{\infty} \left( \sum_{p=0}^{\infty} \overline{\gamma}_{ijk;p}
\right) \vert iH,jV \rangle \, \langle kH, (k+j-i)V \vert +\\ &+
\sum_{i=0}^{\infty} \sum_{j=i}^{\infty} \sum_{k=0}^{\infty} \left(
\sum_{p=j+1-i}^{\infty} \overline{\gamma}_{ijk;p} \right) \vert iH,jV
\rangle \, \langle kH, (k+j-i)V \vert \end{aligned}
\end{equation}
where the coefficients $\overline{\gamma}_{ijk;p}$ are: 
\begin{equation}
\begin{aligned} \overline{\gamma}_{ijk;p} &= \frac{\Gamma^{2p+i+k-2}}{C^{4}}
\sqrt{p+i} \sqrt{p+k} \, T^{k+j} R^{2p+i-1-j} \\ &\left[�\begin{pmatrix} p+i
\\ i \end{pmatrix} \begin{pmatrix} p+i-1 \\ j \end{pmatrix} \begin{pmatrix}
p+k \\ k \end{pmatrix} \begin{pmatrix} p+k-1 \\ k+j-1
\end{pmatrix}\right]^{\frac{1}{2}} \end{aligned}
\end{equation}

\section{Wigner function for the non collinear QI-OPA in absence of decoherence}
\label{app:non_coll_no_losses}

We begin the calculation of the Wigner function, without loss of generality, 
by restricting our attention to the equatorial polarization basis, defined by:
\begin{eqnarray}
\vec{\pi}_{\varphi} &=& \frac{1}{\sqrt{2}} \left( \vec{\pi}_{H} + e^{\imath \varphi}
\vec{\pi}_{V} \right) \\
\vec{\pi}_{\varphi_{\bot}} &=& \frac{1}{\sqrt{2}} \left( - e^{- \imath \varphi} 
\vec{\pi}_{H} + \vec{\pi}_{V} \right)
\end{eqnarray}
As for the collinear case, the injected state over spatial mode $\mathbf{k}_{1}$ 
is the generic Fock-state $\vert \psi_{in} \rangle_{12} = \vert N \varphi, M \varphi_{\bot} 
\rangle_{1} \otimes \vert 0\varphi, 0\varphi_{\bot} \rangle_{2}$. The characteristic
function is then evaluated starting from the definition:
{\small
\begin{equation}
\begin{aligned}
\chi_{N,M} \left\{ \eta, \xi,t \right\} &= \, _{12}\langle \psi_{in} \vert e^{\eta_{1} \hat{a}_{1H}^{\dag}(t) 
- \eta_{1}^{\ast} \hat{a}_{1H}(t)} e^{\eta_{2} \hat{a}_{2V}^{\dag}(t) - \eta_{2}^{\ast} 
\hat{a}_{2V}(t)} \\
&e^{\xi_{1} \hat{a}_{1V}^{\dag}(t) - \xi_{1}^{\ast} \hat{a}_{1V}(t)} 
e^{\xi_{2} \hat{a}_{2H}^{\dag}(t) - \xi_{2}^{\ast} \hat{a}_{2H}(t)} \vert \psi_{in} 
\rangle_{12}
\end{aligned}
\end{equation}
}
Let us apply the following transformations over the $\left\{ \eta, \xi \right\}$ variables,
corresponding to the rotations from the $\left\{ \vec{\pi}_{H}, \vec{\pi}_{V} \right\}$ to the
$\left\{ \vec{\pi}_{\varphi}, \vec{\pi}_{\varphi_{\bot}} \right\}$ polarization basis:
\begin{eqnarray}
\overline{\eta}_{1} &=& \frac{1}{\sqrt{2}} \left( \eta_{1} + e^{- \imath \varphi} \xi_{1} \right) \\
\overline{\xi}_{1} &=& \frac{1}{\sqrt{2}} \left( - e^{\imath \varphi} \eta_{1} + \xi_{1} \right) \\
\overline{\eta}_{2} &=& \frac{1}{\sqrt{2}} \left( \eta_{2} - e^{\imath \varphi} \xi_{2} \right) \\
\overline{\xi}_{2} &=& \frac{1}{\sqrt{2}} \left( e^{- \imath \varphi} \eta_{2} + \xi_{2} \right)
\end{eqnarray}
We then explicitly insert the time evolution of the field operators
(\ref{eq:non_coll_evol_1}-\ref{eq:non_coll_evol_4}). Defining:
\begin{eqnarray}
\overline{\eta}_{1}(t) &=& \overline{\eta}_{1} C - \overline{\eta}_{2}^{\ast} S \\
\overline{\eta}_{2}(t) &=& \overline{\eta}_{2} C - \overline{\eta}_{1}^{\ast} S \\
\overline{\xi}_{1}(t) &=& \overline{\xi}_{1} C + \overline{\xi}_{2}^{\ast} S \\
\overline{\xi}_{2}(t) &=& \overline{\xi}_{2} C + \overline{\xi}_{1}^{\ast} S
\end{eqnarray}
we can re-write the characteristic function in the following form:
{\small
\begin{equation}
\begin{aligned}
\chi_{N,M} \left\{ \eta, \xi,t \right\} &= \, _{12}\langle \psi_{in} \vert e^{\overline{\eta}_{1}(t)
\hat{a}_{1\varphi}^{\dag} - \overline{\eta}_{1}^{\ast}(t) \hat{a}_{1\varphi}} 
e^{\overline{\eta}_{2}(t) \hat{a}_{2\varphi_{\bot}}^{\dag} - \overline{\eta}_{2}^{\ast}(t) 
\hat{a}_{2\varphi_{\bot}}} \\
&e^{\overline{\xi}_{1}(t) \hat{a}_{1\varphi_{\bot}}^{\dag} - \overline{\xi}_{1}^{\ast}(t) 
\hat{a}_{1\varphi_{\bot}}} e^{\overline{\xi}_{2}(t) \hat{a}_{2\varphi}^{\dag} - 
\overline{\xi}_{2}^{\ast}(t) \hat{a}_{2\varphi}} \vert \psi_{in} 
\rangle_{12}
\end{aligned}
\end{equation}
}
The averages can now be evaluated by writing the operators in the  normally ordered
form.
With analogous calculation to the one performed in Section \ref{sec:wigner_no_losses} 
we obtain:
\begin{equation}
\begin{aligned}
\chi_{N,M} \left\{ \eta, \xi,t \right\} &= \exp\left[- \frac{1}{2} \sum_{j=1}^{2} \left( \vert 
\overline{\eta}_{j}(t) \vert^{2} + \vert \overline{\xi}_{j}(t) \vert^{2} \right) \right] \\
& \times \left( \sum_{n=0}^{N} \frac{N! (-1)^{n}}{(N-n)! n!^{2}} \vert \overline{\eta}_{1}(t) 
\vert^{2n} \right) \\
& \times \left( \sum_{m=0}^{M} \frac{M! (-1)^{m}}{(M-m)! m!^{2}} \vert 
\overline{\xi}_{1}(t) \vert^{2m} \right)
\end{aligned}
\end{equation}

The Wigner function of the amplified field can be then expressed as the 8-dimensional
Fourier transform of the characteristic function, according to its definition.
The integration variables are changed according to 
$\left\{ \eta_{j}, \xi_{j} \right\}_{j=1}^{2} \rightarrow 
\left\{ \overline{\eta}_{j}(t), \overline{\xi}_{j}(t) \right\}_{j=1}^{2}$, 
leading to:
\begin{widetext}

\begin{equation}
\begin{aligned}
W_{\vert N \varphi, M \varphi_{\bot} \rangle} \left\{ \alpha, \beta ,t\right\} &= 
\frac{1}{\pi^{8}} \int \int \int \int \left( \prod_{j=1}^{2} d^{2} 
\overline{\eta}_{j} d^{2} \overline{\xi}_{j} \right) \chi_{N,M} \left\{ \eta, \xi,t \right\}
\exp\left\{ \sum_{j=1}^{2} \left[ \overline{\eta}^{\ast}_{j}(t) \overline{\alpha}_{j}(t) - \overline{\eta}_{j}(t) \overline{\alpha}^{\ast}_{j}(t) \right] \right\} \\
&\exp \left\{ \sum_{j=1}^{2} \left[ \overline{\xi}^{\ast}_{j}(t) \overline{\beta}_{j}(t) - \overline{\xi}_{j}(t) \overline{\beta}^{\ast}_{j}(t) \right] \right\}
\end{aligned}
\end{equation}

\end{widetext}
The calculation then proceeds as follows. By inserting the explicit expression of the 
characteristic function, by separating the integrals on each couple of complex conjugate 
variables, and by exploiting the integral relations already used in sec.\ref{sec:wigner_no_losses}
we find:
\begin{equation}
\begin{aligned}
W&_{\vert N \varphi, M \varphi_{\bot} \rangle} \left\{ \alpha, \beta,t \right\} = 
\left( \frac{2}{\pi} \right)^{4} (-1)^{N+M} L_{N} \left( 4 \vert 
\overline{\alpha}_{1}(t) \vert^{2} \right) \\
&L_{M} \left( 4 \vert \overline{\beta}_{1}(t) \vert^{2} \right) 
\exp\left\{ - 2 \sum_{j=1}^{2} \left[ \vert \overline{\alpha}_{j}(t) 
\vert^{2} + \vert \overline{\beta}_{j}(t) \vert^{2} \right] \right\}
\end{aligned}
\end{equation}
As a further step, we can define the following set of squeezed and antisqueezed
variables:
\begin{eqnarray}
\label{eq:squeezing_variables_non_coll_1}
\gamma_{A+} &=& \left( \alpha_{1} + \alpha_{2}^{\ast} \right) e^{g} \\
\gamma_{A-} &=& \imath \left( \alpha_{1} - \alpha_{2}^{\ast} \right) e^{-g} \\
\gamma_{B+} &=& \left( \beta_{1} - \beta_{2}^{\ast} \right) e^{g} \\
\label{eq:squeezing_variables_non_coll_4}
\gamma_{B-} &=& \imath \left( \beta_{1} + \beta_{2}^{\ast} \right) e^{-g}
\end{eqnarray}
and the following set of quadrature variables:
\begin{eqnarray}
\label{eq:quadrature_variables_non_coll_1}
\Delta_{A} \left\{ \alpha \right\} &=& \frac{1}{\sqrt{2}} \left( \gamma_{A+} - 
\imath \gamma_{A-} \right) \\
\Delta_{B} \left\{ \beta \right\} &=& \frac{1}{\sqrt{2}} \left( \gamma_{B+} - 
\imath \gamma_{B-} \right)\\
\label{eq:quadrature_variables_non_coll_3}
\vert \Delta \vert^{2} &=& \frac{1}{2} \left( \vert \gamma_{A+} \vert^{2} + 
\vert \gamma_{A-} \vert^{2} + \vert \gamma_{B+} \vert^{2} + \vert \gamma_{B-} 
\vert^{2} \right)
\end{eqnarray}
The Wigner function can then be simply expressed as:
\begin{equation}
\begin{aligned}
W&_{\vert N \varphi, M \varphi_{\bot} \rangle} \left\{ \alpha, \beta,t \right\} = 
\left( \frac{2}{\pi} \right)^{4} (-1)^{N+M} e^{- 2 \vert \Delta \vert^{2}} \\
&L_{N} \left( \vert \Delta_{A} \left\{ \alpha \right\} + e^{- \imath \varphi} 
\Delta_{B} \left\{ \beta \right\}\vert^{2} \right) \\
&L_{M} \left( \vert - \Delta_{A} \left\{ \alpha \right\} e^{\imath \varphi} +
\Delta_{B} \left\{ \beta \right\}\vert^{2} \right) 
\end{aligned}
\end{equation}


\section{Integral relation}

In this appendix we derive the integral relation used in the text in Section
\ref{sec:non_coll_wigner_losses}. We are interested in evaluating integrals of the form:
\begin{equation}
\label{eq:jnm_def}
\begin{aligned}
J_{n,m}&(\tau, \mu; z, w) = \frac{1}{\pi^{2}} \int \int d^{2}\alpha d^{2}\beta \,
\vert \alpha \vert^{2n} \vert \beta \vert^{2m} \\ 
&e^{\alpha^{\ast} z - \alpha z^{\ast}} e^{\beta^{\ast} w - \beta w^{\ast}} 
e^{- \tau \left( \vert \alpha \vert^{2} + \vert \beta \vert^{2} \right)} 
e^{- \mu \left( \alpha \beta + \alpha^{\ast} \beta^{\ast} \right)}
\end{aligned}
\end{equation}
We first note that:
\begin{equation}
\label{eq:int_deriv_jnm}
\begin{aligned}
J_{n,m}&(\tau, \mu; z, w) = (-1)^{n+m} \\
&\frac{\partial^{2n+2m}}{\partial z^{n} \partial
z^{\ast n} \partial w^{m} \partial w^{\ast m}} \Big[ J_{0,0}(\tau, \mu; z, w) \Big]
\end{aligned}
\end{equation}
According to this result, it is sufficient to evaluate only the $J_{0,0}$ integral. 
We start from its explicit expression:
\begin{equation}
\begin{aligned}
J_{0,0}&(\tau, \mu; z, w) = \frac{1}{\pi^{2}} \int \int d^{2}\alpha d^{2}\beta 
e^{\alpha^{\ast} z - \alpha z^{\ast}} \\ &e^{\beta^{\ast} w - \beta w^{\ast}} 
e^{- \tau \left( \vert \alpha \vert^{2} + \vert \beta \vert^{2} \right)} 
e^{- \mu \left( \alpha \beta + \alpha^{\ast} \beta^{\ast} \right)}
\end{aligned}
\end{equation}
We now apply the following $\vert J \vert = 1$ integration variables rotation:
\begin{eqnarray}
\alpha &=& \cos \theta \, \gamma + \sin \theta \, \delta \\
\beta &=& - \sin \theta \, \gamma + \cos \theta \, \delta
\end{eqnarray}
Choosing $\cos \theta = \sin \theta = \frac{1}{\sqrt{2}}$ we obtain:
\begin{equation}
\begin{aligned}
J_{0,0}&(\tau, \mu; z, w) = \frac{1}{\pi^{2}} \int \int d^{2}\gamma d^{2}\delta 
e^{\gamma^{\ast} \overline{z} - \gamma \overline{z}^{\ast}} \\ &e^{\delta^{\ast} 
\overline{w} - \delta \overline{w}^{\ast}} e^{- \tau \left( \vert \gamma \vert^{2} 
+ \vert \delta \vert^{2} \right)} e^{- \frac{1}{2} \mu \left( \delta^{2} + \delta^{\ast 2}
- \gamma^{2} - \gamma^{\ast 2} \right)}
\end{aligned}
\end{equation}
where we defined the rotated parameters:
\begin{eqnarray}
\overline{z} &=& \frac{1}{\sqrt{2}} \left( z - w \right) \\
\overline{w} &=& \frac{1}{\sqrt{2}} \left( z + w \right)
\end{eqnarray}
The integrals over $\gamma$ and $\delta$ are now uncoupled, and can be evalated
separately by using Eq.(\ref{eq:fond_int_rel}). Exploiting this integral relation, 
and applying the inverse rotation $\left\{\overline{z}, \overline{w} \right\} \rightarrow \left\{ z, w \right\}$, we obtain:
\begin{equation}
\begin{aligned}
J_{0,0}(\tau, \mu; z, w) &= \frac{1}{\tau^{2} - \mu^{2}} \exp\left[ - \frac{ \tau 
\left( \vert z \vert^{2} + \vert w \vert^{2} \right)}{\tau^{2} - \mu^{2}} \right] \\
&\exp\left[ \frac{- \mu \left( z w + z^{\ast} w^{\ast} \right)}{\tau^{2} - \mu^{2}} 
\right]
\end{aligned}
\end{equation}
The latter result allow then to explicitly evaluate the $J_{n,m}$ integrals, by starting
from $J_{0,0}$ and performing the opportune derivatives, according to 
Eq.(\ref{eq:int_deriv_jnm}).


\end{document}